\documentclass[11pt]{article}
\usepackage[a4paper,margin=27mm]{geometry}
\usepackage[T1]{fontenc}
\usepackage[utf8]{inputenc}
\usepackage{lmodern}
\usepackage{microtype}
\usepackage{amsmath,amssymb,mathtools,bm}
\usepackage{mathrsfs}
\usepackage{graphicx}
\usepackage{tikz}
\usepackage{pgfplots}
\pgfplotsset{compat=1.18}
\usepgfplotslibrary{groupplots}
\usepackage{booktabs}
\usepackage{tabularx}
\usepackage{array}
\newcolumntype{L}[1]{>{\raggedright\arraybackslash}p{#1}}
\newcolumntype{Y}{>{\raggedright\arraybackslash}X}
\usepackage{enumitem}
\usepackage[numbers,sort&compress]{natbib}
\usepackage{hyperref}
\hypersetup{hidelinks,pdftitle={Enriched Surface Thermodynamics for Sharp-Interface Fluid Models},pdfauthor={Dieter Bothe}}
\usepackage[nameinlink,noabbrev]{cleveref}
\usepackage{caption}
\usepackage{xcolor}
\usepackage{amsthm}

\newtheorem{proposition}{Proposition}
\newtheorem{result}{Result}
\newtheorem{corollary}{Corollary}
\newtheorem{remark}{Remark}
\newtheorem{definition}{Definition}
\newcommand{\dd}{\,\mathrm d}
\newcommand{\jump}[1]{[\![#1]\!]}
\newcommand{\tr}{\operatorname{tr}}
\newcommand{\divs}{\operatorname{div}_{\!\Sigma}}
\newcommand{\grads}{\nabla_{\!\Sigma}}
\newcommand{\DtS}{D_t^{\Sigma}}
\newcommand{\Vsurf}{\mathcal V^{\Sigma}}
\newcommand{\PiV}{\Pi_{\mathcal V}^{\Sigma}}
\newcommand{\cS}[1]{c_{#1}^{\Sigma}}

\newcommand{\pplus}{\left.p^+\right|_{\Sigma}}
\newcommand{\pminus}{\left.p^-\right|_{\Sigma}}
\newcommand{\Is}{\bm I_{\Sigma}}
\newcommand{\kappas}{\kappa_{\Sigma}}
\newcommand{\Ks}{K_{\Sigma}}
\newcommand{\sigmatp}{\widetilde{\sigma}^{+}}
\newcommand{\sigmatm}{\widetilde{\sigma}^{-}}

\providecommand{\bibcommenthead}{}
\title{Enriched Surface Thermodynamics\\for Sharp-Interface Fluid Models}
\author{Dieter Bothe\\[0.35em]
\small Mathematical Modeling and Analysis, Department of Mathematics,\\
\small Technical University of Darmstadt, Peter-Gr\"unberg-Stra{\ss}e 10, 64287 Darmstadt, Germany\\[0.2em]
\small \href{mailto:bothe@mma.tu-darmstadt.de}{bothe@mma.tu-darmstadt.de}}
\date{31 October 2026}

\begin{document}
\maketitle
\begin{abstract}
A sharp interface can retain material and structural information about an unresolved fluid transition zone. We formulate a local thermodynamic framework for isotropic surfaces carrying constituent surface mass densities and a thickness descriptor. The distinction between densities of additive contents and internal descriptors determines the area conjugate, reversible surface stress, and Gibbs-Duhem relation. A volume-per-area thickness has a pressure-like conjugate, a mechanical work contribution, and a separate evolution equation. Independent, constrained, and rapidly equilibrating thicknesses require different kinematics and reductions. Material storage can be retained without retaining surface inertia.

The balances and entropy production specify admissible exchange and structural relaxation. For a moving surface patch, an available-free-energy identity accounts for mechanical power and for the contributions of material and structural fluxes across its boundary. Equilibrium reduction retains the production and transport needed to follow the reduced state. An explicit mass-thickness free energy has a positive-definite material Hessian and gives coupled adsorption, expansion, compression, and dissipative relaxation. Pressure response yields conditional restrictions on reduced states and demonstrates that distinct independent and constrained states can share an entire pressure-tension curve. Geometric thickness realizations and finite-rate transfer are treated with their associated mechanical and balance requirements. The framework separates state selection, constitutive closure, reduction, and identification from observations.
\end{abstract}


\section{Introduction}
\label{sec:intro}

A fluid transition zone stores matter and energy and has an internal spatial structure. A sharp-interface model suppresses its normal spatial resolution, but this geometric simplification does not prescribe which thermodynamic information must be discarded. The mathematical support and the state carried on that support are separate modeling choices. The purpose of this paper is to formulate a systematic enrichment of the surface state and to carry each retained variable through thermodynamics, mechanics, balances, and constitutive closure.

The development considers two types of retained interfacial information. Surface mass per area represents material storage. In a mixture its constituent resolution also retains composition. A thickness descriptor represents the extent or structure of the transition zone replaced by the sharp surface. These variables represent distinct physical information. The same material content can occupy different structural extents, and the same extent can contain different amounts or compositions. Their relation is constitutive unless a stated reduction imposes a constraint. Thickness remains a field on the mathematical surface and does not require resolving a third spatial domain.

The interface is treated as an autonomous phenomenological continuum subsystem. Its state is specified at the continuum level, without requiring a unique molecular profile, diffuse-interface solution, Gibbs dividing surface, or limiting construction. More resolved theories can guide the selection of retained variables, determine their constitutive parameters, identify discarded information, or test reductions. An operational identification is necessary when comparing a state variable with a measurement, but a unique microscopic derivation is not a prerequisite for formulating its phenomenological thermodynamics.

This viewpoint has substantial antecedents. Gibbs established dividing surfaces and signed excesses \cite{Gibbs1878}. Guggenheim retained finite interfacial-region volume \cite{Guggenheim1940}. Singular-surface and nonequilibrium theories develop their own balance and gauge structures \cite{BedeauxAlbanoMazur1976,Bedeaux1986,OttingerBedeauxVenerus2009,Savin2012Gauge,BedeauxKjelstrup2018}. M\"uller, Moeckel, and Grauel formulate surfaces carrying actual matter, energy, entropy, and exchange processes \cite{Muller1985,Moeckel1975,Grauel1980,Grauel1982I,Grauel1982II,Grauel1988,Grauel1989}. Surface mass is consequently an established enrichment relative to models that omit interfacial material storage.

The immediate theoretical basis is the Class-II mixture thermodynamics of Bothe and Dreyer and subsequent sharp-interface developments \cite{BotheDreyer2015,BothePruss2010,Bothe2022,Bothe2025ClassII}. The present formulation adopts positive constituent surface amount densities, constituent momentum balances, one-sided transfer, and complete transfer cofactors from these theories. Incompressibility approximations retaining the pressure dependence of chemical potentials \cite{BotheDreyerDruet2023} and thermodynamic modeling of interfacial mass-transfer hindrance \cite{BotheTomiyama2026} provide related constitutive developments. Variational mechanics supplies an independent check on the area conjugate and reversible surface forces \cite{Gouin1978Thesis,CasalGouin1988JMech,CasalGouin1988AnnPhys,Gouin1990Mixtures,GouinRuggeri2005,CubisolGouin2009}.

Structural nonequilibrium variables have already been introduced into liquid--vapor interface thermodynamics by Struchtrup and {\"O}ttinger \cite{StruchtrupOttinger2023}. Their formulation already includes the structural conjugate in the Euler and Gibbs--Duhem relations and relates internal-variable relaxation to nonlinear transfer resistivities. The present work specializes the retained structure to material surface densities and a volume-type thickness, and follows the resulting deformation, work, and reduction rules. 

The organizing distinction is between an additive content, its density per area, and an internal descriptor with a separately prescribed deformation relation. This determines which terms enter the Euler relation and which processes require a balance or an evolution equation. The detailed scalar specialization uses a positive volume per area as a thickness coordinate. In the planar geometric realization, it equals the width of the represented layer, with a pressure-like work conjugate. A profile width defined by another convention need not have the same kinematics or the same work interpretation. No general identity equates every length-valued descriptor with geometric volume.

The formulation gives a general enriched area conjugate, a nonredundant mass/composition coordinate system with its storage balance, a structural thickness work contribution, and a coupled, strictly stable mass--thickness example with admissible local relaxation. Independent coordinates, holonomic constraints, and elimination by fast equilibration are distinguished. The available-free-energy identity for a moving surface patch includes mechanical power and the contributions of material and structural fluxes at its boundary. The reduction conditions specify the structural production required by the constraint and the power terms that must remain controlled when a state variable is eliminated. The detailed theory is local and isotropic, with scalar structural variables. Surface gradients, independent bending energies, and tensorial microstructure require further constitutive and mechanical terms.

Pressure response is used subsequently to illustrate consequences and limits of a selected state. Classical pressure invariants and excess-volume observables already have established thermodynamic interpretations \cite{Guggenheim1940,Radke2015,SchmeddingZuend2023,SantosReis2018,Sega2026}. The present application derives conditional restrictions and demonstrates that static tension data need not identify structural independence. The general framework is independent of any particular pressure-sign test or empirical reconstruction. The detailed pressure calculations, assessment of archival measurements, curved geometry, and transfer reductions are collected in appendices so that the main development follows the retained state variables.

\section{Conceptions of Fluid Interfaces}
\label{sec:conceptions}

A continuum theory of fluid interfaces requires an explicit definition of the object represented by the interface. The physical transition region, the mathematical support on which interfacial balances are written, and the thermodynamic coordinates used to represent interfacial properties need not coincide. This section specifies the state spaces and representational assumptions of the principal descriptions.

\subsection{Physical Interphases}

At molecular resolution a fluid-fluid interface is not a mathematical discontinuity. Density, composition, orientational order, stress structure, correlation functions, and other microscopic observables change over a finite transition region. That region contains actual matter, energy, entropy, and volume. Its detailed structure depends on the molecular system and on the thermodynamic state. Close to criticality or under strong confinement its extent can become comparable to other physically relevant length scales. In ordinary macroscopic two-phase flow it is usually extremely small compared with the hydrodynamic scales.

A diffuse-interface continuum model retains a representation of this normal structure. The transition is described by smoothly varying fields, often supplemented by gradient contributions to a free-energy density. Such models permit a description in which the interface is not tracked as a separate geometric object and breakup, coalescence, and changes of topology can be represented without prescribing reconnection rules \cite{AndersonMcFaddenWheeler1998,LowengrubTruskinovsky1998}. Thermodynamically consistent diffuse models for fluids of unequal density and for systems with phase transition have been developed in several forms \cite{AbelsGarckeGrun2012,AkiDreyerGiesselmannKraus2014}.

Recent surfactant-resolving developments include the conservative Allen--Cahn formulation of Liang et al.\ \cite{LiangEtAl2026}, which uses nonlocal multipliers to preserve both phase-field and surfactant contents while dissipating energy. Such resolved formulations offer comparison models for a sharp-state reduction. Their diffuse width is not thereby an independently balanced sharp-surface coordinate.

An established resolved continuum approach is the variational and gradient-capillarity formulation associated with Casal and Gouin. In that approach the interphase is described by a bulk internal energy depending on density, entropy, and their gradients, while conservative equations and natural boundary terms are generated by a d'Alembert--Lagrange or Hamiltonian variation. Gouin's foundational thesis develops the virtual-motion formulation in a geometric form and stresses that discontinuity conditions depend on the admissible variational representation. The later interface papers use the same machinery for thermocapillary fluids, capillary stresses, and surface energies \cite{Gouin1978Thesis,CasalGouin1988JMech,CasalGouin1988AnnPhys,GouinSlemrod1995}. This route is conceptually distinct from an autonomous sharp surface with its own thermodynamic state, but it demonstrates how sharp capillary forces can emerge from resolved normal structure and how variational and balance-law formulations must be checked against one another.

The physical diffuseness of an interphase does not, however, imply that a diffuse-interface continuum description is mandatory. Each continuum model applies at a specified resolution. In a diffuse theory one assumes that continuum thermodynamic fields remain meaningful within the narrow transition layer itself. In a sharp theory one instead places the unresolved interfacial physics into jump relations, constitutive surface quantities, and possibly independent surface state variables. Klika and {\"O}ttinger analyze this distinction in their comparison of sharp and diffuse nonequilibrium interface descriptions \cite{KlikaOttinger2024}.

An experimental width must likewise specify which profile is measured. Fellows et al.\ \cite{FellowsEtAl2024} measured a decay length of about $6$--$8\,\text{\AA}$ for the structural anisotropy of the air--water interface by nonlinear vibrational spectroscopy. This is an observable-specific structural length. Equating it with the volume-per-area variable introduced below requires a constitutive relation between the measured width and that variable. The measurement alone does not establish an independent thickness balance.

Asymptotic consistency imposes a further requirement. When a diffuse model is used as a representation of a transition layer that becomes small on the macroscopic scale, the limit in which the interfacial thickness tends to zero provides a stringent test of the model. Dreyer and Kraus showed for the van der Waals--Cahn--Hilliard theory that recovering the complete equilibrium sharp-interface structure requires attention not only to equality of thermodynamic potentials but also to the mechanical jump condition \cite{DreyerKraus2010}. Related studies recovered different sharp-interface limits under different scalings \cite{AkiEtAl2012,AkiDreyerGiesselmannKraus2014}. Dreyer and Guhlke formulated the corresponding admissibility criterion: the limiting interfacial conditions should be compatible with the sharp-interface balance equations and the second law \cite{DreyerGuhlke2017}. The sharp-interface limit therefore tests whether the diffuse constitutive structure admits a physically consistent macroscopic reduction.

\begin{figure}[t]
\centering
\includegraphics[width=420bp]{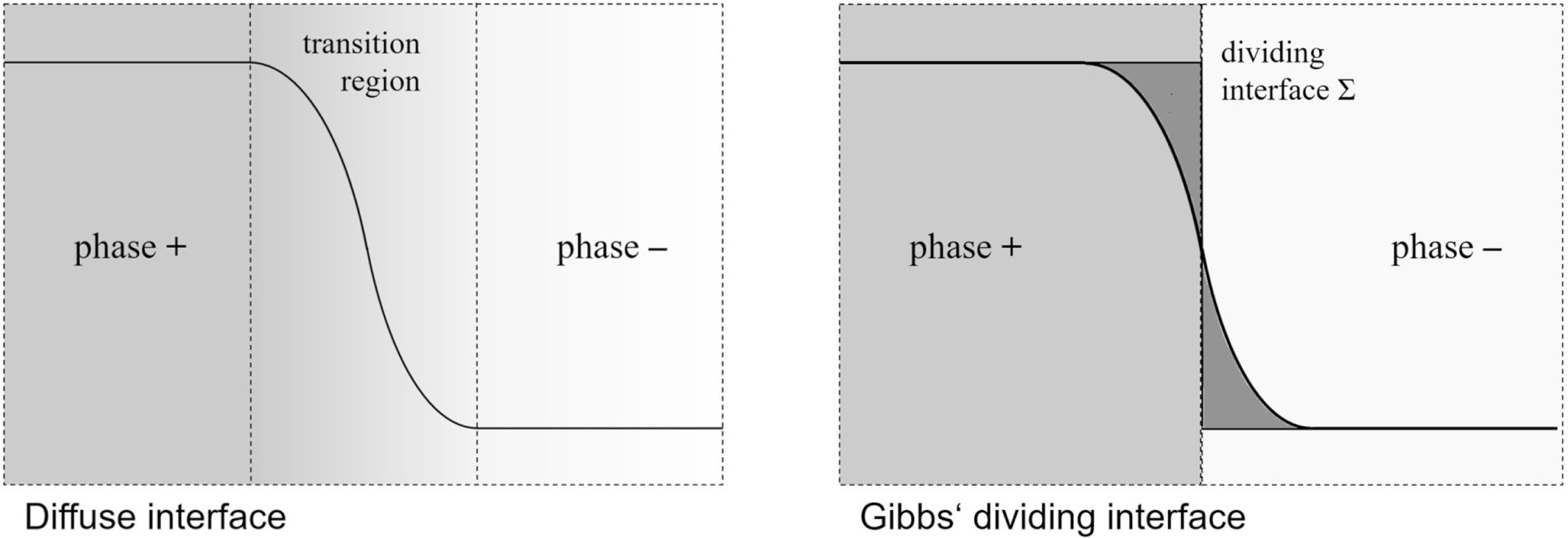}
\caption{Diffuse-interface profile and Gibbs dividing-surface representation, with the shaded surface-excess contribution. Original artwork and phase labels retained. Adapted from \cite[Fig.~2]{Bothe2022}}
\label{fig:diffuse-gibbs}
\end{figure}

The two panels of \cref{fig:diffuse-gibbs} represent distinct continuum descriptions and do not define a universal hierarchy of accuracy. Their respective adequacy depends on the observables and scales of interest, on the constitutive information available, and on the extent to which the discarded or retained interfacial structure matters for the process under consideration.

The interpretation of material surface amounts also requires consideration of molecular residence times. Here, the inventory of a constituent means its total amount in a selected physical region. A molecule that crosses a physical interphase does not in general enter and leave at the same instant. If \(t_{\mathrm{in}}\) and \(t_{\mathrm{out}}\) denote the times at which a molecule enters and leaves a selected transition region, its residence time is
\[
\tau_{\mathrm{res}}=t_{\mathrm{out}}-t_{\mathrm{in}}>0
\]
for every non-instantaneous crossing event. Therefore a transferring species can possess a positive instantaneous amount in the interphase even when it does not accumulate on a macroscopic time scale. A heuristic estimate of the surface amount density from the entry flux and residence time,
\[
\cS{i}\sim J_{i,\mathrm{in}}\tau_i^{\Sigma},
\]
uses a nonnegative gross inward molar flux per unit area \(J_{i,\mathrm{in}}\), measured in moles per square metre per second, into a selected physical region and a corresponding mean residence time. A signed net transfer rate can vanish despite a positive amount in that region. This estimate neither supplies a constitutive relation nor identifies the autonomous surface amount density with a unique integral through the physical region.

\subsection{Dividing Surfaces, Excesses, and Gauge Invariance}

The Gibbs construction replaces the finite physical interphase by a mathematical dividing surface. Homogeneous bulk states are extrapolated up to that surface, and the difference between the actual integral through the transition region and the corresponding extrapolated bulk contributions is assigned to a surface excess \cite{Gibbs1878}. For a planar interface normal to \(z\), a generic excess may be written schematically as
\[
q_G^{\Sigma}(z^{\Sigma})
=
\int_{-\infty}^{\infty}
\left[
q(z)-q^-H(z^{\Sigma}-z)-q^+H(z-z^{\Sigma})
\right]\dd z,
\]
where \(H\) is the Heaviside function, \(q^{\pm}\) are the constant extrapolated bulk values, and the difference in brackets is assumed integrable.

The excess is a difference quantity. It can therefore be positive, zero, or negative without implying negative physical matter. In multicomponent systems, this distinction prevents a quantity with units of amount per area from being identified with a material interfacial amount density solely on dimensional grounds. The nonequilibrium interfacial thermodynamics developed by Bedeaux, Albano, Mazur and coworkers retains the Gibbsian singular-surface structure and derives interfacial entropy production and phenomenological transport relations from it \cite{BedeauxAlbanoMazur1976,Kovac1977,Bedeaux1986}. Modern multicomponent formulations continue to use this excess framework while making the representation dependence explicit \cite{BedeauxKjelstrup2018}.

The location of the dividing surface is a coordinate freedom of the representation. Physical observables and invariant thermodynamic relations must remain independent of this choice.

A displacement of the dividing surface changes the numerical values of excess quantities while leaving the physical interphase unchanged. For an oriented displacement \(z^{\Sigma\prime}=z^{\Sigma}+\ell\), with \(z\) increasing from the minus phase to the plus phase, the excess defined above transforms exactly as
\[
q_G^{\Sigma\prime}
=
q_G^{\Sigma}-\ell\,\Delta q,
\qquad
\Delta q=q^- - q^+,
\]
since \(\partial q_G^{\Sigma}/\partial z^{\Sigma}=q^+-q^-\). The family of dividing surfaces therefore generates an equivalence class of surface-excess representations.

{\"O}ttinger, Bedeaux, and Venerus explicitly formulated this freedom as a gauge transformation in nonequilibrium interfacial thermodynamics \cite{OttingerBedeauxVenerus2009}. Savin and coworkers subsequently analyzed local equilibrium of the Gibbs interface from this gauge perspective and showed how invariance constrains the thermodynamic description \cite{Savin2012Gauge}. {\"O}ttinger and Venerus further developed the implications for moving interfaces, where different excess densities may lead to different associated normal velocities unless the gauge freedom is handled consistently \cite{OttingerVenerus2014}. Bedeaux and Kjelstrup used the same invariance principle to construct dividing-surface-independent intensive properties and relative adsorptions for multicomponent interfaces \cite{BedeauxKjelstrup2018}. Rauscher, {\"O}ttinger, and de Pablo later connected the gauge-invariant sharp-interface framework with nonequilibrium molecular-dynamics measurements \cite{RauscherOttingerDePablo2022}.

For two excesses \(\Gamma_A\) and \(\Gamma_B\), with \(\Delta c_B\ne0\), a relative adsorption of the form
\[
\Gamma_A^{(B)}
=
\Gamma_A
-
\Gamma_B\frac{\Delta c_A}{\Delta c_B}
\]
is invariant under the simultaneous transformations
\[
\Gamma_i^{\prime}=\Gamma_i-\ell\Delta c_i.
\]
Two distinctions are essential for the present work. First, gauge fixing is the selection of one representative from an equivalence class. It is not the selection of a new physical state. Second, a gauge transformation is not a material evolution of the interface. If a chosen dividing-surface convention is maintained during a nonequilibrium process, the gauge position itself may become time dependent, and the associated kinematic terms belong to the evolution of the representation. They must not be confused with balances of actual material surface densities.

Positive-excess conventions are useful in surface thermodynamics because they can permit familiar composition variables and equations of state \cite{LucassenReyndersVanDenTempel1967}. Nevertheless,
\[
\text{positive excess in a selected gauge}
\neq
\text{positive material inventory}.
\]
A positive excess remains a gauge-dependent difference quantity, whereas an actual material surface amount density belongs to a different state representation.

\subsection{Finite Interfacial Regions}

A finite interfacial phase provides another representation. Instead of assigning the interfacial contribution to a zero-volume surface, one selects a thin region bounded by two surfaces and treats it as a thermodynamic subsystem of finite volume. In the classical construction of Guggenheim, the interfacial region carries actual amounts and an explicit volume \cite{Guggenheim1940}. The numerical values assigned to the region depend on the chosen boundaries, but properly constructed observable combinations are boundary independent. Radke has developed an invariant comparison of Gibbs and finite-region formulations for planar fluid-fluid interfaces \cite{Radke2015}. Related surface-phase thermodynamics remains in active use. For example, Santos and Reis formulate a molar surface volume and relate it to the pressure coefficient of surface tension before defining an effective surface thickness for liquid mixtures \cite{SantosReis2018}. Such constructions reinforce the need to distinguish a finite/effective surface-phase volume from an independent structural coordinate assigned to a zero-thickness matter-carrying continuum.

\begin{figure}[t]
\centering
\includegraphics[width=420bp]{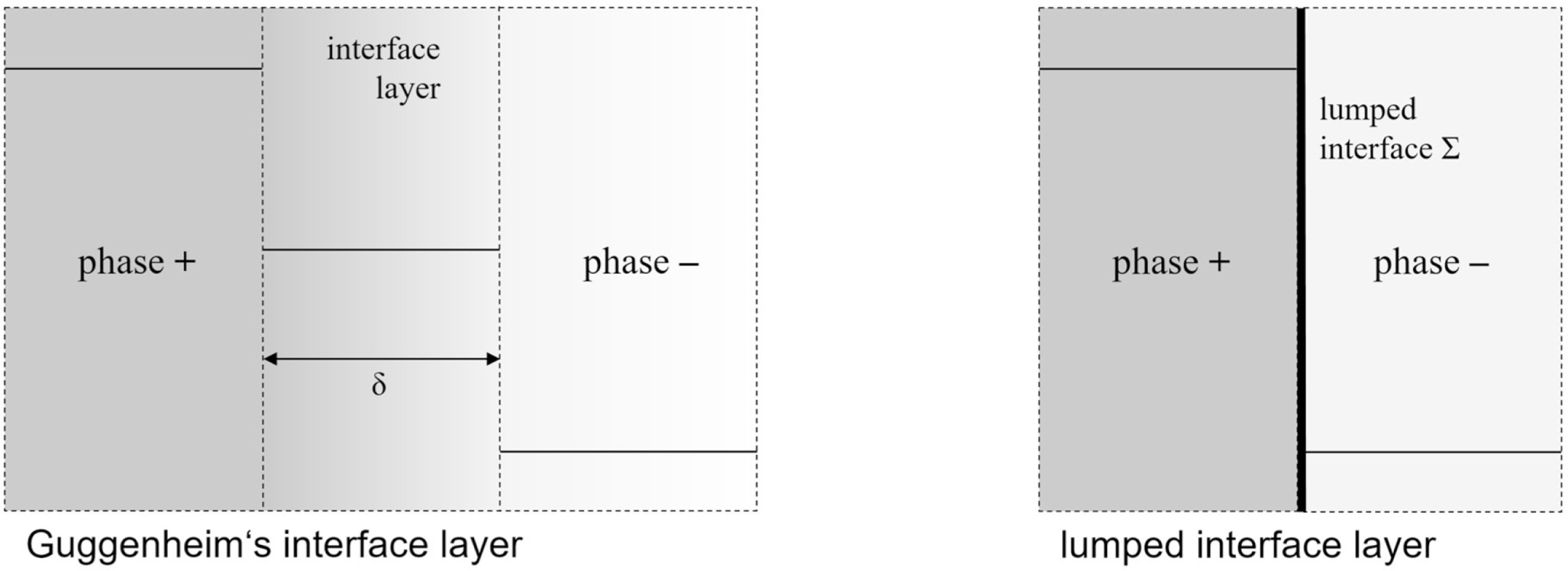}
\caption{Guggenheim's finite interfacial layer (left) and its lumped zero-thickness representation (right). The source width $\delta$ denotes the finite-layer thickness. The diagram motivates actual material surface amounts. The present sharp state remains a separately specified phenomenological subsystem. Original artwork retained. Adapted from \cite[Fig.~3]{Bothe2022}}
\label{fig:guggenheim-lumped}
\end{figure}

The finite-region viewpoint is especially instructive for the present paper because it retains two types of extensive information that will later be separated in the sharp model, namely constituent amounts and volume. If the volume of a planar interfacial region is \(V_I\) and its area is \(A\), then
\[
\frac{V_I}{A}
\]
is an area-specific volume. Its numerical value depends on how the finite region is bounded, but the existence of a volume contribution at this level of description is unambiguous.

The finite-region and Gibbs constructions should not be contrasted as competing physical theories. They are different thermodynamic representations. Their equivalence at the level of properly reduced equilibrium adsorption relations is a principal motivation for invariant formulations. The comparison also demonstrates why equality of observables must not be mistaken for equality of state coordinates.

\subsection{Matter-Carrying Sharp Interfaces}

A further conception is to treat the mathematical surface itself as a continuum subsystem endowed with its own thermodynamic fields. This idea has a substantial history in rational and continuum thermodynamics. M{\"u}ller develops surface matter, surface balances, and a thermodynamics of material surfaces within the same entropy-principle framework used for bulk continua \cite{Muller1985}. Moeckel formulates a thermodynamic theory for a surface embedded in and interacting with a three-dimensional continuum, including surface mass density, surface motion, temperature, balance equations, constitutive equations, objectivity, and an entropy inequality \cite{Moeckel1975}. Grauel subsequently develops the idea into a broad field theory for semipermeable interfaces carrying actual matter, with multicomponent surface balances, chemical reactions, diffusion, heat transport, partial momenta, surface chemical potentials, and equilibrium matching \cite{Grauel1980,Grauel1982I,Grauel1982II,Grauel1988,Grauel1989}. The M{\"u}ller--Moeckel--Grauel formulation is a principal antecedent of the phenomenological description adopted here.

Shikhmurzaev's two-layer model \cite{Shikhmurzaev1993} is another antecedent. It retains interfacial component densities and exchange structure for moving-contact-line modeling. Its representation and closure assumptions must be distinguished from an independently evolving volume-per-area coordinate. More recently, van Sluijs, Stoter, and van Brummelen \cite{VanSluijsStoterVanBrummelen2025} derived nonisothermal binary-fluid/surfactant constitutive restrictions and equilibrium conditions by a Coleman--Noll procedure. Their model retains surfactant concentration and interfacial thermodynamic storage while neglecting total surface mass in the mechanical balances and excluding carrier-fluid phase transfer. It therefore provides a useful comparison with the present constituent-resolved material storage. Neither omission of surface inertia nor retention of surfactant alone fixes the full surface state.

A second, more recent line is technically continuous with the present formulation. The bulk Class-II theory of Bothe and Dreyer uses individual constituent momenta to resolve entropy-production terms and identify physically interpretable constitutive mechanisms \cite{BotheDreyer2015}. The corresponding approach applies at sharp interfaces in the multi-velocity surface theory, where species-specific bulk and surface momenta expose one-sided momentum transmission, slip, energy transmission, and mass-transfer cofactors that can become obscured after premature reduction to a single barycentric velocity \cite{Bothe2025ClassII}. The Class-I interfacial-mass theory developed in between establishes the description by positive surface amount densities and two-step bulk--surface--bulk transfer used here \cite{Bothe2022}. Thus the present theory does not introduce interfacial mass, partial momenta, or one-sided transfer as new modeling assumptions. Its additional question is whether a reduced autonomous sharp surface must retain further structural information and mechanical work to obtain closed predictions for a specified set of observables, and, if so, whether that information is best represented by an independent additive coordinate or by a lower-dimensional constrained state.

A terminological distinction is necessary. In Grauel's work a \emph{material surface} has the strict kinematic meaning of a surface moving with the material. The semipermeable interfacial surfaces of principal interest are generally not material in that sense, because matter can cross them. They nevertheless carry actual material surface amounts. We therefore describe Grauel's construction here as a \emph{matter-carrying semipermeable sharp interface} rather than using the expression material surface without qualification. In the present paper, by contrast, the phrase \emph{material surface amount} denotes an actual non-negative amount assigned to the phenomenological surface subsystem. Its density per unit area is a local thermodynamic variable.

Grauel's finite-region-to-sharp construction is particularly instructive for state selection. He begins from a finite three-dimensional interfacial region that is part of an exact spatial decomposition and that may exchange matter, momentum, and energy with the adjoining phases. Surface quantities are then obtained from limiting normal integrals as the geometrical thickness tends to zero. In schematic form, an additive volumetric density $q_{\varepsilon}$ gives a surface density
\[
q^{\Sigma}
=
\lim_{\varepsilon\to0}
\int_{-\varepsilon/2}^{\varepsilon/2}
F_{\varepsilon}(\xi)q_{\varepsilon}(\xi)\,\dd\xi,
\]
provided the limiting integral exists and remains finite. A nonzero limit therefore corresponds to a singular concentration scaling rather than to a uniformly bounded volumetric field with uniformly bounded normal Jacobian, whose integral vanishes with the thickness. Vanishing higher normal moments requires integrability control in addition to convergence of the signed zeroth integral. A sufficient condition is
\[
\sup_{\varepsilon}\int_{-\varepsilon/2}^{\varepsilon/2}
 |F_{\varepsilon}q_{\varepsilon}|\dd\xi<\infty.
\]
For every integer \(j\ge1\), this implies
\[
\left|\int_{-\varepsilon/2}^{\varepsilon/2}
 \xi^jF_{\varepsilon}q_{\varepsilon}\dd\xi\right|
\le(\varepsilon/2)^j\int_{-\varepsilon/2}^{\varepsilon/2}
 |F_{\varepsilon}q_{\varepsilon}|\dd\xi\longrightarrow0.
\]
For nonnegative mass densities, bounded integrated mass supplies this condition. Signed quantities require their corresponding total-variation bound. These are explicit sufficient assumptions for the stated schematic reduction. The final sharp theory retains surface mass, momentum, energy, entropy, and constituent amounts but no independent geometrical thickness or normal-volume coordinate.

Grauel's construction preserves selected integrated quantities while eliminating independent normal-extent and higher-moment information. It does not retain a finite area-specific volume in the zero-thickness limit. The more resolved theory therefore identifies the information discarded by this reduction. Reintroducing a volume-type coordinate in the present phenomenological model is a new state-selection decision, not a continuation of Grauel's limiting thickness.

Grauel's multicomponent theory is also more resolved than an amount-only single-velocity surface mixture. It introduces partial surface densities and partial constituent velocities, a surface barycentric velocity, diffusion velocities, partial momentum balances, and interconstituent interaction forces. In the hierarchy used in the present work this is structurally close to a Class-II surface theory. 

Dreyer's 2003 WIAS preprint on jump conditions, reprinted in the memorial volume in 2025 \cite{Dreyer2003Jump,Dreyer2025Jump}, provides a related example. The general singular-surface balances contain surface mass, partial surface masses, internal surface energy, and surface entropy before simplifying restrictions are imposed. In particular, neglecting interfacial inertia does not force the surface internal energy or surface entropy to vanish. Mechanical reduction, storage reduction, and thermodynamic state reduction are distinct operations.

The constitutive meaning of surface tension also depends on the chosen interface conception. In the local matter-carrying theory Grauel obtains an isotropic surface-stress coefficient as a constitutive function of surface state variables. In a later specialization without additional surface density distributions he instead formulates a massless interface whose interfacial tension is tied to the adjoining bulk states. These distinct model classes demonstrate that interpretation of an equation of state for surface tension requires prior specification of the interface representation and state space.

A complementary variational result is important for the mechanics developed below. Cubisol and Gouin consider a material surface carrying an actual surfactant concentration and show, by virtual work, that the mechanically acting tension is not in general the surface-energy density itself. In their notation the surface energy and surface tension are denoted by different symbols. For the corresponding isothermal free-energy variation, the one-component relation in the notation used here is \(\sigma=a^{\Sigma}-c^{\Sigma}\mu^{\Sigma,\mathrm m}\), and their calculation extends to several attached surface species \cite{CubisolGouin2009}. The relevance of this result is its independent mechanical derivation of the surface Legendre structure and its explicit relation to normal and tangential surface forces.

The phenomenological sharp surface adopted below belongs to this broad family but is not identified with any one historical derivation. It is postulated directly as a lower-dimensional subsystem. If constituent $i$ is represented as material carried by the surface, its area-specific amount satisfies
\[
\cS{i}\ge 0.
\]
The positivity is part of the state domain. It is not inferred from a Gibbs gauge and it does not assert that a corresponding Gibbs excess must be positive.

The lumped layer in \cref{fig:guggenheim-lumped} motivates finite material surface amounts without identifying $\cS{i}$ with a unique integral of a prescribed profile. The bulk continua continue to be extrapolated to the sharp surface, while the surface measure is an independent lower-dimensional measure. This interpretation does not define an exact spatial partition. Adding the finite-region content to unmodified extrapolated bulk contents would count the same material twice. The schematic motivates the existence and sign of the surface variable without defining its quantitative relation to a resolved profile.

The state-selection question is therefore which discarded information must be restored for thermodynamic consistency and predictive closure.
\begin{table}[t]
\centering
\small
\caption{Selected interface conceptions and the thermodynamic meaning of their lower-dimensional variables. The rows describe different model objects, not an ordering of accuracy.}
\label{tab:conceptions}
\begin{tabularx}{\textwidth}{L{0.18\textwidth}L{0.20\textwidth}L{0.25\textwidth}Y}
\toprule
Conception & Geometric support & Characteristic interfacial variables & Representational issue \\
\midrule
Physical or diffuse interphase & finite three-dimensional transition region & local densities, composition, gradients, stress structure, possibly order parameters & normal structure is explicitly resolved at continuum level \\
Gibbs dividing surface & zero-volume mathematical surface & signed surface excesses & excess coordinates change under displacement of the dividing surface. Invariant combinations do not \\
Finite interfacial region & finite three-dimensional subsystem & actual amounts, entropy, energy, and volume & numerical assignments depend on the selected boundaries. Complete observables are boundary independent \\
Matter-carrying sharp interface & lower-dimensional autonomous continuum & non-negative constituent surface amount densities, surface energy, momentum and additional selected states & variables are phenomenological state coordinates, not Gibbs excesses. Their evolution is physical rather than a gauge transformation \\
Gradient-capillarity continuum & finite transition represented by nonuniform bulk fields & density and entropy gradients and their energetic conjugates & sharp quantities arise through integration, limiting arguments, or natural boundary terms \\
\bottomrule
\end{tabularx}
\end{table}

\section{Surface Thermodynamic Structure}
\label{sec:thermo}

\subsection{State Variables, Units, and Additivity}
\label{sec:notation}

We use the mass-specific convention of \cite{BotheDreyer2015,Bothe2025ClassII,BotheRuggeri2026}. Here $\rho_i$ and $\rho_i^\Sigma$ are partial mass densities, $M_i$ is the constant molar mass, $\mu_i$ and $\mu_i^\Sigma$ are chemical potentials per unit mass, and $y_i^\Sigma=\rho_i^\Sigma/\rho^\Sigma$ is a mass fraction. Amount densities satisfy $c_i^\Sigma=\rho_i^\Sigma/M_i$. A roman superscript $\mathrm m$ always marks a molar chemical potential,
\[
\mu_i^{\alpha,\mathrm m}=M_i\mu_i^\alpha
\quad(\alpha=+, -, \Sigma),
\qquad
[\mu_i^\alpha]=\mathrm{J\,kg^{-1}},\quad
[\mu_i^{\alpha,\mathrm m}]=\mathrm{J\,mol^{-1}}.
\]
The amount-coordinate pressure calculations retain this explicit marker. Constitutive hats distinguish a function from its value. They do not indicate a mass-to-mole conversion.

All temperatures are absolute and strictly positive: \(T^\Sigma>0\) and \(T^\pm>0\) wherever the corresponding state is used.

The surface free-energy and entropy densities per area are denoted $a^\Sigma$ and $\eta^\Sigma$. Where $\rho^\Sigma>0$, they correspond to the usual specific quantities through $a^\Sigma=\rho^\Sigma\psi^\Sigma$ and $\eta^\Sigma=\rho^\Sigma s^\Sigma$. The area-density notation also applies to models omitting material storage, where division by $\rho^\Sigma$ is unavailable. The surface pressure is $p^\Sigma=-\sigma$ and the reversible isotropic stress is $\bm S^{\Sigma,\mathrm{rev}}=\sigma\bm I_\Sigma$. The structural conjugate $\Pi_{\mathcal V}^\Sigma$ has pressure units, Pa, whereas $p^\Sigma$ has units N\,m$^{-1}$.

The normal $\bm n_\Sigma$ points from the minus phase to the plus phase. We use $\bm I_\Sigma=\bm I-\bm n_\Sigma\otimes\bm n_\Sigma$ and $\kappa_\Sigma=-\divs\bm n_\Sigma$. The one-sided outward bulk normals are $\bm n^-=\bm n_\Sigma$ and $\bm n^+=-\bm n_\Sigma$. Positive $\dot m_i^{\alpha,\Sigma}$ denotes transfer from bulk $\alpha$ into the surface. These conventions fix the signs of the mechanical and exchange relations below.

\label{sec:general-enrichment}

Let \(\Sigma(t)\) be the selected sharp support. A state enrichment retains information about the represented interface by introducing additional independent thermodynamic coordinates. Replacing one coordinate system by another does not itself enrich the state. A prescribed parameter in an equation of state likewise supplies no evolution equation for an additional state variable.

For a uniform patch of area \(A\), consider the local scalar state
\[
a^\Sigma=\widehat a^\Sigma
 (T^\Sigma,\bm c^\Sigma,\bm q^\Sigma,\bm\xi^\Sigma),
\qquad c_i^\Sigma=N_i^\Sigma/A,\qquad q_\alpha^\Sigma=Q_\alpha^\Sigma/A.
\]
The totals \(N_i^\Sigma\) and \(Q_\alpha^\Sigma\) add over disjoint patches. For a heterogeneous interface the free energy is the integral of the local density. The single expression \(F^\Sigma=Aa^\Sigma\) applies to uniform states. The descriptors \(\xi_a^\Sigma\) are not densities of specified additive contents. All quantities must have an admissible domain, and all constraints must be imposed before independent variations are counted.

Define
\[
\eta^\Sigma=-a^\Sigma_{T},\qquad
\mu_{i}^{\Sigma,\mathrm m}=a^\Sigma_{c_i},\qquad
\varpi_\alpha^\Sigma=a^\Sigma_{q_\alpha},\qquad
\mathcal A_a^\Sigma=-a^\Sigma_{\xi_a}.
\]
The subscripts denote partial derivatives holding the other independent coordinates fixed. The free energy is assumed to be \(C^2\) on the interior of the constitutive state domain wherever mixed derivatives or Hessians are used. Thus
\[
\dd a^\Sigma=-\eta^\Sigma\dd T^\Sigma
+\sum_i\mu_{i}^{\Sigma,\mathrm m}\dd c_i^\Sigma
+\sum_\alpha \varpi_\alpha^\Sigma\dd q_\alpha^\Sigma
-\sum_a\mathcal A_a^\Sigma\dd\xi_a^\Sigma.
\]
Positivity is determined by physical interpretation. Actual constituent surface amount densities are nonnegative. A signed excess is a different object. Additivity alone does not impose positivity on an arbitrary structural content.

\subsection{Area Conjugacy and Consistent Enrichment}
\label{app:additive}

An isothermal area variation at fixed additive totals gives
\[
\delta c_i^\Sigma=-c_i^\Sigma\,\delta A/A,
\qquad \delta q_\alpha^\Sigma=-q_\alpha^\Sigma\,\delta A/A.
\]
Holding the descriptors \(\bm\xi^\Sigma\) fixed as well, differentiation of \(Aa^\Sigma\) yields
\begin{equation}
\sigma=a^\Sigma-\sum_i c_i^\Sigma\mu_{i}^{\Sigma,\mathrm m}
-\sum_\alpha q_\alpha^\Sigma \varpi_\alpha^\Sigma.
\label{eq:general-enriched-euler}
\end{equation}
For an isotropic reversible local fluid surface this is the scalar area conjugate and the coefficient of its tangential surface stress. It is generally different from \(a^\Sigma\). Differentiating it gives
\begin{equation}
\dd\sigma=-\eta^\Sigma\dd T^\Sigma
-\sum_i c_i^\Sigma\dd\mu_{i}^{\Sigma,\mathrm m}
-\sum_\alpha q_\alpha^\Sigma\dd \varpi_\alpha^\Sigma
-\sum_a\mathcal A_a^\Sigma\dd\xi_a^\Sigma.
\label{eq:general-enriched-gd}
\end{equation}

Descriptor kinematics changes the area conjugate. If an admissible area deformation instead induces
\(\delta\xi_a^\Sigma=b_a^\Sigma\delta A/A\), the coefficient of that variation acquires the additional contribution
\(-\sum_a\mathcal A_a^\Sigma b_a^\Sigma\). A descriptor without an additive total can therefore contribute to mechanical work. Its transport and deformation rule must be specified. Gradient-dependent or curvature-dependent energies require further virtual-work terms. Equation~\eqref{eq:general-enriched-euler} is not their complete stress law.

For each proposed variable, the model must specify its represented physical information and units, independent coordinates and admissible domain, transformation under surface motion, free-energy conjugate, and balance or evolution equation. It must also specify the external work or exchange that drives the variable and a constitutive closure consistent with entropy production and the state domain. These requirements apply whether the variable is motivated by a molecular model, a finite region, a measurement, or a direct continuum hypothesis.

A retained coordinate can be eliminated in different ways. A holonomic constraint restricts the admissible variations and changes the reduced derivatives. Fast equilibration eliminates a coordinate by a stationarity condition and generally changes the remaining susceptibilities through a Schur complement. A frozen variable remains a parameter of a specified time-scale reduction. These operations can give different reduced tensions and kinetics even if their equilibrium predictions coincide at one state.

\subsection{Surface Transport, Balances, and Dissipation}
\label{sec:transport-kinematics}

The evolving support determines a normal speed \(V_\Sigma\) and the purely normal representative \(\bm w_\Sigma=V_\Sigma\bm n_\Sigma\). A transport field \(\bm u\) on the support must satisfy \((\bm u-\bm w_\Sigma)\cdot\bm n_\Sigma=0\). Its tangential component is an additional modeling choice. If positive material surface mass is retained, \(\bm u\) may be chosen as the surface barycentric velocity \(\bm v^\Sigma\). Structural content need not be convected by the same field. For a structural transport field \(\bm u_q\) with the same normal speed, the flux relative to \(\bm u\) is, for example,
\[
\bm J_q=q(\bm u_q-\bm u)+\bm J_q^{\mathrm{diff}},
\qquad \bm J_q\cdot\bm n_\Sigma=0.
\]
A common convective field and vanishing structural drift are constitutive restrictions rather than consequences of additivity. If material storage is omitted, the structural transport field must be specified independently. The barycentric velocity is undefined when the corresponding mass density vanishes. These choices also determine the fluxes relative to a moving patch boundary.

\Cref{cor:structural-flux-reference} gives the accompanying transformation of interior and boundary power when the structural flux is expressed relative to another compatible transport field.

We use $\partial_t^\Sigma$ for differentiation along the purely normal motion, as in \cite{Bothe2025ClassII}. Thus $D_t^\Sigma f=\partial_t^\Sigma f+\bm v_\parallel^\Sigma\cdot\grads f$ denotes the surface material derivative when the transport field is $\bm v^\Sigma$. For a different compatible field $\bm u$, replace $\bm v_\parallel^\Sigma$ by $\bm u_\parallel$. Geometric motion and tangential material transport are thereby kept distinct.

Spatial additivity does not imply conservation. An additive structural density can have a source and a tangential flux:
\[
\DtS q_\alpha^\Sigma+q_\alpha^\Sigma\divs\bm v^\Sigma
+\divs\bm j_\alpha^\Sigma=r_\alpha^\Sigma.
\]
A non-additive descriptor can instead be assigned a deformation and relaxation rule, for example
\[
\DtS\xi_a^\Sigma=b_a^\Sigma\divs\bm v^\Sigma+R_a^\Sigma.
\]
Here the specified \(b_a^\Sigma\) produces the reversible area-work correction described above, and \(R_a^\Sigma\) is the independent internal rate. This is a local scalar kinematics. Additional transport mechanisms require their own terms.

For an isothermal homogeneous setting, suppose external work on the retained additive modes has the rate \(\sum_\alpha\ell_\alpha^\Sigma r_\alpha^\Sigma\), with \(\ell_\alpha^\Sigma\) defined in this signed work-on-the-system convention. With no additional external driving of the descriptors, subtracting the reversible free-energy storage gives the structural contribution
\begin{equation}
T^\Sigma\zeta_{\mathrm{str}}
=\sum_\alpha(\ell_\alpha^\Sigma-\varpi_\alpha^\Sigma)r_\alpha^\Sigma
+\sum_a\mathcal A_a^\Sigma R_a^\Sigma.
\label{eq:general-enrichment-dissipation}
\end{equation}
The dilation terms are already in the area conjugate and must not be counted again. For volume-type thickness, \(\varpi=-\PiV\) and \(\ell=-\Pi_{\mathrm{load}}\), so this formula yields \((\PiV-\Pi_{\mathrm{load}})r_h\). A structural coordinate without external work has \(\ell=0\). At nonuniform state, tangential transport introduces additional flux--gradient contributions. Equation~\eqref{eq:general-enrichment-dissipation} states only the homogeneous structural part, not the complete entropy production.

A positive-semidefinite force--rate closure supplies admissible local kinetics. Preservation of its admissible state domain and satisfaction of its constraints must still be checked, and cross-coupling with constituent exchange requires the full set of forces. Thus every enrichment requires both a thermodynamic coordinate and a specified set of transport, exchange, and evolution processes. The mass and thickness examples below implement this general rule explicitly.

\subsection{Mass--Thickness Thermodynamics and Surface Tension}

The two examples developed below retain constituent amounts and a volume-type structural content,
\[
\bm z^\Sigma=(T^\Sigma,\bm c^\Sigma,\Vsurf,\bm\xi^\Sigma),
\qquad \Vsurf=V^\Sigma/A>0.
\]
For a planar geometric realization we write \(h=\Vsurf\), so that \(h\) represents transition-zone thickness. More general thickness descriptors are discussed separately in \cref{sec:thickness-meaning}. Define
\[
\PiV=-\frac{\partial a^\Sigma}{\partial\Vsurf}.
\]
Then
\[
\dd a^\Sigma=-\eta^\Sigma\dd T^\Sigma
+\sum_i\mu_{i}^{\Sigma,\mathrm m}\dd c_i^\Sigma
-\PiV\dd\Vsurf-\sum_a\mathcal A_a^\Sigma\dd\xi_a^\Sigma.
\]
The pressure-like convention gives the intrinsic expansion term \(-\PiV\,\dd\Vsurf\). A positive external compressive load opposes expansion. Bulk pressure is not a primitive argument of this autonomous surface free energy. A mechanical realization determines how the structural conjugate is related to the surrounding tractions.

The total free energy of a uniform surface patch is
\[
F^{\Sigma}
=
A\,a^{\Sigma}
\left(
T^{\Sigma},\frac{N_1^{\Sigma}}A,\ldots,
\frac{N_N^{\Sigma}}A,
\frac{V^{\Sigma}}A,\bm\xi^{\Sigma}
\right).
\]
The scalar mechanical tension of an isotropic reversible fluid surface is the area conjugate at fixed temperature and fixed additive surface contents,
\[
\sigma
=
\left.
\frac{\partial F^{\Sigma}}{\partial A}
\right|_{T^{\Sigma},N_i^{\Sigma},V^{\Sigma},\bm\xi^{\Sigma}}.
\]
Direct differentiation yields
\[
\sigma
=
a^{\Sigma}
-\sum_i c_i^{\Sigma}\mu_{i}^{\Sigma,\mathrm m}
+\PiV\Vsurf .
\]
This intrinsic surface Legendre relation is independent of any direct constitutive dependence on bulk pressure.

Introduce the intrinsic thermodynamic surface pressure \(p^{\Sigma}\) by
\[
p^{\Sigma}=-\sigma.
\]
The Euler relation is therefore
\[
a^{\Sigma}+p^{\Sigma}+\PiV\Vsurf
=
\sum_i c_i^{\Sigma}\mu_{i}^{\Sigma,\mathrm m}.
\]
For an amount-only surface the \(\PiV\Vsurf\) term is absent. More generally, every additional additive surface density contributes its own Euler term, whereas descriptors held fixed in this area variation do not acquire dilution terms.

The reversible surface stress tensor is
\[
\bm S^{\Sigma}_{\mathrm{rev}}
=-p^{\Sigma}\Is
=\sigma\Is,
\]
where
\[
\Is=\bm I-\bm n_\Sigma\otimes\bm n_\Sigma
\]
is the tangential identity. A general surface-fluid model may add viscous or otherwise dissipative surface stress, for example in a Boussinesq--Scriven surface-fluid closure \cite{BothePruss2010}. The scalar \(\sigma\) is then the isotropic reversible part, not the complete surface stress tensor.

The differential consequence of the Euler relation is
\[
\dd\sigma
=
-\eta^{\Sigma}\dd T^{\Sigma}
-\sum_i c_i^{\Sigma}\dd\mu_{i}^{\Sigma,\mathrm m}
+\Vsurf\dd\PiV
-\sum_a\mathcal A_a^{\Sigma}\dd\xi_a^{\Sigma}.
\]
On an equilibrium manifold the affinities of independently relaxing internal variables vanish. The formula distinguishes the surface free-energy density from the mechanical surface tension.

The mixed derivatives also give constitutive couplings before any kinetic closure is introduced. For example,
\[
\frac{\partial\mu_{i}^{\Sigma,\mathrm m}}{\partial\Vsurf}
=
-\frac{\partial\PiV}{\partial c_i^{\Sigma}}.
\]
A nonzero mixed derivative couples composition and structural volume thermodynamically even for diagonal kinetic coefficients. A separable free energy gives the uncoupled special case.

Notation for fluid surfaces varies among disciplines. The present paper therefore defines each physical quantity explicitly. IUPAC accepts both \(\gamma\) and \(\sigma\) as symbols for surface or interfacial tension of fluid phases and defines the static quantity through reversible surface work \cite{IUPACColloid1972,IUPACSurfaceTerms2025}. Consequently, a chemical-surface-science paper using \(\gamma\) and a continuum-mechanics paper using \(\sigma\) may be referring to the same scalar physical tension. The identification depends on the definitions in the source, since these symbols also denote surface free energies, stress components, and other quantities.

The chemical-surface-science term \emph{surface pressure} requires special care. For a covered fluid interface, the IUPAC convention is the decrease of interfacial tension relative to a reference clean interface,
\[
\pi^{\mathrm s}
=\gamma^0-\gamma.
\]
In the notation of the present paper, when the literature \(\gamma\) is the same scalar fluid tension as our \(\sigma\),
\[
\pi^{\mathrm s}
=\sigma_0-\sigma
=p^{\Sigma}-p_0^{\Sigma}.
\]
Thus \(\pi^{\mathrm s}\) is a reference-state difference. It is not the intrinsic thermodynamic surface pressure \(p^{\Sigma}\) itself \cite{IUPACSurfaceTerms2025}.

The term \emph{surface stress} is likewise not synonymous with surface free-energy density. IUPAC defines surface stress mechanically, and for solid surfaces the familiar strain dependence of the surface free energy leads to a distinction between surface free energy and surface stress of Shuttleworth type \cite{IUPACSurfaceTerms2025,Shuttleworth1950,KramerWeissmuller2007}. The present paper concerns fluid interfaces. Nevertheless, the solid-surface example demonstrates that identical dimensions of energy per area do not establish equality between a free-energy density and a mechanical stress.

The same distinction arises in fluid surface thermodynamics when the surface carries conserved matter. Cubisol and Gouin use \(\sigma\) for the superficial energy and \(\gamma\) for the superficial tension. For a single attached surfactant concentration \(c\) they obtain, in their notation, a Legendre-transform relation. Translating the objects rather than the symbols into the present notation gives
\[
\sigma_{\mathrm{ours}}
=a^{\Sigma}-c^{\Sigma}\mu^{\Sigma,\mathrm m}.
\]
They emphasize that a small change in surfactant concentration can strongly change the mechanical tension while the surface-energy density changes little \cite{CubisolGouin2009}. We therefore retain \(\sigma\) exclusively for our scalar fluid surface tension and use \(\gamma\) only when discussing literature that uses that symbol.

\begin{table}[t]
\centering
\small
\caption{Notation used for the principal surface and bulk mechanical quantities.}
\label{tab:notation}
\begin{tabularx}{\textwidth}{L{0.14\textwidth}L{0.27\textwidth}Y}
\toprule
Symbol & Object & Meaning and exclusions \\
\midrule
\(a^{\Sigma}\) & surface Helmholtz free-energy density & intrinsic constitutive scalar per area. It is not generally equal to surface tension \\
\(p^{\Sigma}\) & intrinsic thermodynamic surface pressure & defined by \(p^{\Sigma}=-\sigma\) for an isotropic reversible fluid surface. It is neither a bulk pressure nor the monolayer surface-pressure increment \\
\(\sigma\) & scalar fluid surface tension & isotropic reversible area conjugate and coefficient of \(\Is\) in \(\bm S^{\Sigma}_{\mathrm{rev}}\) \\
\(\gamma\) & literature symbol & used only when translating sources that denote fluid surface tension by \(\gamma\) \\
\(\pi^{\mathrm s}\) & surface-pressure increment & reference difference \(\sigma_0-\sigma\) used in chemical surface science \\
\(\PiV\) & structural pressure conjugate & thermodynamic conjugate of the additive surface volume \(\Vsurf\) \\
\(\pminus,\pplus\) & one-sided bulk thermodynamic pressure traces & enter bulk chemical potentials. They are generally unequal at a curved equilibrium interface \\
\(\pi_n^-,\pi_n^+\) & one-sided normal mechanical compressions & defined from the full bulk Cauchy stresses. They equal the corresponding thermodynamic pressure only in a simple isotropic equilibrium bulk state \\
\(P\) & bulk mechanical pressure & reserved exclusively for \(P=-\tfrac13\tr\bm S\). No other use of capital \(P\) is made \\
\bottomrule
\end{tabularx}
\end{table}

\subsection{Material Stability and Relaxed Susceptibility}

Each additional coordinate requires an assessment of thermodynamic admissibility. The conditions in this subsection concern the primitive enriched model in which \(V^{\Sigma}\) is an independently variable additive coordinate. A later holonomic amount--volume constraint restricts the admissible tangent space further and must be tested using the corresponding constrained/open-system second variation rather than by reusing the primitive structural variation. A \emph{constrained} convexity test must be distinguished from full equilibrium stability. After restricting to the independent additive coordinates permitted by any composition constraints, let
\[
\bm q^{\Sigma}=(c_1^{\Sigma},\ldots,c_N^{\Sigma},\Vsurf).
\]
At an interior state, with \(T^{\Sigma}\) and the non-additive internal variables \(\bm\xi^{\Sigma}\) fixed, a necessary local single-phase stability condition is
\[
\bm H_{qq}^{\Sigma}
=
\frac{\partial^2 a^{\Sigma}}
{\partial\bm q^{\Sigma}\partial\bm q^{\Sigma}}
\succeq0
\]
on the admissible tangent subspace. Define
\[
\bm M^{\Sigma}
=
\left(\frac{\partial\mu_{i}^{\Sigma,\mathrm m}}{\partial c_j^{\Sigma}}\right),
\qquad
b_i
=
\frac{\partial\mu_{i}^{\Sigma,\mathrm m}}{\partial\Vsurf}
=-\frac{\partial\PiV}{\partial c_i^{\Sigma}}.
\]
Then
\[
\bm H_{qq}^{\Sigma}
=
\begin{pmatrix}
\bm M^{\Sigma} & \bm b\\
\bm b^{\mathsf T} & -\dfrac{\partial\PiV}{\partial\Vsurf}
\end{pmatrix}
\succeq0.
\]
In particular,
\[
\left(\frac{\partial\PiV}{\partial\Vsurf}\right)_{T^{\Sigma},\bm c^{\Sigma},\bm\xi^{\Sigma}}
\le0,
\]
so an invertible constrained structural equation of state has the expected positive compressibility in the sense that increasing the compressive conjugate decreases \(\Vsurf\). If \(\bm M^{\Sigma}\) is positive definite, the Schur complement gives the stronger restriction
\[
-\frac{\partial\PiV}{\partial\Vsurf}
\ge
\bm b^{\mathsf T}(\bm M^{\Sigma})^{-1}\bm b.
\]
Thus composition--structure coupling cannot be selected independently of the constrained structural stiffness.

These inequalities are not, by themselves, a complete stability theorem when internal variables are allowed to relax. At thermodynamic equilibrium the corresponding affinities must first vanish,
\[
\bm{\mathcal A}^{\Sigma}=\bm0,
\]
and the complete second variation with respect to the admissible variations of \((\bm q^{\Sigma},\bm\xi^{\Sigma})\) must be non-negative. If the internal-variable block
\[
\bm H_{\xi\xi}^{\Sigma}
=
\frac{\partial^2a^{\Sigma}}{\partial\bm\xi^{\Sigma}\partial\bm\xi^{\Sigma}}
\]
is positive definite and the internal variables relax locally to \(\bm\xi_*^{\Sigma}(T^{\Sigma},\bm q^{\Sigma})\), then the Hessian of the relaxed free energy is the Schur complement
\[
\bm H_{\mathrm{rel}}^{\Sigma}
=
\bm H_{qq}^{\Sigma}
-
\bm H_{q\xi}^{\Sigma}
(\bm H_{\xi\xi}^{\Sigma})^{-1}
\bm H_{\xi q}^{\Sigma}.
\]
With \(\bm H_{\xi\xi}^{\Sigma}\succ0\), nonnegativity of this Schur complement is equivalent to nonnegativity of the complete material--internal-variable Hessian. It does not follow from positivity of the internal-variable block alone.
Accordingly, the structural response measured on a time scale on which \(\bm\xi^{\Sigma}\) is frozen is governed by the constrained derivative above, whereas the fully equilibrated response is governed by the corresponding derivative of the relaxed free energy. Loss of the relevant material-coordinate convexity signals a constitutive surface instability, phase separation/spinodal behavior, or the need for a different state description. These Hessian tests concern interior states at fixed local area and geometry. If boundary states are admitted, stability requires the corresponding one-sided constrained variations and first-order conditions.

The block test assumes affine restrictions on the additive densities. A nonlinear material constraint requires the second variation of the constrained loaded potential, including the corresponding constraint-curvature terms. See \cref{eq:holonomic-second-stable}. Stability with respect to area or shape perturbations is a different second-variation problem and must include the capillary area term, bulk mechanical constraints, curvature dependence, and any additional geometric energies. It is not implied by the material-state convexity conditions above.

\subsection{Variational Surface Mechanics}

Size- and shape-dependent interface thermodynamics also requires a distinction between integral and differential surface tensions. Chassagne, Bedeaux, and Kjelstrup \cite{ChassagneBedeauxKjelstrup2026} formulate this distinction using Hill's subdivision potential and obtain a common setting for Shuttleworth and Lippmann relations. That finite-system description addresses a different extensivity assumption. The variation below assumes a local additive surface free energy and does not include a subdivision potential or an independent curvature energy.

The preceding area derivative can be recovered without assuming a finite-patch homogeneity argument. Let \(\bm\zeta\) be a virtual surface displacement. Then
\[
\delta(\dd A)
=\divs\bm\zeta\,\dd A.
\]
Take a material virtual transport preserving every infinitesimal constituent amount and structural content, \(\delta(c_i^{\Sigma}\dd A)=0\) and \(\delta(\Vsurf\dd A)=0\), with temperature and internal descriptors fixed. No independent redistribution is included in this geometric mode. Then
\[
\delta c_i^{\Sigma}
=-c_i^{\Sigma}\divs\bm\zeta,
\qquad
\delta\Vsurf
=-\Vsurf\divs\bm\zeta.
\]
Varying
\[
F^{\Sigma}=\int_{\Sigma}a^{\Sigma}\,\dd A
\]
therefore gives the purely geometric contribution
\[
\delta F^{\Sigma}
=\int_{\Sigma}
\left(
 a^{\Sigma}-\sum_i c_i^{\Sigma}\mu_{i}^{\Sigma,\mathrm m}+\PiV\Vsurf
\right)
\divs\bm\zeta\,\dd A.
\]
Hence the coefficient of area change is again \(\sigma\).

Surface integration by parts gives
\[
\int_{\Sigma}\sigma\divs\bm\zeta\,\dd A
=
-\int_{\Sigma}\divs(\sigma\Is)\cdot\bm\zeta\,\dd A
+\int_{\partial\Sigma}\sigma\bm\nu_{\Sigma}\cdot\bm\zeta\,\dd l.
\]
Thus the same scalar \(\sigma\) produces both the normal capillary force and the tangential Marangoni force. This is the direct mechanical content of the distinction between \(a^{\Sigma}\) and \(\sigma\).

This calculation follows the logic of the d'Alembert--Lagrange surface variations used by Gouin and collaborators \cite{Gouin1978Thesis,CubisolGouin2009,Gouin2009CR}. Its role here is a consistency check and a mechanical interpretation. It does not define the phenomenological state and, alone, it does not determine the mechanical conjugate of an independent change of \(\Vsurf\). Determination of that conjugate requires an additional structural virtual mode, developed in \cref{sec:volume}.
\section{Material Enrichment: Surface Mass and Composition}
\label{sec:mass-enrichment}

Surface mass per area is a first enrichment relative to a sharp-interface model that neglects interfacial material storage. For constituent masses \(m_i^\Sigma\), set
\[
\rho_i^\Sigma=m_i^\Sigma/A\ge0,
\qquad \rho^\Sigma=\sum_i\rho_i^\Sigma.
\]
These partial surface mass densities represent actual material assigned to the autonomous surface subsystem, rather than Gibbs excesses. Their introduction is established in the matter-carrying sharp-interface theories cited above \cite{Muller1985,Moeckel1975,Grauel1980,Bothe2022,Bothe2025ClassII}. The present formulation makes these modeling assumptions explicit and couples material storage consistently to structural enrichment.

On the interior \(\rho^\Sigma>0\), an equivalent coordinate choice is
\[
(\rho_1^\Sigma,\ldots,\rho_N^\Sigma)
\quad\longleftrightarrow\quad
(\rho^\Sigma,y_1^\Sigma,\ldots,y_{N-1}^\Sigma),
\qquad y_i^\Sigma=\rho_i^\Sigma/\rho^\Sigma,
\quad y_N^\Sigma=1-\sum_{i<N}y_i^\Sigma.
\]
For constant molar masses $M_i$, the relation $\rho_i^\Sigma=M_i c_i^\Sigma$ shows that total mass is not an additional independent variable when all constituent amounts have been retained. The chemical potentials in mass and amount coordinates satisfy
\[
\mu_i^\Sigma=\frac{\partial a^\Sigma}{\partial\rho_i^\Sigma},
\qquad
\mu_i^{\Sigma,\mathrm m}=\frac{\partial a^\Sigma}{\partial c_i^\Sigma}
=M_i\mu_i^\Sigma,
\qquad
\sum_i\rho_i^\Sigma\mu_i^\Sigma
=\sum_i c_i^\Sigma\mu_i^{\Sigma,\mathrm m}.
\]
The units are J\,kg$^{-1}$ for $\mu_i^\Sigma$ and J\,mol$^{-1}$ for $\mu_i^{\Sigma,\mathrm m}$. All derivatives hold temperature and the other independent structural and material coordinates fixed. Thus the Euler relation is unchanged by the coordinate conversion.

At fixed structure and temperature, the mass part of the differential becomes
\[
\sum_i\mu_i^\Sigma\dd\rho_i^\Sigma
=\left(\sum_iy_i^\Sigma\mu_i^\Sigma\right)\dd\rho^\Sigma
+\rho^\Sigma\sum_{i<N}(\mu_i^\Sigma-\mu_N^\Sigma)\dd y_i^\Sigma.
\]
Total storage and composition are distinct coordinates. Only \(N-1\) mass fractions are independent. This form makes the composition constraint explicit without introducing an extra thermodynamic degree of freedom.

Let \(\bm v^\Sigma\) be the surface barycentric velocity and \(\bm j_i^\Sigma\) the tangential diffusive mass fluxes, with \(\sum_i\bm j_i^\Sigma=0\). Let \(\dot m_i^{\pm,\Sigma}\) denote signed mass supplies from each adjoining phase into the surface, positive for inward transfer, and let \(r_i^\Sigma\) be the reaction production. Then
\begin{equation}
\DtS\rho_i^\Sigma+\rho_i^\Sigma\divs\bm v^\Sigma
+\divs\bm j_i^\Sigma=\dot m_i^{-,\Sigma}+\dot m_i^{+,\Sigma}+r_i^\Sigma.
\label{eq:mass-enrichment-balance}
\end{equation}
Mass-conserving reactions satisfy \(\sum_i r_i^\Sigma=0\). Summation gives
\[
\DtS\rho^\Sigma+\rho^\Sigma\divs\bm v^\Sigma
=\sum_i(\dot m_i^{-,\Sigma}+\dot m_i^{+,\Sigma}).
\]
Area dilation changes the mass per area, tangential diffusion redistributes constituents, and one-sided exchange changes the stored constituent amounts. A surface can thus store material even when its velocity is not that of either bulk phase. Global mass conservation requires opposite supplies in the corresponding bulk balances and compatible boundary fluxes. An exact bulk--surface partition requires a definition independent of the finite-layer schematic.

Neglecting surface inertia is a reduction of the momentum balance. It does not set the densities in equation~\eqref{eq:mass-enrichment-balance} to zero and does not remove their free-energy dependence. Likewise, assuming fast exchange replaces finite-rate transfer relations by appropriate equilibrium constraints. It does not identify the remaining densities with arbitrary excesses or constants.

Neglecting the inertial term does not, by itself, remove the momentum carried by active material transfer. Omitting that contribution from a reduced traction balance requires a separate estimate relative to the retained mechanical forces.

With no structural coordinate and no deforming internal descriptor, the local fluid-surface relation is
\[
\sigma=a^\Sigma-\sum_i\rho_i^\Sigma\mu_i^\Sigma.
\]
An adsorption or composition change can therefore alter tension through the constitutive state. Determination of the densities requires balances, initial data, and transfer or equilibrium conditions in addition to the Euler relation. Retaining thickness modifies the area conjugate and introduces additional mechanical and kinetic processes.

The admissible domain of the surface partial mass densities must be preserved by finite-rate closures. Linear constitutive relations for exchange and diffusion are local interior approximations unless their extension to the boundary is specified. Physical material storage, independently chosen surface inertia, and an operationally defined measurement of adsorption should therefore be kept distinct throughout reduction and calibration.

\section{Structural Enrichment: Transition-Zone Thickness}
\label{sec:structural}
\label{sec:volume}

\subsection{Thickness Coordinates and State Reductions}
\label{sec:thickness-meaning}

The normal extent of a transition zone is unresolved in a sharp-interface description. Retaining a thickness field restores selected information about that extent while preserving the lower-dimensional support. Such a field is a thermodynamic descriptor of the represented zone. It is not the geometric thickness of the mathematical surface itself.

There are at least two different realizations. A volume per current surface area coordinate \(\Vsurf=V^\Sigma/A\) retains an additive structural content. In a planar layer interpretation, \(h=\Vsurf\) is its thickness. A profile width, defined for example by a specified measure of variation across a resolved profile, can instead be retained as an internal variable \(\xi\). That choice needs its own transport, deformation, and mechanical-work relations.

Throughout the subsequent local fluid-surface formulas, internal descriptors are held fixed in reversible area variations unless their deformation relation is explicitly retained. The common unit of length does not equate the two models. A change of coordinate normalization must preserve every conjugate work product.

The principal realization developed here uses \(h=V^\Sigma/A>0\) locally in planar geometry. Together with mass storage it gives
\[
a^\Sigma=\widehat a^\Sigma(T^\Sigma,\bm\rho^\Sigma,h),
\qquad
\PiV=-a^\Sigma_h,
\qquad
\sigma=a^\Sigma-\sum_i\rho_i^\Sigma\mu_i^\Sigma+h\PiV.
\]
The ratio \(\rho^\Sigma/h\) is an effective density of the represented material in this realization. Its interpretation as a measured average requires a consistent choice of represented mass and width. The theory does not automatically identify it with either adjacent bulk density.

For a curved reference surface, volume per area and the sum of the two normal extents differ by geometric factors. The exact tubular expression, its admissible neighborhood, and the limitations of its shape mechanics are developed in \cref{app:geometric}. The present scalar framework does not supply an independent bending theory merely by rewriting volume in curved coordinates.

\label{sec:state-hierarchy}
\label{sec:thickness-reductions}

Independent \((\bm\rho^\Sigma,h)\) coordinates permit changes of material content at fixed extent and changes of extent at fixed content. An equation such as \(h=\rho^\Sigma/\rho_{\mathrm{ref}}\), with prescribed positive \(\rho_{\mathrm{ref}}\), instead constrains the effective density and removes an independent structural mode. A nonlinear relation must be applied locally on heterogeneous patches. Arbitrary unions are described by integrals, not generally by inserting total amounts into the same nonlinear constitutive relation.

Even a constrained thickness carries work. For a local constraint \(h=f(\bm\rho^\Sigma)\), define \(\bar a(\bm\rho^\Sigma)=a^\Sigma(\bm\rho^\Sigma,f(\bm\rho^\Sigma))\). Under a prescribed scalar compressive structural load \(\Pi_{\mathrm{load}}\), the loaded mass derivatives are
\[
\frac{\partial\bar a}{\partial\rho_i^\Sigma}
+\Pi_{\mathrm{load}}\frac{\partial f}{\partial\rho_i^\Sigma}.
\]
They enter the corresponding exchange equilibrium. In contrast, elimination by fast structural equilibration solves \(\PiV=\Pi_{\mathrm{load}}\) for \(h\) where possible, and transmits the structural compliance to the remaining state through the relaxed Hessian. A frozen thickness defines a separate time-scale reduction. The detailed distinctions and finite-rate transfer consequences are retained in \cref{sec:holonomic,sec:transfer}.

\label{sec:omitted-material-storage}

Omission of resolved surface mass storage does not imply that the physical transition region contains no matter. It specifies which information is retained by the sharp-interface model. A structural free energy and a structural transport equation may remain after material storage variables and their exchange balances have been removed. Such a constitutive branch uses the general phenomenological interpretation of \(\Vsurf\). It does not automatically inherit the interpretation of a conserved film volume or of the effective density \(\rho^\Sigma/h\).

A constitutive zero-capacity restriction and a singular zero-capacity limit are distinct operations. In the first, the structural free energy and remaining equations are defined directly, or by a finite continuous restriction of a parent energy when such a restriction exists. Derivatives with respect to omitted variables are not invoked. In the second, a family of positive-capacity models must additionally control exchange, momentum, kinetic storage, and the associated energy and boundary powers. A vanishing surface amount density alone implies neither vanishing through-transfer nor impermeability. Exact impermeability requires suppression of the corresponding material-exchange processes. A literal geometric-thickness realization remains admissible when its material, geometric, and mechanical balances are specified consistently.

\subsection{Structural Content, Balance, and Mechanical Power}

The structural \emph{volume} denotes an extensive, patch-additive structural content with the dimensions and work conjugacy of volume. In the general zero-thickness theory, \(V^{\Sigma}\) is not automatically the Euclidean measure of a third spatial subdomain and therefore is not, by definition, inserted into a global identity such as \(|\Omega|=|\Omega^-|+|\Omega^+|+V^{\Sigma}\). If a model is to impose such a global volume partition, that is an additional constitutive/geometric identification and the adjacent bulk representation must be modified consistently. This distinction is especially important for a positive-material sharp surface, for which the finite transition material represented on the interface may still be counted in the extrapolated bulk regions at sharp-interface resolution \cite{Bothe2022,Bothe2025ClassII}.

At this abstract level the normalization of the coordinate is not unique. For any constant \(\alpha>0\), the reparametrization
\[
\mathcal V^{\Sigma\prime}=\alpha\Vsurf,
\qquad
\Pi_{\mathcal V}^{\Sigma\prime}=\PiV/\alpha
\]
leaves the work product \(\PiV\Vsurf\), the Euler relation, and the Gibbs--Duhem contribution \(\Vsurf\,\dd\PiV\) invariant. Thermodynamics alone therefore does not determine the numerical scale of an otherwise abstract additive scalar. An operational definition, constitutive convention, or mechanical/geometric realization must fix that scale before \(\Vsurf\) can be inferred as a physical volume-like quantity. The planar geometric-thickness realization removes this freedom by identifying \(V^{\Sigma}\) with a specified swept Euclidean volume.

Its minimal balance is
\[
\DtS\Vsurf
+\Vsurf\divs\bm v^{\Sigma}
=r_{\mathcal V}^{\Sigma},
\]
when tangential transport of the structural quantity is neglected. The production \(r_{\mathcal V}^{\Sigma}\) represents intrinsic structural expansion or contraction relative to pure area dilation. It is independent, at the general level, of the constituent mass-transfer rates. Because the state domain requires \(\Vsurf\ge0\), any kinetic closure for \(r_{\mathcal V}^{\Sigma}\) must also preserve that domain. If the boundary state \(\Vsurf=0\) is admitted, the constitutive evolution cannot point outward into negative structural content. In the abstract zero-thickness theory this source is a rate of the structural coordinate, not literal creation or destruction of Euclidean space. In a geometric-thickness specialization it must instead be generated by compatible side displacements and, if global volume conservation is imposed, by the corresponding neighboring-bulk kinematics.

For \(\varpi=-\PiV\), equations~\eqref{eq:general-enriched-euler} and \eqref{eq:general-enriched-gd} supply the structural contributions \(\PiV\Vsurf\) to the area conjugate and \(\Vsurf\,\dd\PiV\) to its differential. These follow from additivity independently of an experimental comparison.

The thermodynamic conjugate \(\PiV\) does not alone specify the mechanical power associated with an intrinsic change of \(\Vsurf\). At the most general level let
\[
\mathcal W_{\mathcal V,\mathrm{mech}}^{\Sigma}
\]
denote the mechanical power expenditure per unit area associated with the chosen structural mode, counted positive when a compressive load resists positive structural expansion. The sign of the thermodynamic term follows directly from the free-energy rate. Separating the remaining balance contributions, the structural part of the material free-energy rate contains
\[
-\PiV\DtS\Vsurf.
\]
Using \(\DtS\Vsurf=r_{\mathcal V}^{\Sigma}-\Vsurf\divs\bm v^{\Sigma}\) gives
\[
-\PiV r_{\mathcal V}^{\Sigma}
+\PiV\Vsurf\divs\bm v^{\Sigma}.
\]
The dilation term is already absorbed into the complete area conjugate through the Euler relation \(\sigma=a^{\Sigma}-\sum_i c_i^{\Sigma}\mu_{i}^{\Sigma,\mathrm m}+\PiV\Vsurf\). It must not be counted again as an independent structural dissipation. Consequently, mechanical power minus reversible free-energy storage leaves the primitive structural contribution
\[
T^{\Sigma}\zeta_{\mathcal V}^{\Sigma}
=
\PiV r_{\mathcal V}^{\Sigma}
-
\mathcal W_{\mathcal V,\mathrm{mech}}^{\Sigma}.
\]
This form applies before specification of structural kinematics and permits distinct contributions to the structural rate to act against different mechanical loads.

For a single scalar structural mode whose mechanical power factorizes as
\[
\mathcal W_{\mathcal V,\mathrm{mech}}^{\Sigma}
=
\Pi_{\mathrm{mech}}^{\Sigma}r_{\mathcal V}^{\Sigma},
\]
one obtains
\[
\zeta_{\mathcal V}^{\Sigma}
=
\frac{\PiV-\Pi_{\mathrm{mech}}^{\Sigma}}{T^{\Sigma}}
\,r_{\mathcal V}^{\Sigma}.
\]
For an independently variable interior structural mode, reversible stationarity requires
\[
\PiV=\Pi_{\mathrm{mech}}^{\Sigma}.
\]
The equality is a constitutive jump relation only after the mechanical conjugate has been expressed in terms of the one-sided mechanical state. It must then be solved together with the normal momentum jump balance.

The area variation in \cref{sec:thermo} holds the total \(V^{\Sigma}\) fixed. Determining \(\Pi_{\mathrm{mech}}^{\Sigma}\) therefore requires the additional variation of structural content at fixed area.

For a factorized single-mode mechanics, a simple \emph{local interior} near-equilibrium relaxation relation is
\[
r_{\mathcal V}^{\Sigma}
=L_{\mathcal V}
\left(\PiV-\Pi_{\mathrm{mech}}^{\Sigma}\right),
\qquad
L_{\mathcal V}\ge0.
\]
A vanishing production rate implies vanishing structural force only when \(L_{\mathcal V}>0\). Zero mobility can arrest a state away from thermodynamic stationarity.
This linear constitutive relation is intended as a linearization about an equilibrium with \(\Vsurf_*>0\). It must not be extrapolated unchanged to the boundary \(\Vsurf=0\). Indeed, because the balance gives \(\DtS\Vsurf=r_{\mathcal V}^{\Sigma}\) at that boundary when tangential structural transport is absent, a closure defined there must satisfy the inward-pointing condition
\[
r_{\mathcal V}^{\Sigma}\big|_{\Vsurf=0}\ge0.
\]
If the structural force has a finite one-sided boundary limit, a possible entropy-compatible choice is
\[
r_{\mathcal V}^{\Sigma}
=L_{\mathcal V}(\Vsurf)
\left(\PiV-\Pi_{\mathrm{mech}}^{\Sigma}\right),
\qquad L_{\mathcal V}\ge0,
\qquad L_{\mathcal V}(0)=0.
\]
For the local ordinary differential equation, assume additionally that the complete rate field, including the dilution term \(-\Vsurf\divs\bm v^\Sigma\), is continuous in time and locally Lipschitz in the state up to the boundary. Its vanishing boundary value then ensures local forward invariance of \(\Vsurf\ge0\). Positive initial values remain positive on finite time intervals on which the solution and the required Lipschitz bounds exist. A finite force limit and vanishing mobility, without regularity of their product, do not by themselves establish uniqueness or forward invariance. Singular forces, tangential transport, and coupled systems require the corresponding domain-invariance analysis. Projected, complementarity, or other nonlinear closures are also possible.

The unfactorized structural power is required when independent and transfer-induced structural increments act against different loads. See Appendix~\ref{sec:transfer}.

\subsection{Planar Thickness Work and Equilibrium}
\label{sec:planar-thickness-main}

For the geometric realization at fixed planar area, let \(h=h^-+h^+\). A structural virtual variation displaces the two notional sides. The resisting normal-traction work per reference area is
\[
\delta w_{\mathrm{res}}=\pi_n^-\delta h^-+\pi_n^+\delta h^+.
\]
At planar hydrostatic equilibrium the isotropic one-sided normal loads have a common value \(p\), giving \(\delta w_{\mathrm{res}}=p\delta h\). Thus the scalar structural load is \(p\), and stationarity against the intrinsic term \(-\PiV\delta h\) gives \(\PiV=p\). This conclusion uses a specified geometric variation, the normal momentum balance, and isotropic equilibrium stresses. It is not an identity for every phenomenological width coordinate.

For this planar realization the fixed-area potential for exchange with reservoirs is
\[
\omega^\Sigma=a^\Sigma+p h-\sum_i\mu_i^{\mathrm{res}}\rho_i^\Sigma.
\]
The reservoir chemical potentials and pressure must lie on an admissible bulk state or coexistence path. This notation does not assert that they can all be varied independently in a given fluid experiment. Stationarity yields \(\mu_i^\Sigma=\mu_i^{\mathrm{res}}\) and \(\PiV=p\). At that state \(\omega^\Sigma=\sigma\). The ensemble and closed-control-volume balances are given in \cref{app:geometric}. Both routes count the displaced bulk work exactly once.

The structural balance and its work term remain meaningful away from structural equilibrium when the corresponding mode and external load are retained. Surface mass and thickness then evolve through their own rates, with possible kinetic coupling. Their thermodynamic forces need not vanish on the same time scale.

\section{Coupled Exchange, Mechanics, and Evolution}
\label{sec:equilibrium}

\subsection{Coupled Storage and Structural Evolution}
\label{sec:coupled-evolution}

The essential finite-rate structure can be displayed in a local isothermal setting with no tangential gradients, reactions, relative kinetic-energy terms, or irreversible normal-traction corrections. Those corrections are retained in \cref{sec:transfer} when the mechanical specialization is relaxed. Let \(\dot m_i^{\pm,\Sigma}\) be the inward mass supplies of \cref{sec:mass-enrichment}, let \(r_h\) be the independently retained thickness production, and let \(\Pi_{\mathrm{load}}\) be the factorized structural load. The local contribution to entropy production is
\begin{equation}
T\zeta_{\mathrm{loc}}=
\sum_{i,\pm}(\mu_i^\pm-\mu_i^\Sigma)\dot m_i^{\pm,\Sigma}
+(\PiV-\Pi_{\mathrm{load}})r_h.
\label{eq:coupled-mass-thickness-dissipation}
\end{equation}
Each term is a binary product of a rate and its thermodynamic driving force. In particular, the same constituent can exchange through two sides with different driving forces. Transfer-induced structural production and an independently relaxing structural rate must be separated before adding their work. Otherwise the structural mechanical power can be counted twice.

Collect these forces and rates into \(\bm X\) and \(\bm J\), with consistent dimensional coefficients. A local interior closure
\[
\bm J=\bm L\bm X,\qquad
\tfrac12(\bm L+\bm L^{\mathsf T})\succeq0,
\]
gives \(T\zeta_{\mathrm{loc}}=\bm X^{\mathsf T}\bm L\bm X\ge0\).
Symmetry of \(\bm L\) is a possible additional constitutive choice. The entropy inequality alone only constrains its symmetric part. Cross-coefficients permit exchange to drive structural evolution and structural forces to affect exchange. Positive semidefiniteness does not alone ensure preservation of \(\rho_i^\Sigma\ge0\) and \(h>0\), so a global closure also needs an invariant state domain.

At fixed area, constant temperature, and constant structural load, neglect tangential transport and write
\[
\phi^\Sigma=a^\Sigma+\Pi_{\mathrm{load}}h,
\qquad
\dot\rho_i^\Sigma=\sum_\alpha \dot m_i^{\alpha,\Sigma},
\qquad \dot h=r_h,
\]
where \(\alpha\) indexes the accessible reservoir sides. Equation~\eqref{eq:coupled-mass-thickness-dissipation} then gives the exact chemical-power balance
\[
\dot\phi^\Sigma
=\sum_{i,\alpha}\mu_i^\alpha \dot m_i^{\alpha,\Sigma}-T\zeta_{\mathrm{loc}}.
\]
If all accessible reservoirs for each constituent have a common time-independent potential \(\mu_i^{\mathrm{res}}\), the surface grand potential
\[
\Psi^\Sigma=a^\Sigma+\Pi_{\mathrm{load}}h
-\sum_i\mu_i^{\mathrm{res}}\rho_i^\Sigma
\]
satisfies \(\dot\Psi^\Sigma=-T\zeta_{\mathrm{loc}}\le0\). If opposite reservoirs maintain different chemical potentials for the same constituent, the chemical power cannot generally be absorbed into a function of the instantaneous surface partial mass densities alone. For constant reservoir potentials define the cumulative inward transfers per unit fixed area by
\[
Q_i^\alpha(t)=\int_0^t \dot m_i^{\alpha,\Sigma}(\tau)\dd\tau.
\]
The extended reservoir-accounting potential
\[
\Psi_{\mathrm{ext}}^\Sigma
=a^\Sigma+\Pi_{\mathrm{load}}h
-\sum_{i,\alpha}\mu_i^\alpha Q_i^\alpha
\]
then satisfies \(\dot\Psi_{\mathrm{ext}}^\Sigma=-T\zeta_{\mathrm{loc}}\). It depends on cumulative transfers and need not be bounded below under maintained nonequilibrium forcing. It is not a Lyapunov function of the surface state. The surface balances and bulk jump conditions determine the bulk conditions compatible with the surface evolution. Heat transport, diffusion, reactions, slip, and additional mechanical transfer terms require their corresponding power and entropy-production contributions.

\subsection{Loaded Affinity and Free Energy of an Open Moving Patch}
\label{sec:open-patch-energy}

For a surface patch with a boundary, the free-energy balance must include transport across that boundary and mechanical power exerted there, in addition to local structural dissipation. Consider an isothermal local-energy specialization with one material mass density \(\rho\), one additive structural density \(q\), and scalar descriptors \(\xi_a\). The descriptors are held fixed in reversible area variations. Surface superscripts are suppressed within this subsection. Let \(a=a(\rho,q,\bm\xi)\), and define the intrinsic conjugate \(\Pi=-a_q\). For a prescribed compressive load \(\Pi_{\mathrm{load}}\), define
\begin{equation}
\phi=a+\Pi_{\mathrm{load}}q,\qquad
\mu=\phi_\rho=a_\rho,\qquad
\mathcal A_q=-\phi_q=\Pi-\Pi_{\mathrm{load}},\qquad
\mathcal B_a=-\phi_{\xi_a}.
\label{eq:loaded-patch-potential}
\end{equation}
The load is independent of the state coordinates in these partial derivatives. Its linear contribution cancels from the area conjugate at fixed state:
\[
\sigma=\phi-\rho\mu+q\mathcal A_q
=a-\rho a_\rho+q\Pi.
\]
The load can nevertheless change the equilibrium state at which the tension is evaluated. The potential \(\phi\) includes the prescribed-load contribution and is not an intrinsic dependence of \(a\) on bulk pressure.

If a state-dependent relation \(\Pi_{\mathrm{load}}=\Pi_{\mathrm{load}}(\rho,q,\bm\xi)\) is substituted before differentiation, the resulting derivatives generally contain additional chain-rule terms. Such a substitution requires a derivation of the coupled surface and loading model rather than an unchanged use of \eqref{eq:loaded-patch-potential}.

Let \(\Sigma_p(t)\) be a smooth moving patch, let \(\bm u\) be a transport field as in \cref{sec:transport-kinematics}, and let \(D_t^{\bm u}\) denote its surface material derivative. Assume
\begin{align}
D_t^{\bm u}\rho+\rho\divs\bm u+\divs\bm j_m&=s_m,
\label{eq:open-patch-mass-balance}\\
D_t^{\bm u}q+q\divs\bm u+\divs\bm J_q&=r_q,
\qquad D_t^{\bm u}\xi_a=R_a.
\label{eq:open-patch-structure-balance}
\end{align}
The fluxes \(\bm j_m\) and \(\bm J_q\) are tangential and relative to \(\bm u\). In a one-component barycentric description \(\bm j_m=\bm0\). The more general notation permits another reference transport field. Multiple constituents are included by summing their chemical terms. Material supply \(s_m\) has an opposite contribution in the bulk balances, whereas intrinsic structural production need not transport material mass.

Let \(\bm w\) be the edge velocity, with normal component compatible with the support, and let \(\bm N\) be the outward unit surface co-normal. Define the outward relative fluxes
\[
j_m^\partial=[\rho(\bm u-\bm w)+\bm j_m]\cdot\bm N,
\qquad
j_q^\partial=[q(\bm u-\bm w)+\bm J_q]\cdot\bm N.
\]
The reversible local surface stress is \(\sigma\Is\).

\begin{proposition}[Open-patch available-free-energy identity]
\label{prop:open-patch-energy}
Suppose that the fields, support, and patch boundary are sufficiently smooth for the surface transport and integration-by-parts formulas, the temperature is constant, and the prescribed load is spatially and temporally constant. Then
\begin{align}
\frac{\dd}{\dd t}\int_{\Sigma_p(t)}\phi\dd A
={}&-\int_{\Sigma_p}\bm u\cdot\divs(\sigma\Is)\dd A
+\int_{\Sigma_p}\mu s_m\dd A
+\int_{\Sigma_p}\bm j_m\cdot\grads\mu\dd A\notag\\
&-\int_{\Sigma_p}\left(
\bm J_q\cdot\grads\mathcal A_q+
\mathcal A_qr_q+\sum_a\mathcal B_aR_a
\right)\dd A\notag\\
&+\int_{\partial\Sigma_p}
\left(\sigma\bm w\cdot\bm N
-\mu j_m^\partial+\mathcal A_qj_q^\partial\right)\dd\ell.
\label{eq:open-patch-energy-identity}
\end{align}
This is a transport and free-energy identity. It does not replace the coupled momentum, thermal-energy, or contact-line balances.
\end{proposition}

\begin{proof}
The chain rule and \cref{eq:open-patch-mass-balance,eq:open-patch-structure-balance} yield
\[
D_t^{\bm u}\phi+\phi\divs\bm u
=\sigma\divs\bm u+\mu s_m-\mu\divs\bm j_m
+\mathcal A_q\divs\bm J_q-\mathcal A_qr_q
-\sum_a\mathcal B_aR_a.
\]
Apply the transport theorem for a patch whose edge moves with \(\bm w\), then integrate the flux and surface-stress terms by parts. The boundary integrand is initially
\[
\sigma\bm u\cdot\bm N-\mu\bm j_m\cdot\bm N
+\mathcal A_q\bm J_q\cdot\bm N
+\phi(\bm w-\bm u)\cdot\bm N.
\]
Substitution of \(\phi=\sigma+\rho\mu-q\mathcal A_q\) gives the boundary contribution in \eqref{eq:open-patch-energy-identity}.
\end{proof}

The mechanical power at the patch boundary is determined by the surface tension \(\sigma\). Replacing this coefficient by \(\phi\), while retaining the material and structural flux contributions, violates the identity. At a junction, the boundary contributions from all incident surfaces must be added with consistent co-normal orientations and combined with the junction's momentum balance and any retained line balances. The identity itself imposes neither an independent line phase nor a constitutive relation for line dissipation.

Eliminating a surface field in a spatial model must retain its initial and boundary conditions. In particular, imposing \(\mathcal A_q=0\) at an edge makes the structural affinity contribution to boundary power vanish, but does not remove the boundary flux or its balance. A response calculated on an unbounded surface does not generally satisfy prescribed endpoint conditions.

For a prescribed nonuniform or time-dependent load, the same definitions give the additional exact power
\begin{equation}
\int_{\Sigma_p}qD_t^{\bm u}\Pi_{\mathrm{load}}\dd A
\label{eq:variable-load-power}
\end{equation}
on the right-hand side of \eqref{eq:open-patch-energy-identity}. The gradient \(\grads\mathcal A_q\) already contains \(-\grads\Pi_{\mathrm{load}}\). Here \(D_t^{\bm u}\Pi_{\mathrm{load}}\) is the derivative of the prescribed scalar field along the moving support. For a spatially uniform time-dependent load it reduces to \(\partial_t\Pi_{\mathrm{load}}\). A dynamical loading subsystem requires its own energy and power balance rather than a prescribed-load potential.

A relation that expresses the structural load in terms of another subsystem's stress must therefore be accompanied by the kinematics and work of that subsystem. A finite reaction force can remain in a rigid limit while its conjugate velocity vanishes. The force alone then does not supply a finite structural power.

\begin{corollary}[Change of structural flux reference velocity]
\label{cor:structural-flux-reference}
Under the hypotheses of \cref{prop:open-patch-energy}, let \(\bm v_R\) be another sufficiently smooth transport field with the same normal velocity as the support. Define the tangential flux
\begin{equation}
\bm Q_q=\bm J_q+q(\bm u-\bm v_R).
\label{eq:structural-flux-reference}
\end{equation}
The structural balance is equivalently
\begin{equation}
D_t^{\bm v_R}q+q\divs\bm v_R+\divs\bm Q_q=r_q.
\label{eq:structural-reference-balance}
\end{equation}
Keeping the material transport and mechanical terms in the convention of \cref{prop:open-patch-energy}, the structural transport and boundary contributions are rewritten using
\begin{align}
-\bm J_q\cdot\grads\mathcal A_q
&=-\bm Q_q\cdot\grads\mathcal A_q
  +q(\bm u-\bm v_R)\cdot\grads\mathcal A_q,
\label{eq:structural-reference-power}\\
j_q^\partial
&=[q(\bm v_R-\bm w)+\bm Q_q]\cdot\bm N.
\label{eq:structural-reference-edge}
\end{align}
Thus changing the reference velocity of the structural flux changes its contribution to the interior power identity as well as its expression at a moving edge.
\end{corollary}

\begin{proof}
Since \(\bm u-\bm v_R\) is tangential,
\[
D_t^{\bm u}q-D_t^{\bm v_R}q
=(\bm u-\bm v_R)\cdot\grads q.
\]
The surface product rule and \eqref{eq:structural-flux-reference} give \eqref{eq:structural-reference-balance}. Substituting
\(\bm J_q=\bm Q_q-q(\bm u-\bm v_R)\)
in the bulk term and in
\(j_q^\partial=[q(\bm u-\bm w)+\bm J_q]\cdot\bm N\)
gives \cref{eq:structural-reference-power,eq:structural-reference-edge}.
\end{proof}

This transformation concerns the velocity relative to which a flux is defined. It does not change the surface state or the total structural transport. The additional term in \eqref{eq:structural-reference-power} must nevertheless be retained when combining the free-energy identity with a momentum balance. A constitutive relation such as
\(\bm Q_q=M_q\grads\mathcal A_q\), with \(M_q\ge0\), makes its own flux--force product nonnegative. It does not alone establish nonnegative dissipation of the coupled system if an earlier mechanical constitutive relation is left unchanged. The combined power must be evaluated before the transformed constitutive relations are selected. For a stationary wall used as the reference field, \(\bm v_R=\bm0\), the additional free-energy-rate term is \(q\bm u\cdot\grads\mathcal A_q\).

For a prescribed variable load, \eqref{eq:variable-load-power} remains part of the identity in this representation. Its material derivative is still \(D_t^{\bm u}\), because only the structural flux has been rewritten. Replacing that derivative by one along \(\bm v_R\) requires a corresponding transformation of the remaining power terms.

\subsection{Equilibrium Reduction, Constraint Production, and Power Control}
\label{sec:equilibrium-reduction-power}

At fixed prescribed load, suppose that \(\phi\) has a smooth interior stationary branch \(q_*(\rho)\) satisfying \(\phi_q(\rho,q_*)=0\) and \(\phi_{qq}(\rho,q_*)>0\). Suppress any descriptors or hold them fixed. The implicit-function theorem and the chain rule give
\[
q_*'=-\frac{\phi_{q\rho}}{\phi_{qq}},\qquad
\bar\phi(\rho)=\phi(\rho,q_*(\rho)),\qquad
\bar\phi'=\phi_\rho,\qquad
\bar\phi''=\phi_{\rho\rho}-\frac{\phi_{\rho q}^2}{\phi_{qq}}.
\]
All derivatives on the right-hand sides are evaluated on the branch. This equilibrium reduction gives the envelope derivative and relaxed stiffness. Positivity of \(\phi_{qq}\) alone does not imply \(\bar\phi''\ge0\).

The structural balance remains necessary on the branch and determines its required production:
\begin{equation}
r_*=D_t^{\bm u}q_*+q_*\divs\bm u+\divs\bm J_{q,*}.
\label{eq:equilibrium-constraint-production}
\end{equation}
For \(q_*=q_0+\upsilon\rho\), with constants \(q_0\) and \(\upsilon\),
\[
r_*=\upsilon(s_m-\divs\bm j_m)
+q_0\divs\bm u+\divs\bm J_{q,*}.
\]
Thus an equilibrated structural coordinate may require nonzero production. The algebraic constraint \(\mathcal A_q=0\) cannot generally be combined with \(r_q=0\) in a finite-mobility constitutive relation. In a controlled fast-relaxation limit, \(\mathcal A_q\to0\) while the mobility diverges, and their product can approach the finite rate \(r_*\). A dynamical limiting theorem additionally requires control of initial and inflow preparation, boundary transport, and possible initial-layer energy loss.

For a prescribed variable load, let
\(q_*=q_*(\rho,\Pi_{\mathrm{load}})\)
be a smooth interior stationary branch with \(\phi_{qq}>0\), with any descriptors held fixed. Write \(q_{*,\rho}=\partial q_*/\partial\rho\) and \(q_{*,\Pi_{\mathrm{load}}}=\partial q_*/\partial\Pi_{\mathrm{load}}\). Substituting the material balance into \eqref{eq:equilibrium-constraint-production} gives
\begin{align}
r_*={}&q_{*,\rho}(s_m-\divs\bm j_m)
 +(q_*-\rho q_{*,\rho})\divs\bm u\notag\\
&+q_{*,\Pi_{\mathrm{load}}}D_t^{\bm u}\Pi_{\mathrm{load}}
 +\divs\bm J_{q,*}.
\label{eq:loaded-branch-production}
\end{align}
The third term is the structural production required by the changing prescribed load. With
\[
\bar\phi(\rho,\Pi_{\mathrm{load}})
=\phi(\rho,q_*(\rho,\Pi_{\mathrm{load}});\Pi_{\mathrm{load}}),
\qquad
\bar\sigma=\bar\phi-\rho\bar\phi_\rho,
\]
stationarity gives
\(\bar\phi_{\Pi_{\mathrm{load}}}=q_*\).
At fixed temperature and fixed descriptors, differentiation therefore yields
\begin{equation}
\grads\bar\sigma
=-\rho\bar\phi_{\rho\rho}\grads\rho
 +(q_*-\rho q_{*,\rho})\grads\Pi_{\mathrm{load}}.
\label{eq:loaded-branch-tension-gradient}
\end{equation}
A spatial reduction derived at uniform load must retain the second term when that restriction is removed.

A corresponding qualification applies to suppression of material capacity. If a chemical conjugate diverges, a flux tending to zero need not have vanishing chemical power. For example, in dimensionless variables, the product of a logarithmic conjugate and a flux of magnitude \(1/|\log\rho|\) need not vanish as \(\rho\downarrow0\). A limit without residual energy defects must control the integrated chemical powers, relative kinetic-energy and transfer-momentum terms, and edge flux powers, as well as the surface partial mass densities. A constitutive branch defined directly without material storage does not establish this singular limit.

Boundary constraints can also survive removal of material capacity. For example, let surfaces indexed by \(k\) meet at a line with velocity \(\bm v_\Gamma\), and let \(\bm N_k\) point out of each surface toward that line. In the absence of line storage, reaction, external supply, and tangential line transport, constituent conservation requires
\begin{equation}
\sum_k
\left[\rho_i^k(\bm u_k-\bm v_\Gamma)+\bm J_i^k\right]
\cdot\bm N_k=0,
\label{eq:capacity-limit-line-balance}
\end{equation}
where \(\rho_i^k\) is the surface mass density and \(\bm J_i^k\) its tangential flux relative to \(\bm u_k\). If the material fluxes from surfaces whose capacities are removed tend to zero, the remaining terms still impose a boundary condition on the retained surfaces. Vanishing density alone does not establish that flux limit. Relative velocities and relative fluxes must also be controlled, and any retained junction kinetic relation requires its own limiting analysis.

Finally, partition a linear force--rate relation as
\[
\begin{pmatrix}\bm J_1\\\bm J_2\end{pmatrix}
=
\begin{pmatrix}\bm L_{11}&\bm L_{12}\\\bm L_{21}&\bm L_{22}\end{pmatrix}
\begin{pmatrix}\bm X_1\\\bm X_2\end{pmatrix},
\qquad
\tfrac12(\bm L+\bm L^{\mathsf T})\succeq0.
\]
The restriction \(\bm X_2=\bm0\) leaves \(\bm J_1=\bm L_{11}\bm X_1\), whereas \(\bm J_2=\bm0\), when \(\bm L_{22}\) is invertible, yields
\[
\bm J_1=
(\bm L_{11}-\bm L_{12}\bm L_{22}^{-1}\bm L_{21})\bm X_1.
\]
Both reduced relations satisfy the entropy inequality. Under either restriction, the full product \(\bm X^{\mathsf T}\bm J\) equals the retained product \(\bm X_1^{\mathsf T}\bm J_1\ge0\). This argument also applies to a nonsymmetric mobility with positive-semidefinite symmetric part. Symmetry is an additional constitutive assumption. The two restrictions nevertheless impose different conditions. Reduction of the free energy by equilibrium conditions, kinematic restriction, and kinetic elimination must therefore be specified separately.

\subsection{One-Sided Transfer and Jump Conditions}

A material sharp surface turns bulk-to-bulk transfer into two one-sided exchange processes. For constituent \(i\),
\[
A_i^-
\rightleftharpoons
A_i^{\Sigma}
\rightleftharpoons
A_i^+.
\]
The two steps are separately bidirectional. A transferring species can therefore possess a positive instantaneous surface amount density even if the net surface storage is small on the macroscopic time scale. This two-step transfer representation is central to the phenomenological interpretation adopted here and is also the structure used in modern sharp-interface mass-transfer thermodynamics \cite{Bothe2022,Bothe2025ClassII}. In the surfactant-hindrance problem, the same representation provides a local thermodynamic mechanism by which interfacial composition changes the chemical potential of a transferring species. The resulting model has been quantitatively validated against dissolution data for carbon-dioxide bubbles in non-ionic surfactant solutions \cite{BotheTomiyama2026}. This example illustrates the state-selection principle used below. A positive interfacial amount density can affect the interfacial thermodynamic state even when the inertia of interfacial matter is negligible.

\begin{figure}[t]
\centering
\includegraphics[width=175bp]{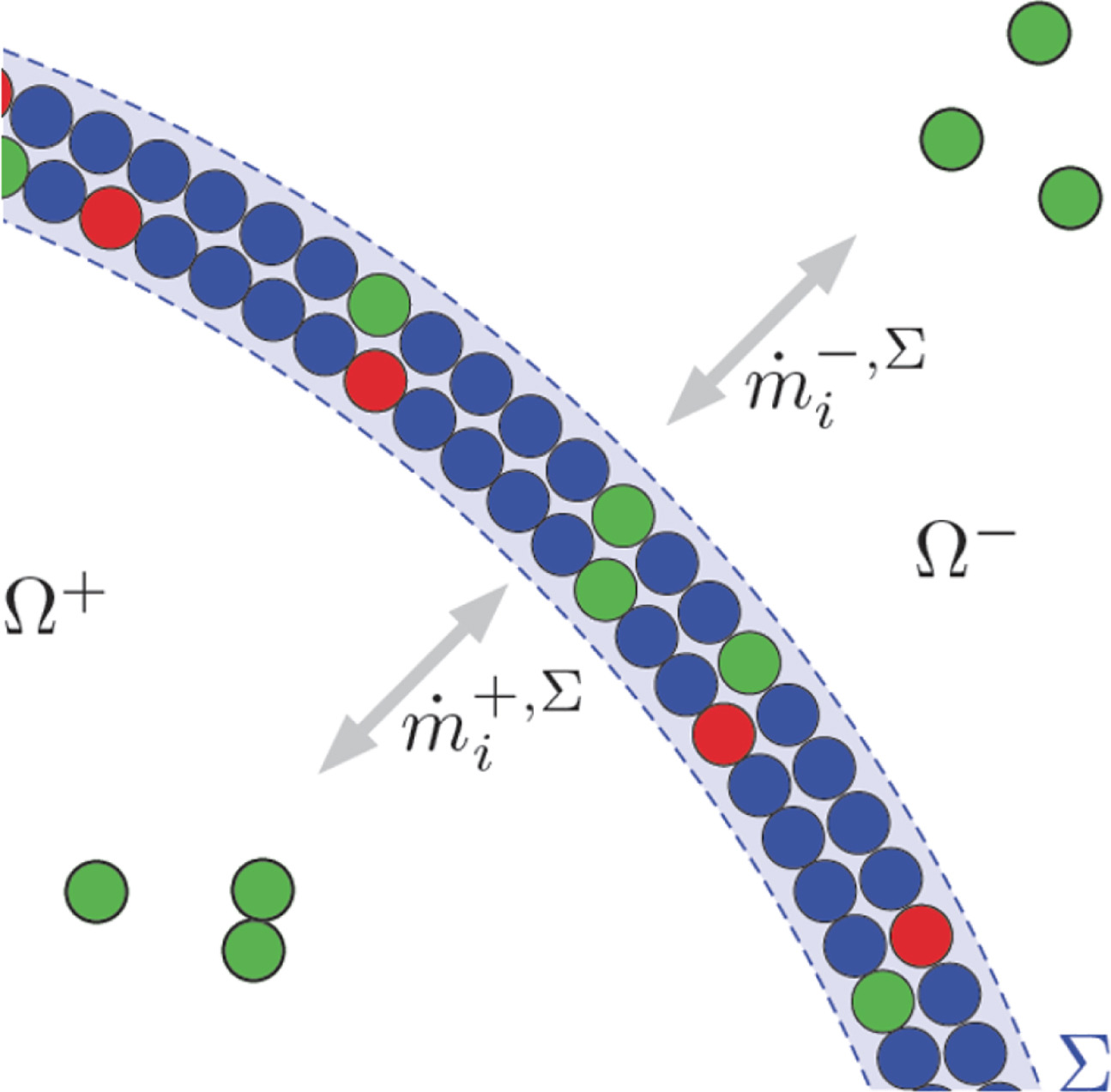}
\caption{Two one-sided bulk--surface mass-transfer processes. The drawn band depicts the unresolved interphase, while the continuum support is the zero-thickness surface $\Sigma$. Original artwork retained. Adapted from \cite[Fig.~4]{Bothe2022}}
\label{fig:two-sided-transfer}
\end{figure}

The finite-rate one-sided mass-transfer cofactor contains more than a chemical-potential difference. In the established Class-II formulation it also contains a relative kinetic-energy term and a one-sided normal irreversible-traction term \cite{Bothe2025ClassII}. Those terms will be retained in Appendix~\ref{sec:transfer}. For the equilibrium reduction considered first, velocities relative to the interface and irreversible stresses vanish, and equality of chemical potentials is recovered under this mechanical and thermal specialization.

A fast-exchange limit is not the same as a quasi-steady storage limit. If surface storage becomes fast while exchange resistances remain finite, the two one-sided rates can balance without either affinity vanishing. If the exchange law has the scaling \(\bm J=\varepsilon^{-1}\bm L\bm X\), bounded rates imply \(\bm X\to\bm0\) on the accessible force subspace provided \(\bm L\) has a uniformly bounded inverse there. Constitutive null modes and constrained exchange processes require separate treatment. Compatible reservoirs and solvability of the coupled thermodynamic and mechanical constraints are additionally required to obtain an equilibrium manifold.

The term \emph{jump condition} includes balance-derived and constitutive interface relations. A \emph{jump balance} is a relation obtained from an integral balance law by localization at the interface. A \emph{constitutive jump relation} is a constitutive relation connecting one-sided and surface quantities. The distinction is essential because a constitutive jump relation must be interpreted together with the associated jump balances.

With the outward-normal convention of \cref{sec:notation}, the inward mass supply on an active Class-II transfer process with $\rho_i^\alpha>0$ is
\[
\dot m_i^{\alpha,\Sigma}
=\rho_i^\alpha(\bm v_i^\alpha-\bm v_i^\Sigma)
 \cdot\bm n^\alpha.
\]
The surface constituent velocities have the normal speed of the common
support. Cofactors containing $1/\rho_i^\alpha$ are interior-state
expressions. A missing-species limit requires a separately defined
constitutive extension of the complete transfer product.

A standard example is one-sided tangential slip. A Navier-type constitutive jump relation may have the schematic form
\[
\alpha_i^{\alpha}
\left(\bm v_i^{\alpha}-\bm v_i^{\Sigma}\right)_{\parallel}
+
\left(\bm S_i^{\alpha}\bm n^{\alpha}\right)_{\parallel}
=\bm0,
\qquad
\alpha_i^{\alpha}\ge0.
\]
Its consequences are obtained only after it is combined with the tangential part of the momentum jump balance. The same requirement applies to the normal direction: the structural constitutive relation for \(\PiV\) must be combined with the normal momentum jump balance, not interpreted alone.

With the normal and curvature conventions of \cref{sec:notation}, a spherical minus phase of radius \(R\), with \(\bm n_\Sigma\) directed outward, has \(\kappas=-2/R\). Let
\[
\pi_n^{\pm}
=-\bm n_\Sigma\cdot\bm S^{\pm}\bm n_\Sigma
\]
denote the one-sided normal mechanical compressions. For a static interface with isotropic reversible surface stress and no additional surface force density, the momentum jump balance has the form
\[
\jump{\bm S}\bm n_\Sigma+\divs(\sigma\Is)=\bm0,
\qquad
\jump{\bm S}=\bm S^+-\bm S^-.
\]
Its normal projection is
\[
\pi_n^--\pi_n^++\sigma\kappas=0.
\]
In simple isotropic equilibrium bulks,
\[
\bm S^{\pm}=-p^{\pm}\bm I,
\]
so that
\[
\pi_n^{\pm}=\left.p^{\pm}\right|_{\Sigma},
\]
and the normal jump balance becomes
\[
\pminus-\pplus=-\sigma\kappas.
\]
For the spherical convention above this gives the usual positive inner-minus-outer pressure difference.

The equality \(\pi_n^{\pm}=p^{\pm}|_{\Sigma}\) is a specialization, not a notation convention. In nonequilibrium the normal mechanical compression contains the complete one-sided stress, whereas the thermodynamic pressure is a state variable of the bulk constitutive theory. Likewise,
\[
P=-\frac13\tr\bm S
\]
is the bulk mechanical pressure and is not used as a synonym for thermodynamic \(p\).

\subsection{Fast Transfer and Reduced Equilibrium Response}
\label{app:ift}

At local thermal and mechanical equilibrium, and for a constituent that can exchange with both bulk phases, the zero-resistance conditions reduce to
\[
\mu_{i}^{-,\mathrm m}
\left(T,\pminus,\bm x^-\right)
=
\mu_{i}^{\Sigma,\mathrm m}
\left(T,\bm c^{\Sigma},\Vsurf,\bm\xi^{\Sigma}\right)
=
\mu_{i}^{+,\mathrm m}
\left(T,\pplus,\bm x^+\right).
\]
The two bulk pressures remain distinct if the interface is curved. The equality of chemical potentials does not require equality of pressures because the compositions and pressure arguments enter the bulk chemical potentials separately. The corresponding curvature dependence of equilibrium solubility is familiar from Gibbs--Thomson and Kelvin-type relations. Modern sharp-interface thermodynamics makes the two-pressure structure explicit \cite{Bothe2025ClassII}.

For the idealized pure-reservoir benchmark used later, component \(A\) is supplied by the minus phase and component \(B\) by the plus phase. The reduced conditions are then
\[
\mu_{A}^{\Sigma,\mathrm m}=\mu_{A}^{-,\mathrm m}
\left(T,\pminus\right),
\qquad
\mu_{B}^{\Sigma,\mathrm m}=\mu_{B}^{+,\mathrm m}
\left(T,\pplus\right).
\]
These are one-sided constraints. A common pressure can be introduced only after the planar mechanical specialization has been imposed.

The complete reduced state is determined by a coupled set of equations. Let \(\lambda\) be a loading parameter and collect the remaining variables into \(\bm y\). We write
\[
\bm G(\bm y,\lambda)=\bm0,
\]
where \(\bm G\) contains the bulk coexistence relations, one-sided chemical-equilibrium conditions, intrinsic surface constitutive equations, normal momentum jump balance, structural equilibrium condition when \(\Vsurf\) is present, and any geometric relation determining curvature. After dependent equations and redundant coordinates have been removed, assume that \(\bm G\) is \(C^1\), its Jacobian in \(\bm y\) is square, and this Jacobian is nonsingular at a solution. The implicit-function theorem then gives a local reduced manifold
\[
\bm y=\widetilde{\bm y}(\lambda),
\qquad
\frac{\dd\widetilde{\bm y}}{\dd\lambda}
=-
\left(
\frac{\partial\bm G}{\partial\bm y}
\right)^{-1}
\frac{\partial\bm G}{\partial\lambda}.
\]

For the surface-state block $\bm z^\Sigma$ of $\bm y$ and the intrinsic law
$\sigma=\widehat\sigma^\Sigma(\bm z^\Sigma)$, the reduced response is
\[
\frac{\dd\widetilde\sigma}{\dd\lambda}
=\frac{\partial\widehat\sigma^\Sigma}{\partial\bm z^\Sigma}
 \frac{\dd\widetilde{\bm z}^\Sigma}{\dd\lambda},
\]
where the state derivative is the corresponding block of
$-\bm G_{\bm y}^{-1}\bm G_\lambda$. Thus pressure enters through the coupled
equilibrium state, not as a primitive argument of the surface law. This is a
local finite-dimensional result. A field-level reduction requires an
implicit-function theorem in specified function spaces.

\section{A Coupled Mass--Thickness Constitutive Model}
\label{sec:mass-thickness-model}

\subsection{Free Energy, Stability, and Loaded Equilibrium}

An explicit constitutive example permits independent variations of stored mass and represented width while coupling their thermodynamic forces. Consider an isothermal local planar surface with one retained material mass density \(m=\rho^\Sigma>0\) and independent geometric thickness \(h>0\). The model is a scalar constitutive illustration, not a complete one-component two-phase coexistence model. Reference parameters satisfy
\[
\rho_*>0,\quad h_*>0,\quad K>0,\quad k>0,
\qquad m_*=\rho_*h_*.
\]
Here \(\rho_*\) is a reference effective volumetric density, \(K\) has units Pa, and \(k\) has units Pa\,m\(^{-1}\). Define
\begin{equation}
a(m,h)=\sigma_*+\mu_*m-\Pi_*h
+\frac{K}{2h}\left(\frac{m}{\rho_*}-h\right)^2
+\frac{k}{2}(h-h_*)^2.
\label{eq:mass-thickness-energy}
\end{equation}
All parameters are fixed constitutive reference quantities. In particular \(\Pi_*\) and \(\mu_*\) are not varying imposed loads. The first quadratic term penalizes effective-density deviations and the second supplies an independent thickness stiffness. Both are needed for strict material-state stability in the two independent coordinates.

Writing \(n=m/(\rho_*h)\), differentiation gives
\begin{align}
\mu&=a_m=\mu_*+\frac K{\rho_*}(n-1),\label{eq:mass-thickness-mu}\\
\Pi&=-a_h=\Pi_*+\frac K2(n^2-1)-k(h-h_*),\label{eq:mass-thickness-pi}\\
\sigma&=a-m\mu+h\Pi
=\sigma_*+\frac k2(h_*^2-h^2).\label{eq:mass-thickness-sigma}
\end{align}
The effective-density energy is homogeneous of degree one in \((m,h)\), so its contribution to this area derivative cancels. This cancellation is specific to the chosen closure. Mass affects the structural force. Relaxation of thickness then changes tension. Additional constitutive terms can give a direct dependence of tension on mass at fixed thickness.

For any variation \((\delta m,\delta h)\), the material Hessian satisfies
\begin{equation}
\delta^2a=\frac K h
\left(\frac{\delta m}{\rho_*}-n\delta h\right)^2
+k(\delta h)^2>0
\quad\text{if }(\delta m,\delta h)\ne(0,0).
\label{eq:mass-thickness-hessian}
\end{equation}
Equivalently,
\[
a_{mm}=\frac K{\rho_*^2h},\qquad
a_{mh}=-\frac{Kn}{\rho_*h},\qquad
a_{hh}=\frac{Kn^2}{h}+k,
\qquad \det\nabla^2a=\frac{Kk}{\rho_*^2h}>0.
\]
Thus the free energy is strictly convex on the positive material domain. This is a fixed-area material stability statement. It is not a stability theorem for shape, bulk-phase, or arbitrary finite-area perturbations. The illustration is further restricted to positive tension, which requires
\(h^2<h_*^2+2\sigma_*/k\) when \(\sigma_*>0\).

Let \(\mu_{\mathrm{res}}\) and \(\Pi_{\mathrm{load}}\) denote the chemical and structural loading coordinates. The open potential
\[
\omega=a-\mu_{\mathrm{res}}m+\Pi_{\mathrm{load}}h
\]
has the same positive Hessian. Its stationary state is explicitly
\begin{align}
n_{\mathrm{eq}}&=1+\frac{\rho_*}{K}(\mu_{\mathrm{res}}-\mu_*),\\
h_{\mathrm{eq}}&=h_*+
\frac{\Pi_*-\Pi_{\mathrm{load}}+\tfrac K2(n_{\mathrm{eq}}^2-1)}{k},\\
m_{\mathrm{eq}}&=\rho_* n_{\mathrm{eq}}h_{\mathrm{eq}}.
\label{eq:mass-thickness-equilibrium}
\end{align}
Where \(n_{\mathrm{eq}}>0\) and \(h_{\mathrm{eq}}>0\), strict convexity on the convex positive quadrant makes this stationary state the unique global minimizer over that open material-state domain. No boundary minimizer is asserted when these positivity conditions fail. The positive-tension restriction can be imposed on the selected branch. The two loading coordinates describe constitutive partial responses. An actual experiment must supply a compatible path through them. In a planar hydrostatic realization the structural load is \(p\), and reservoir thermodynamics constrains the accompanying chemical-potential change.

The partial derivatives quantify this coupling. At fixed structural load,
\[
\frac{\partial h_{\mathrm{eq}}}{\partial\mu_{\mathrm{res}}}
=\frac{\rho_* n_{\mathrm{eq}}}{k}>0,
\]
so increasing the chemical load increases density, expands the represented zone, and increases stored mass. At fixed chemical load,
\[
\frac{\partial h_{\mathrm{eq}}}{\partial\Pi_{\mathrm{load}}}=-\frac1k,
\qquad
\frac{\partial m_{\mathrm{eq}}}{\partial\Pi_{\mathrm{load}}}
=-\frac{\rho_* n_{\mathrm{eq}}}{k}.
\]
Open exchange keeps the effective density fixed on the latter partial path while compression reduces thickness and stored mass. With mass held fixed, the local structural stiffness is \(Kn^2/h+k\). After mass equilibration it is the Schur complement \(k\). This distinction is a direct consequence of retaining both state variables.

\subsection{Dissipative Relaxation and Reduction}

At fixed area and fixed loads, a diagonal local interior closure is
\[
\dot m=\ell_m(\mu_{\mathrm{res}}-\mu),\qquad
\dot h=\ell_h(\Pi-\Pi_{\mathrm{load}}),\qquad
\ell_m,\ell_h>0.
\]
The corresponding dissipation is
\[
-\dot\omega
=\ell_m(\mu_{\mathrm{res}}-\mu)^2
+\ell_h(\Pi-\Pi_{\mathrm{load}})^2\ge0.
\]
This closure is intended near a positive equilibrium. It is not asserted to preserve every state-domain boundary under arbitrary forcing. For constant loads and positive constant mobilities, local exponential convergence follows directly. In fixed dimensionless coordinates, choose a compact convex neighborhood of the positive equilibrium with Hessian bounds \(\lambda\bm I\preceq\nabla^2\omega\preceq\Lambda\bm I\) and mobility bound \(\bm D_{\mathrm{sc}}\succeq d\bm I\) for the mobility expressed in those scaled coordinates, where \(\lambda,\Lambda,d>0\). A sufficiently small sublevel component of \(E=\omega-\omega_{\mathrm{eq}}\) is contained in this neighborhood and is forward invariant. Strong convexity and the dissipation identity give
\[
\dot E\le-d\|\nabla\omega\|^2
\le-\frac{2d\lambda^2}{\Lambda}E.
\]
Thus the local state perturbation converges exponentially.

For perturbations \(\delta\bm z=(\delta m,\delta h)\), the linearized relaxation is
\[
\delta\dot{\bm z}=-\bm D\bm H\delta\bm z,
\qquad \bm D=\operatorname{diag}(\ell_m,\ell_h),\quad
\bm H=\nabla^2a|_{\mathrm{eq}}.
\]
Because \(\bm D\bm H\) is similar to the positive-definite symmetric matrix
\(\bm D^{1/2}\bm H\bm D^{1/2}\), both relaxation rates are positive. The modes generally mix mass and thickness even for diagonal mobilities. Fast mass equilibration reduces the remaining structural stiffness to \(k\). Freezing mass retains the larger stiffness.

This reduction assumes the specified diagonal kinetics, a separation of mass-exchange and structural time scales, and states on the chemical-equilibrium manifold after any initial exchange layer. A general coupled mobility requires its own kinetic elimination. The energetic Schur complement alone determines only stiffness. See also \cref{sec:equilibrium-reduction-power}. Imposing \(h=m/\rho_*\) as a constitutive constraint removes a coordinate and requires the constrained work calculation of \cref{sec:thickness-reductions}. It is not equivalent to either of these finite-stiffness dynamical reductions.

\subsection{Dimensioned Illustration and Measurement Meaning}

Choose synthetic reference parameters
\[
\rho_*=10^3\ \mathrm{kg\,m^{-3}},\quad h_*=1\ \mathrm{nm},\quad
K=10^9\ \mathrm{Pa},\quad k=10^{17}\ \mathrm{Pa\,m^{-1}},
\]
\[
\sigma_*=50\ \mathrm{mN\,m^{-1}},\quad
\Pi_*=0.1\ \mathrm{MPa},\quad \mu_*=0.
\]
The reference mass is \(1\ \mathrm{mg\,m^{-2}}\). A structural-load increase of \(10\ \mathrm{MPa}\) at fixed chemical load gives \(h=0.9\ \mathrm{nm}\), \(m=0.9\ \mathrm{mg\,m^{-2}}\), and \(\sigma=59.5\ \mathrm{mN\,m^{-1}}\). A chemical-load increase of \(10\ \mathrm{kJ\,kg^{-1}}\) at fixed structural load instead gives \(n=1.01\), \(h=1.1005\ \mathrm{nm}\), \(m=1.111505\ \mathrm{mg\,m^{-2}}\), and \(\sigma\simeq39.445\ \mathrm{mN\,m^{-1}}\). These are constitutive partial paths, not measured pressure or adsorption series.

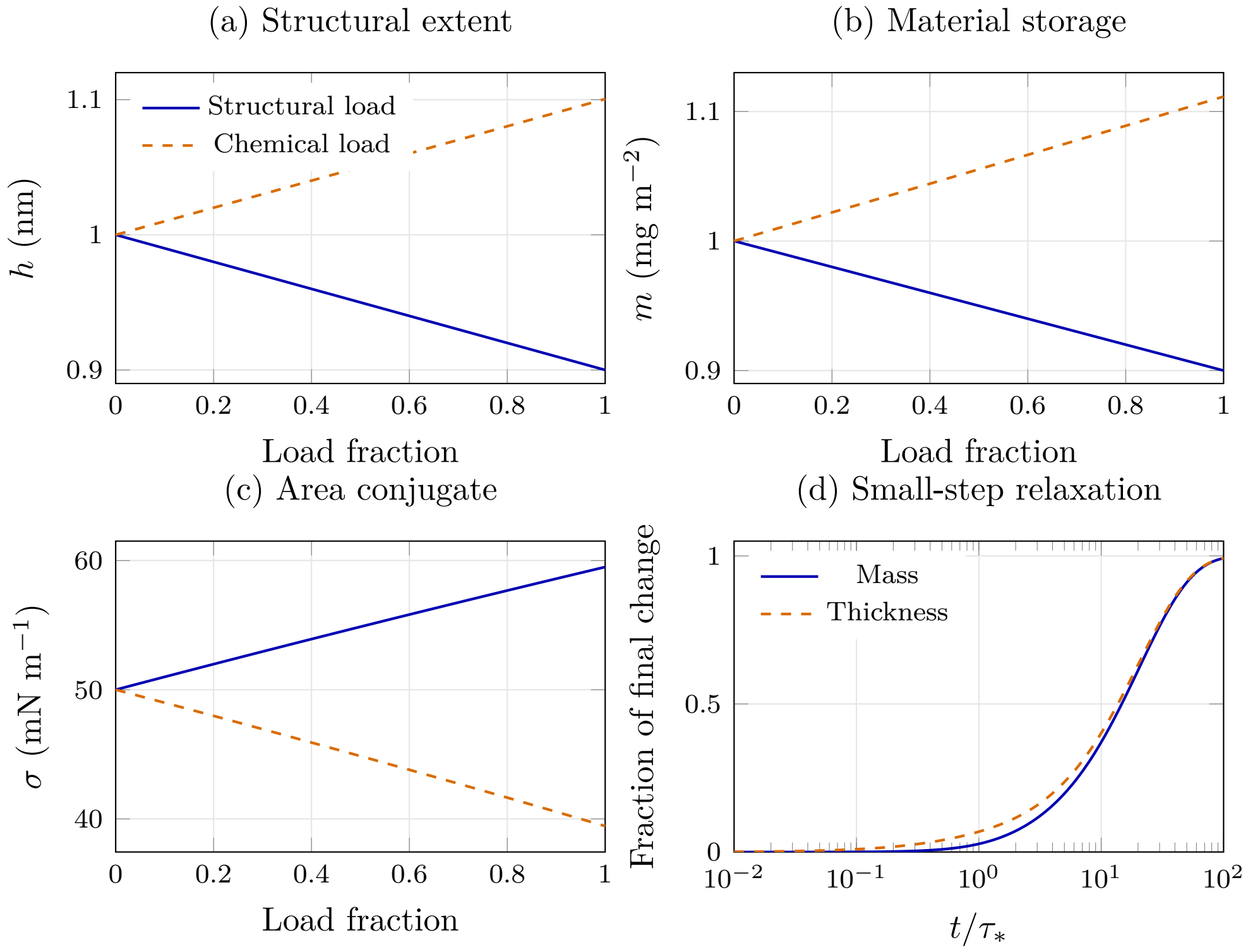
\begin{figure}[htbp]
\centering
\begin{tikzpicture}
\begin{groupplot}[
 group style={group size=2 by 2,horizontal sep=1.35cm,vertical sep=1.65cm},
 width=.43\textwidth,height=.31\textwidth,
 tick label style={font=\scriptsize},label style={font=\small},title style={font=\small},
 grid=major,grid style={black!10},samples=81,
 legend style={font=\scriptsize,draw=none,fill=white},
]
\nextgroupplot[title={(a) Structural extent},xlabel={Load fraction},ylabel={$h$ (nm)},xmin=0,xmax=1,ymin=.89,ymax=1.12,legend pos=north west]
\addplot[blue!70!black,thick,domain=0:1]{1-.1*x};
\addlegendentry{Structural load}
\addplot[orange!85!black,thick,dashed,domain=0:1]{1+.1*x+.0005*x*x};
\addlegendentry{Chemical load}
\nextgroupplot[title={(b) Material storage},xlabel={Load fraction},ylabel={$m$ (mg m$^{-2}$)},xmin=0,xmax=1,ymin=.89,ymax=1.13]
\addplot[blue!70!black,thick,domain=0:1]{1-.1*x};
\addplot[orange!85!black,thick,dashed,domain=0:1]{(1+.01*x)*(1+.1*x+.0005*x*x)};
\nextgroupplot[title={(c) Area conjugate},xlabel={Load fraction},ylabel={$\sigma$ (mN m$^{-1}$)},xmin=0,xmax=1]
\addplot[blue!70!black,thick,domain=0:1]{50+50*(1-(1-.1*x)^2)};
\addplot[orange!85!black,thick,dashed,domain=0:1]{50+50*(1-(1+.1*x+.0005*x*x)^2)};
\nextgroupplot[title={(d) Small-step relaxation},xlabel={$t/\tau_*$},ylabel={Fraction of final change},xmode=log,xmin=.01,xmax=100,ymin=0,ymax=1.05,legend pos=north west]
\addplot[blue!70!black,thick,domain=.01:100,samples=160]{1-(2.0512492197*exp(-.0487507803*x)-.0487507803*exp(-2.0512492197*x))/2.0024984394};
\addlegendentry{Mass}
\addplot[orange!85!black,thick,dashed,domain=.01:100,samples=160]{1-(1.9512492197*exp(-.0487507803*x)+.0512492197*exp(-2.0512492197*x))/2.0024984394};
\addlegendentry{Thickness}
\end{groupplot}
\end{tikzpicture}
\caption{Synthetic mass--thickness enrichment from equation~\eqref{eq:mass-thickness-energy}. For load fraction \(f\in[0,1]\), the structural path has \(\Delta\Pi_{\mathrm{load}}=10f\,\mathrm{MPa}\) at fixed \(\mu_{\mathrm{res}}=\mu_*\), and the chemical path has \(\Delta\mu_{\mathrm{res}}=10f\,\mathrm{kJ\,kg^{-1}}\) at fixed \(\Pi_{\mathrm{load}}=\Pi_*\). These paths change stored mass, thickness, and tension. The relaxation panel resolves the coupled linear response to a small structural-load step in units of the unspecified time scale \(\tau_*\). No experimental data are fitted, and the partial loading paths are not asserted to be bulk coexistence paths}
\label{fig:mass-thickness}
\end{figure}

Figure~\ref{fig:mass-thickness} displays these two responses and the two local relaxation modes. For the latter, use scales \(m_*,h_*\), energy per area \(E_*=Kh_*\), and an arbitrary positive time scale \(\tau_*\). Choosing
\(\ell_m=m_*^2/(E_*\tau_*)\) and
\(\ell_h=h_*^2/(E_*\tau_*)\) gives the dimensionless reference Hessian
\[
\bm H_{\mathrm{d}}=
\begin{pmatrix}1&-1\\-1&1.1\end{pmatrix},
\]
whose positive eigenvalues are approximately \(0.04875\) and \(2.05125\). No dimensional relaxation time is assigned. The plotted transient is the linear response to a small structural-load step. The equilibrium panels use the exact nonlinear stationary formulas.

Comparison with observations requires definitions of both the constituent surface amount densities and the represented thickness. Tension alone may constrain only combinations of them. The separate construction in \cref{sec:constructive} demonstrates exact coincidence of static pressure--tension curves for independent and holonomic models with different state responses. This construction characterizes the limits of inference from a single observable.

\section{Pressure Response as an Application of State Enrichment}
\label{sec:pressure-application}

Pressure response illustrates how retention or suppression of a mechanical-work contribution changes model predictions. Consider a planar isothermal surface retaining positive constituent amounts but no structural work, with equilibrium exchange into pure incompressible reservoirs. Its Gibbs--Duhem relation gives
\[
\frac{\dd\sigma}{\dd p}=-\sum_i c_i^\Sigma v_i<0
\]
when the molar volumes \(v_i\) and the retained amounts are strictly positive. This is a statement about a specified reduced state and loading path. The detailed binary definition, finite-solubility extension, and curved-interface conditions are given in \cref{sec:binary}. A positive slope outside these assumptions does not contradict the result.

For the independent planar geometric-thickness realization, structural equilibrium instead gives \(\PiV=p\). On an equilibrium path without unequilibrated internal descriptors,
\[
\frac{\dd\sigma}{\dd p}=h-\sum_i c_i^\Sigma\frac{\dd\mu_{i}^{\mathrm m}}{\dd p}.
\]
The added term follows from the structural coordinate and its conjugate work. In a mixture it must be evaluated on the coupled bulk--surface equilibrium path. For a curved interface that path also contains the two one-sided pressure traces and the appropriate momentum jump. These conditions are given explicitly in \cref{app:pressure-loading}.

Classical finite-region and dividing-surface thermodynamics already represent either sign of the pressure coefficient through an invariant combination. The detailed comparison in \cref{sec:invariant} distinguishes physical width, actual region inventories, and signed Gibbs excesses. Equality of an observable coefficient is a useful consistency check, but does not identify those coordinates with positive autonomous surface amount densities or uniquely determine an enrichment.

The constructive pressure models in \cref{sec:constructive} provide two further consistency results. First, regular stable independent-volume states have
\[
\sigma''=-\bm r'^{\mathsf T}\bm H^{-1}\bm r'
-\sum_i c_i(\mu_{i}^{\mathrm m})'',\qquad
\bm r=(\bm\mu^{\mathrm m},-p),\quad\bm H\succ0.
\]
Thus an affine chemical-potential path has strictly concave tension in that material-stability subclass. A real-fluid path requires the second-derivative correction. Second, explicit stable independent and holonomic states can have identical complete local pressure--tension curves while their surface amount densities and structural volumes differ. Static tension can constrain constitutive combinations without identifying the number of independent state variables. Dynamic observations can add information when the relevant modes are excited and their competing mechanisms are controlled.

These results establish requirements for the application of enrichment: define the retained physical information, derive complete predictions, and identify which independent observations constrain the resulting state. The mass--thickness model in \cref{sec:mass-thickness-model} provides an explicit mass--thickness coupling independent of the sign of the pressure coefficient.

Appendix~\ref{sec:selection} develops a prospective water--\(n\)-decane test and specifies the preparation, timing, loading-path, and independent-observable information required for validation. No calibration or experimental falsification is claimed.

\section{Discussion and Further Enrichment}
\label{sec:discussion}
\label{sec:outlook}

The principal modeling decision is which information about the interface survives on a sharp support. Surface mass retains material storage and composition. Thickness retains selected normal-structure information. Neither choice is specified by the geometry of the support alone. Their independence is a constitutive decision, and the observable consequences depend on their balances, mechanical work, and relaxation time scales.

The general area-conjugate derivation establishes the dependence of the mechanical stress on descriptor kinematics. Densities of additive contents dilute under an area variation at fixed totals and enter the Euler relation with their conjugates. Internal descriptors require a prescribed deformation rule. Consequently two quantities with the same units, or two free energies written with similarly named variables, need not describe the same model. An apparent discrepancy in surface tension may arise from different admissible variations rather than from an algebraic error.

The model in \cref{sec:mass-thickness-model} shows that thermodynamically coupled state coordinates can affect tension indirectly. Material exchange changes thickness through the structural force. More general composition-dependent terms can supply direct adsorption effects.

The pressure calculations illustrate the limits of reduced states and of inference from a single observable. They do not establish the uniqueness of thickness, invalidate Gibbs excess thermodynamics, or require every enrichment to be justified by a positive pressure slope. The same distinction is required for curvature. A coordinate transformation alone does not introduce independent bending mechanics. The complete mechanical-power balance determines which forces remain after specification of a geometric realization.

Further enrichment should follow the same procedure. An additional scalar descriptor of transition-zone asymmetry, order, or another retained structural feature needs a physical interpretation, an admissible domain, independent variations, conjugate forces, and an evolution or constraint law. More elaborate gradient or tensorial states additionally require objectivity and their associated stresses and boundary actions. These are extensions of the present framework, not consequences already established by its scalar examples.

An extension to fluid--solid contact must likewise specify the substrate's retained mechanical variables and work. Strain-dependent solid surface stresses and configurational forces at a moving junction require the corresponding substrate and line mechanics. They are not determined by the scalar fluid-surface area conjugate.

Further constitutive development requires specifying the relation between measurable quantities and the surface mass densities and thickness descriptor for a particular interface, extending the local free energy over a justified temperature and composition range, and comparing the predicted coupled state responses with independent observations. Archival pressure reconstruction is one available validation route. A theory allowing independent variations of curved-interface shape and thickness and the detailed thermal and kinetic treatment of phase change require separate development.

Two recent developments indicate additional requirements for such extensions. Bothe and Ruggeri \cite{BotheRuggeri2026} retain constituent energy balances and separate temperatures in a Class-III bulk-mixture theory. Adapting that state selection to a surface would require partial surface energies and complete energy-exchange relations. It is not covered by the present single-temperature surface state. Hopf and Tang \cite{HopfTang2026} prove existence and weak--strong uniqueness for entropy-dissipating reaction--diffusion systems with nonlinear interface transmission. Their result concerns a different, fixed-interface PDE class and supplies no well-posedness theorem for the moving, mechanically enriched model developed here.

\section{Conclusions}
\label{sec:conclusions}

Enriched surface thermodynamics retains selected material and structural information on a sharp-interface support. The distinction between additive contents and internal descriptors determines the area conjugate, while the balance and entropy-production structure determines how the state can evolve. The framework developed here makes these choices explicit for local scalar enrichments of isotropic fluid surfaces.

The free-energy identity for a moving surface patch includes the mechanical power at its boundary and the contributions of material and structural transport across that boundary. At a junction of interfaces, these contributions from all incident surfaces must be combined consistently with the junction's mechanical balances and any balances for quantities assigned to the junction line.

Changing the reference velocity of a structural flux requires the corresponding transformation of its interior and boundary powers. The nonnegativity of separately chosen flux--force products does not replace the dissipation check for the resulting coupled mechanical and structural model.

For moving contact lines, the contact-region generalized Navier condition of Fullana et al.\ \cite{FullanaEtAl2026} illustrates the need to check a mechanical closure against contact-angle transport. Its contact-region width is a distinct model parameter, not the normal transition-zone thickness used here. Equilibrium reduction retains the production required to follow the selected stationary branch, while a singular suppression of material capacity additionally requires control of the associated energy and boundary powers.

Surface mass per area is an established and physically significant storage variable. In a mixture, constituent surface mass densities and the total surface mass density together with independent mass fractions are equivalent coordinates. Thickness supplies an additional structural choice when the normal extent of the represented transition zone is retained independently. In the volume-per-area realization it has a pressure-like conjugate and a separate contribution to mechanical power. Constrained thickness, frozen thickness, and fast structural equilibration give distinct reductions.

An explicit mass--thickness free energy demonstrates strict material stability, coupled adsorption and structural response, and dissipative local relaxation. Pressure response and the accompanying comparison models show how selected reductions can be tested and why one static observable need not identify structural independence. These applications complement the central result that a consistent enrichment carries every additional surface variable through its meaning, kinematics, thermodynamic conjugacy, mechanics, balance or evolution, and constitutive closure.

\clearpage
\appendix

\section{Binary Pressure-Response Benchmarks}
\label{sec:binary}

\subsection{Surface Mixture and Amount-Only Benchmark}

Consider two immiscible or partially miscible fluid phases whose interfacial transition contains matter from both sides. A phenomenological surface that retains those actual constituent amounts therefore requires at least two constituent amount densities,
\[
c_A^{\Sigma}>0,
\qquad
c_B^{\Sigma}>0.
\]
The term \emph{amount-only} denotes the restricted model class defined below. It specifies more than the number of independent coordinates.
\begin{definition}[Restricted autonomous amount-only class]
\label{def:restricted-amount-only}
The surface is an autonomous sharp subsystem with positive constituent amount densities. Apart from temperature, its primitive equilibrium free-energy state consists only of these densities. In particular, the primitive surface free energy has no independent additive structural coordinate, no retained holonomic amount--volume relation, and no corresponding contribution of mechanical work. Neighboring bulk pressure, normal traction, curvature, surface strain, electric field, and other externally loaded mechanical variables are also excluded as primitive constitutive arguments. Such effects may enter only through the equilibrium constraints and jump conditions used to reduce the coupled bulk--surface system.
\end{definition}
A later holonomically structured model can have the same number of independent amount coordinates while lying outside \cref{def:restricted-amount-only} because it retains additional amount--volume mechanics. For the restricted binary model,
\[
a^{\Sigma}
=
\widehat a^{\Sigma}(T,c_A^{\Sigma},c_B^{\Sigma}),
\]
and
\[
\sigma
=
a^{\Sigma}
-c_A^{\Sigma}\mu_{A}^{\Sigma,\mathrm m}
-c_B^{\Sigma}\mu_{B}^{\Sigma,\mathrm m}.
\]
The surface composition may be represented by
\[
x_A^{\Sigma}
=
\frac{c_A^{\Sigma}}{c_A^{\Sigma}+c_B^{\Sigma}},
\qquad
x_B^{\Sigma}=1-x_A^{\Sigma}.
\]
The positivity of these mole fractions follows from the positive constituent amount densities. It imposes no restriction on the sign of a Gibbs surface excess.

Fast exchange does not remove the surface thermodynamics. It removes the independent slow evolution of the composition by constraining the surface chemical potentials. When the relevant constrained surface Hessian is locally nonsingular, the Hessian and the equilibrium constraints determine the response of \(c_A^{\Sigma}\) and \(c_B^{\Sigma}\) to changes in reservoir chemical potentials, while the Gibbs--Duhem relation determines the first differential of the tension along that manifold.

\label{sec:amountonly}

Consider two idealized pure incompressible reservoirs at fixed temperature. Component \(A\) exchanges with the minus phase and component \(B\) with the plus phase, so that
\[
\mu_{A}^{\Sigma,\mathrm m}=\mu_{A}^{-,\mathrm m}(T,\pminus),
\qquad
\mu_{B}^{\Sigma,\mathrm m}=\mu_{B}^{+,\mathrm m}(T,\pplus).
\]
For pure incompressible bulks,
\[
\dd\mu_{A}^{-,\mathrm m}=v_A^-\dd\pminus,
\qquad
\dd\mu_{B}^{+,\mathrm m}=v_B^+\dd\pplus,
\]
where \(v_A^-\) and \(v_B^+\) are the molar volumes. Incompressibility removes the leading pressure dependence of density, not the pressure dependence of chemical potential. The pressure-affine limit of chemical potential is therefore retained \cite{BotheDreyerDruet2023}. Retaining this pressure contribution is also essential in thermodynamically consistent models of dissolution and growth of gas bubbles under pressure control with an incompressible-solvent approximation \cite{BotheSoga2016}.

Choose \(\pplus\) as the parameter and define
\[
s_+
=
\frac{\dd\sigmatp}{\dd\pplus},
\qquad
r_p
=
\frac{\dd\pminus}{\dd\pplus}.
\]
The amount-only Gibbs--Duhem relation gives
\[
s_+
=-c_A^{\Sigma}v_A^-\,r_p
-c_B^{\Sigma}v_B^+.
\]
This relation retains the distinct one-sided pressure traces of the pure incompressible benchmark.

For a planar equilibrium interface,
\[
\kappas=0,
\qquad
\pminus=\pplus=p,
\qquad
r_p=1.
\]
This specialization permits introduction of a common thermodynamic pressure and a single planar reduced function \(\widetilde\sigma_{\mathrm{pl}}(p)\).

\begin{proposition}[Planar amount-only sign]
\label{prop:amount-only-sign}
Consider a regular differentiable planar equilibrium branch in the restricted autonomous amount-only class of \cref{def:restricted-amount-only} at fixed temperature, with pure incompressible reservoirs, fast exchange, \(c_A^{\Sigma},c_B^{\Sigma}>0\), and \(v_A^-,v_B^+>0\). Then
\[
\frac{\dd\widetilde\sigma_{\mathrm{pl}}}{\dd p}
=-\left(c_A^{\Sigma}v_A^-+c_B^{\Sigma}v_B^+\right)<0.
\]
\end{proposition}

\begin{proof}
Planarity gives a common equilibrium pressure and hence \(r_p=1\). Substitution into the amount-only Gibbs--Duhem response \(s_+=-c_A^{\Sigma}v_A^-r_p-c_B^{\Sigma}v_B^+\) gives the displayed expression. Strict negativity follows from positivity of both constituent surface amount densities and both molar volumes.
\end{proof}

The proposition is a theorem about this restricted model class and loading path. It is not a general theorem about liquid--liquid interfacial tension.

\subsection{Mutual Solubility}
\label{sec:mutual}
\label{app:solubility}

Finite mutual solubility modifies the pure-reservoir approximation. At fixed temperature, chemical equilibrium of a binary two-phase system gives common chemical-potential differentials \(\dd\mu_{A}^{\mathrm m}\) and \(\dd\mu_{B}^{\mathrm m}\), while the two phase Gibbs--Duhem relations read
\[
x_A^-\dd\mu_{A}^{\mathrm m}+x_B^-\dd\mu_{B}^{\mathrm m}
=v^-\dd\pminus,
\]
\[
x_A^+\dd\mu_{A}^{\mathrm m}+x_B^+\dd\mu_{B}^{\mathrm m}
=v^+\dd\pplus.
\]
Assume \(x_A^-\ne x_A^+\) so that the two bulk constraints are linearly independent. With \(\pplus\) as parameter and
\[
r_p=\frac{\dd\pminus}{\dd\pplus},
\]
inversion gives
\[
\frac{\dd\mu_{A}^{\mathrm m}}{\dd\pplus}
=
\frac{x_B^+v^-r_p-x_B^-v^+}
{x_A^- - x_A^+},
\]
\[
\frac{\dd\mu_{B}^{\mathrm m}}{\dd\pplus}
=
\frac{x_A^-v^+-x_A^+v^-r_p}
{x_A^- - x_A^+}.
\]
The amount-only surface response is therefore
\[
\frac{\dd\sigmatp}{\dd\pplus}
=
-c_A^{\Sigma}\frac{\dd\mu_{A}^{\mathrm m}}{\dd\pplus}
-c_B^{\Sigma}\frac{\dd\mu_{B}^{\mathrm m}}{\dd\pplus}.
\]
For a curved interface this relation is implicit because \(r_p\) is coupled to the surface-tension response by the normal momentum jump balance. In the planar limit \(r_p=1\), and the coexistence-path sign can be examined without introducing a primitive pressure dependence of the surface tension.

For the water--\(n\)-decane benchmark, let phase \(-\) be water rich and phase \(+\) decane rich, and write
\[
x_{\mathrm D}^-=\varepsilon,
\qquad
x_{\mathrm W}^+=\delta.
\]
Thus
\[
x_{\mathrm W}^-=1-\varepsilon,
\qquad
x_{\mathrm D}^+=1-\delta,
\qquad
D=1-\varepsilon-\delta>0.
\]
Along a planar isothermal common-pressure coexistence family,
\[
\frac{\dd\mu_{\mathrm W}^{\mathrm m}}{\dd p}
=
\frac{(1-\delta)v^- -\varepsilon v^+}{D},
\qquad
\frac{\dd\mu_{\mathrm D}^{\mathrm m}}{\dd p}
=
\frac{(1-\varepsilon)v^+ -\delta v^-}{D}.
\]
The exact positivity conditions are collected in \cref{res:finite-solubility-sign}.

Near room temperature the molar-volume ratio is of order \(v^-/v^+\simeq9\times10^{-2}\). The IUPAC--NIST compilation gives the tentative decane-in-water mole fraction \(\varepsilon=2.5\times10^{-9}\) near \(293\,\mathrm K\). The water-in-decane data have greater uncertainty. The older compilation lists \(\delta=2.3\times10^{-3}\) as an order-of-magnitude datum, whereas a later evaluator expects approximately \(6\times10^{-4}\) at \(293\,\mathrm K\) \cite{NISTDecaneWater,NISTDecaneWaterExpected}. These alternative values have different evidential status and should not be interpreted as two recommended measurements at the same state. Broader water--alkane evaluations support the same scale separation \cite{Maczynski2003,Tsonopoulos1999}.

Using the representative phase volumes \(v^-=18.05\,\mathrm{cm^3\,mol^{-1}}\) and \(v^+=195.05\,\mathrm{cm^3\,mol^{-1}}\), either alternative composition gives positive derivatives with a substantial separation from zero. For \(\delta=2.3\times10^{-3}\), for example, the water sign-change threshold is \(\varepsilon\simeq0.0923\), over seven orders of magnitude above the tentative input. The resulting derivatives are approximately \(18.05\) and \(195.46\,\mathrm{cm^3\,mol^{-1}}\). These calculations establish the sign at the representative states. They do not determine a coexistence family over a pressure interval.

\begin{result}[Finite-solubility sign robustness]
\label{res:finite-solubility-sign}
On any regular differentiable planar isothermal common-pressure water--\(n\)-decane coexistence branch, at a state with \(D=1-\varepsilon-\delta>0\), the chemical-potential derivatives are both positive if and only if
\[
\varepsilon<(1-\delta)\frac{v^-}{v^+},
\qquad
\delta<(1-\varepsilon)\frac{v^+}{v^-}.
\]
At either representative low-temperature composition above these inequalities hold with a substantial separation from zero. Consequently every binary amount-only surface with positive constituent amount densities at such a state with
\(
c_{\mathrm W}^{\Sigma}>0
\)
and
\(
c_{\mathrm D}^{\Sigma}>0
\)
has
\[
\frac{\dd\widetilde\sigma_{\mathrm{pl}}}{\dd p}<0.
\]
The strict-positivity hypothesis is the same interior-state assumption used in \cref{prop:amount-only-sign}. If a constituent is absent from the surface state, the reduced one-component boundary case must be formulated separately rather than inferred by continuity of a binary chemical potential at zero amount. Finite mutual solubility therefore cannot reverse the restricted amount-only sign at any interior binary surface state satisfying these explicit inequalities. Applying the result over an extended pressure interval requires verifying the inequalities along that coexistence path.
\end{result}

The magnitude of the pressure coefficient can be calculated using the following reference-volume decomposition, which introduces no additional physical constraint. Set
\[
\bm X=\begin{pmatrix}x_A^-&x_B^-\\x_A^+&x_B^+\end{pmatrix},\qquad
\bm v=\begin{pmatrix}v^-\\v^+\end{pmatrix}.
\]
Let \(\bm V^{\mathrm{ref}}=(V_A^{\mathrm{ref}},V_B^{\mathrm{ref}})^\mathsf T\) be fixed independently of the two coexistence equations, for example by a prescribed partial-molar-volume model or a specified reference-state convention, and write
\[
\bm v=\bm X\bm V^{\mathrm{ref}}+\bm e.
\]
Then
\[
\begin{pmatrix}
\dd\mu_{A}^{\mathrm m}/\dd p\\
\dd\mu_{B}^{\mathrm m}/\dd p
\end{pmatrix}
=
\bm V^{\mathrm{ref}}+\bm X^{-1}\bm e.
\]
This identity separates the chosen reference contribution from the residual only because \(\bm V^{\mathrm{ref}}\) has been specified independently. If \(\bm V^{\mathrm{ref}}\) is instead defined by \(\bm X^{-1}\bm v\) at the state under study, the residual vanishes identically. In that case the cancellation is an algebraic identity and provides no independent constitutive test. Quantitative calculations for real fluids must therefore use independently defined phase or partial-molar-volume information and state the reference convention explicitly.

Tsonopoulos reports weak pressure dependence of mutual solubilities in his low-temperature analysis of alkane--water phase behavior \cite{Tsonopoulos1999}. This observation is consistent with the solubilities remaining far from their sign-change thresholds near the representative states, but provides no measured or proved bound over Susnar's full \(0\)--\(35\,\mathrm{MPa}\) interval. Application over that interval requires verification of the inequalities in \cref{res:finite-solubility-sign} at each state on the coexistence path. Comparing the magnitudes of predicted and measured pressure coefficients additionally requires the actual pressure traces, volumetric properties, and curvature along that path.

\subsection{Curved-Interface Response}

\begin{result}[Pressure-ratio criterion independent of surface composition]
\label{res:curved-pressure-ratio}
Consider a differentiable isothermal binary coexistence branch with
\(0<x_A^+<x_A^-<1\), positive phase molar volumes, and \(\pplus\) a
regular local parameter. Set
\[
D=x_A^--x_A^+>0,
\qquad
r_p=\frac{\dd\pminus}{\dd\pplus},
\]
and define the dimensionless bounds
\[
R_{\min}=\frac{x_B^-v^+}{x_B^+v^-},
\qquad
R_{\max}=\frac{x_A^-v^+}{x_A^+v^-}.
\]
Then \(R_{\min}<R_{\max}\), and both coexistence-path derivatives
\(\dd\mu_{i}^{\mathrm m}/\dd\pplus\) are strictly positive if and only if
\[
R_{\min}<r_p<R_{\max}.
\]
For the restricted amount-only response, the slightly stronger statement
\[
\frac{\dd\sigmatp}{\dd\pplus}<0
\quad\hbox{for every }c_A^{\Sigma},c_B^{\Sigma}>0
\quad\Longleftrightarrow\quad
R_{\min}\le r_p\le R_{\max}
\]
holds as a statement about the Gibbs--Duhem response over all positive
vectors of constituent amount densities at the specified bulk state and along the specified pressure variation.
In particular, the closed-interval condition is sufficient for strict
negative response on every admissible restricted amount-only equilibrium
branch with positive constituent amount densities in the interior of the state domain. No tie-line-interior surface
composition or curvature sign hypothesis is needed.
\end{result}

\begin{proof}
The bulk Gibbs--Duhem inversion gives
\[
\chi_A^+:=\frac{\dd\mu_{A}^{\mathrm m}}{\dd\pplus}
=\frac{x_B^+v^-}{D}(r_p-R_{\min}),
\qquad
\chi_B^+:=\frac{\dd\mu_{B}^{\mathrm m}}{\dd\pplus}
=\frac{x_A^+v^-}{D}(R_{\max}-r_p).
\]
Moreover,
\[
R_{\max}-R_{\min}
=\frac{v^+D}{v^-x_A^+x_B^+}>0.
\]
The open-interval statement follows immediately. At either endpoint one
derivative vanishes and the other is positive. Equivalently, the plus-phase
Gibbs--Duhem equation requires
\(x_A^+\chi_A^++x_B^+\chi_B^+=v^+>0\).
Thus \(-c_A^{\Sigma}\chi_A^+-c_B^{\Sigma}\chi_B^+<0\) for every
vector of strictly positive constituent amount densities throughout the closed interval.
Outside that interval one derivative is negative and the other is
positive. Increasing the positive constituent amount density multiplying the negative
derivative relative to the other amount density then makes the algebraic
response positive. This proves sharpness over vectors of positive constituent amount densities.
It does not assert the existence of a constitutive equilibrium branch
realizing every such vector.
\end{proof}

For water--\(n\)-decane, these bounds become
\[
R_{\min}=\frac{\varepsilon v^+}{(1-\delta)v^-},
\qquad
R_{\max}=\frac{(1-\varepsilon)v^+}{\delta v^-}.
\]
Using the representative ambient phase volumes
\(v^-=18.05\,\mathrm{cm^3\,mol^{-1}}\),
\(v^+=195.05\,\mathrm{cm^3\,mol^{-1}}\), and
\(\varepsilon=2.5\times10^{-9}\), the alternative water-in-decane
compositions \(\delta=6\times10^{-4}\) and \(\delta=2.3\times10^{-3}\)
give, respectively,
\[
[R_{\min},R_{\max}]
\simeq[2.703\times10^{-8},1.801\times10^4],
\qquad
[2.708\times10^{-8},4.698\times10^3].
\]
These calculations concern representative individual states and do not provide measured bounds on
an archival high-pressure loading path. They show that a conservative
bound on the local pressure-derivative ratio can suffice to fix the sign
without determining the material surface composition. Prediction of the magnitude of the pressure coefficient still requires the surface state and constitutive
closure. The criterion uses only the two bulk Gibbs--Duhem relations and
the amount-only surface differential. It does not use the Laplace relation.
Consequently, the quantities entering this criterion can be obtained from pressure or curvature measurements
only after specifying the appropriate mechanical balance and the relation between the apparatus pressure
and the local pressure traces. If the closed-interval condition holds
at every state of a regular path, the tension decreases strictly along
that path.

For the pure incompressible benchmark the normal jump balance gives, along the plus-pressure parameterization,
\[
r_p
=1-\kappas s_+-\sigma
\frac{\dd\kappas}{\dd\pplus}.
\]
Let
\[
a=c_A^{\Sigma}v_A^-,
\qquad
b=c_B^{\Sigma}v_B^+.
\]
Substitution into the amount-only relation yields
\[
\left(1-a\kappas\right)s_+
=-(a+b)
+a\sigma\frac{\dd\kappas}{\dd\pplus}.
\]
At fixed curvature, provided the reduced parameterization is regular in the sense that \(1-a\kappas\neq0\),
\[
s_+
=-\frac{a+b}{1-a\kappas}.
\]
At fixed curvature, if \(1-a\kappas=0\), this particular implicit reduction is singular and \(p^+|_{\Sigma}\) cannot be used as a regular local parameter without returning to the unreduced equilibrium system. More generally, at any regular state with \(a>0\) and \(1-a\kappas>0\), the same identity gives the exact pointwise criterion
\[
\frac{\dd\sigmatp}{\dd\pplus}<0
\quad\Longleftrightarrow\quad
\sigma\frac{\dd\kappas}{\dd\pplus}
<1+\frac{b}{a}.
\]
Thus reversal of the amount-only sign requires either a sufficiently large curvature derivative or loss of the regular branch on which the denominator is positive. For a convex minus-phase drop with the orientation fixed in \cref{sec:equilibrium}, \(\kappas<0\) and hence \(1-a\kappas>0\) automatically. At fixed curvature the response is strictly negative. This criterion quantifies the curvature evolution required for sign reversal within the pure-reservoir benchmark.

\begin{corollary}[Curvature-reversal lower bound]
\label{cor:curvature-reversal}
Under the hypotheses of the curved pure incompressible amount-only benchmark, consider a regular convex minus-phase branch with \(a>0\), \(b>0\), \(\sigma>0\), and \(1-a\kappas>0\). If
\[
\frac{\dd\sigmatp}{\dd\pplus}\ge0,
\]
then necessarily
\[
\frac{\dd\kappas}{\dd\pplus}
>\frac1{\sigma}.
\]
If the nonnegative slope persists on an interval \(p_0^+<p_1^+\) and \(0<\sigma\le\sigma_{\max}\) there, then
\[
\kappas(p_1^+)-\kappas(p_0^+)
>\frac{p_1^+-p_0^+}{\sigma_{\max}}.
\]
If the branch remains convex throughout that interval, so that \(\kappas<0\), this further implies the necessary pressure-span bound
\[
p_1^+-p_0^+
<\sigma_{\max}\left|\kappas(p_0^+)\right|.
\]
\end{corollary}

\begin{proof}
Because the denominator is positive, a nonnegative \(s_+\) in the identity above requires
\[
a\sigma\frac{\dd\kappas}{\dd\pplus}
\ge a+b.
\]
Since \(a,b,\sigma>0\), this gives
\[
\frac{\dd\kappas}{\dd\pplus}
\ge\frac{1+b/a}{\sigma}
>\frac1{\sigma}.
\]
Since \(\dd\kappas/\dd\pplus>1/\sigma_{\max}\) on the interval, the mean value theorem gives the interval bound. If the branch remains convex, \(\kappas(p_1^+)<0\), so
\[
\kappas(p_1^+)-\kappas(p_0^+)
< -\kappas(p_0^+)
=\left|\kappas(p_0^+)\right|,
\]
which yields the pressure-span bound.
\end{proof}

The corollary eliminates the unknown positive constituent amount densities from a necessary condition for sign reversal. Even with the conservative bound \(\sigma_{\max}=0.1\,\mathrm{N\,m^{-1}}\), sustaining a nonnegative amount-only slope over \(1\,\mathrm{MPa}\) of local plus-side pressure would require an increase of curvature exceeding \(10^7\,\mathrm{m^{-1}}\). Over \(35\,\mathrm{MPa}\) the lower bound is \(3.5\times10^8\,\mathrm{m^{-1}}\). Equivalently, a convex branch beginning with \( |\kappas|=10^3\,\mathrm{m^{-1}}\), representative only as an order-of-magnitude macroscopic-drop curvature, could not sustain such a nonnegative slope over \(100\,\mathrm{Pa}\) under that conservative tension bound without ceasing to be convex. Application still requires a relation between apparatus pressure and \(\pplus\). The finite-solubility equations nevertheless allow the curvature bound to be extended under one explicit surface-composition condition.

\begin{result}[Finite-solubility curvature bound for tie-line-interior surface composition]
\label{res:finite-solubility-curvature}
Consider the binary curved amount-only surface on a regular differentiable isothermal coexistence branch satisfying the hydrostatic isotropic normal balance \(\pminus-\pplus=-\sigma\kappas\), with no additional normal-force or momentum-flux contribution. At a state with \(x_A^- > x_A^+\), positive phase molar volumes, positive total surface amount density, \(\sigma>0\), \(\kappas<0\), and with \(\pplus\) a regular local parameter, define the surface mole fraction \(x_A^{\Sigma}=c_A^{\Sigma}/(c_A^{\Sigma}+c_B^{\Sigma})\). If
\[
x_A^+<x_A^{\Sigma}<x_A^-,
\]
then a nonnegative local pressure response at that state implies
\[
\frac{\dd\kappas}{\dd\pplus}>\frac{1}{\sigma}.
\]
If all of these hypotheses---including convexity \(\kappas<0\), regularity of the \(\pplus\) parameterization, the tie-line-interior condition, and nonnegative response---persist on an interval and \(0<\sigma\le\sigma_{\max}\) there, then the integrated curvature and convex-branch pressure-span conclusions of \cref{cor:curvature-reversal} follow.
\end{result}

\begin{proof}
Let \(D=x_A^- - x_A^+>0\), with \(x_B^{\pm}=1-x_A^{\pm}\). Using the finite-solubility derivatives above with \(r_p=\dd\pminus/\dd\pplus\), the amount-only response can be collected as
\[
s_+ = -\alpha_{\Sigma} r_p-\beta_{\Sigma},
\]
where
\[
\alpha_{\Sigma}
=\frac{v^-}{D}
\left(c_A^{\Sigma}x_B^+-c_B^{\Sigma}x_A^+\right),
\qquad
\beta_{\Sigma}
=\frac{v^+}{D}
\left(c_B^{\Sigma}x_A^--c_A^{\Sigma}x_B^-\right).
\]
The inequalities \(x_A^+<x_A^{\Sigma}<x_A^-\) are equivalent to \(\alpha_{\Sigma}>0\) and \(\beta_{\Sigma}>0\). Substitution of
\[
r_p=1-\kappas s_+-\sigma\frac{\dd\kappas}{\dd\pplus}
\]
gives
\[
\left(1-\alpha_{\Sigma}\kappas\right)s_+
=-(\alpha_{\Sigma}+\beta_{\Sigma})
+\alpha_{\Sigma}\sigma\frac{\dd\kappas}{\dd\pplus}.
\]
Because the branch is convex, \(\kappas<0\), and hence \(1-\alpha_{\Sigma}\kappas>0\). If \(s_+\ge0\), then
\[
\alpha_{\Sigma}\sigma\frac{\dd\kappas}{\dd\pplus}
\ge\alpha_{\Sigma}+\beta_{\Sigma}
>\alpha_{\Sigma},
\]
which yields the stated strict bound. The integrated consequences follow exactly as in \cref{cor:curvature-reversal}.
\end{proof}

Unlike the composition-independent planar criterion in \cref{res:finite-solubility-sign}, this curvature bound uses an explicit tie-line-interior surface-composition hypothesis and does not require the planar criterion.

For water--\(n\)-decane near room temperature, the tie-line-interior
condition covers a broad range of mixed material surface compositions,
but it is not automatic. The two results for curved interfaces therefore use different hypotheses. The tie-line-interior result yields
a necessary bound on the curvature derivative for nonnegative
response that is independent of the constituent amount densities, whereas \cref{res:curved-pressure-ratio} fixes the response sign
from a bound on the local pressure ratio without a surface-composition
assumption. Application to archival measurements additionally requires verification that the measurements represent an
equilibrium branch. The integrated curvature bound additionally assumes
pointwise nonnegative response throughout its interval. A positive
regression slope alone does not establish that premise.

For an isothermal enriched binary surface in bulk--surface chemical equilibrium, with internal descriptors absent, fixed, or equilibrated, the same Gibbs--Duhem elimination gives
\[
s_+=\Vsurf\frac{\dd\PiV}{\dd\pplus}
      -\alpha_{\Sigma}r_p-\beta_{\Sigma}.
\]
Under the isotropic normal balance
\(\pminus-\pplus=-\sigma\kappas\), this becomes
\begin{equation}
(1-\alpha_{\Sigma}\kappas)s_+
=\Vsurf\frac{\dd\PiV}{\dd\pplus}
 -(\alpha_{\Sigma}+\beta_{\Sigma})
 +\alpha_{\Sigma}\sigma\frac{\dd\kappas}{\dd\pplus}.
\label{eq:enriched-finite-solubility-curved-response}
\end{equation}
The coefficients \(\alpha_{\Sigma}\) and \(\beta_{\Sigma}\) are those
defined above. In the pure incompressible limit they reduce to \(a\)
and \(b\), respectively. The structural loading derivative must be
determined from the structural mechanical closure and all retained jump
conditions. Additional normal forces or momentum-flux contributions
must be included in the normal balance before it is differentiated.

\section{Classical Invariant Pressure Thermodynamics}
\label{sec:invariant}

The pressure response of an equilibrium liquid--liquid interface has a classical thermodynamic description that predates any present state-enrichment proposal. The pressure coefficient is an invariant observable, although the intermediate surface coordinates depend on the interface representation. This section derives the invariant coefficient with an explicit orientation convention for comparison with the surface theory with positive constituent amount densities.
Throughout this section, \(\gamma\) denotes the scalar equilibrium fluid interfacial tension in the notation conventional to the cited surface-chemistry literature. For the isotropic fluid interfaces considered here it represents the same physical area conjugate denoted by \(\sigma\) elsewhere in this paper. The symbol is retained to preserve the notation of the classical formulas.

\subsection{Finite-Region Representation}

Consider first a planar finite interfacial region of arbitrary physical width \(\tau\). Its two bounding planes are taken in the homogeneous bulk portions on either side and far enough from the transition zone that the selected region contains the complete inhomogeneous interfacial region. Admissible changes of the finite-region width below keep each bounding plane in its corresponding homogeneous bulk portion. In a Guggenheim-type construction the selected region contains actual constituent amounts per area \(\Gamma_{i\tau}\) and actual volume per area \(\tau\). These quantities depend on the arbitrary placement of the two finite-region boundaries and therefore cannot themselves be the final coefficients in an invariant adsorption equation.

The comparison requires an explicit derivation of the finite-region invariant pressure coefficient. Radke's finite-region differential, following Guggenheim, is \cite{Guggenheim1940,Radke2015}
\[
-\dd\gamma
=s_{\tau}^{\gamma}\dd T
-\tau\dd p
+\sum_i\Gamma_{i\tau}\dd\mu_{i}^{\mathrm m}.
\]
For a binary system at fixed temperature, assume the two bulk Gibbs--Duhem constraints are linearly independent, equivalently
\(
D_c\neq0
\), and define
\[
D_c=c_1^-c_2^+-c_1^+c_2^-,
\qquad
\chi_1=\frac{c_2^+-c_2^-}{D_c},
\qquad
\chi_2=\frac{c_1^- -c_1^+}{D_c}.
\]
The two bulk Gibbs--Duhem equations give \(\dd\mu_{i}^{\mathrm m}/\dd p=\chi_i\). Hence
\begin{equation}
\label{eq:tauHcorrected}
\frac{\dd\widetilde\gamma_{\mathrm{pl}}}{\dd p}
=
\tau
-\Gamma_{1\tau}\chi_1
-\Gamma_{2\tau}\chi_2
=: \tau_{\mathrm H}^{\mathrm{inv}}.
\end{equation}
The physical width is therefore part of the invariant combination. The constituent-amount contribution \(\Gamma_{1\tau}\chi_1+\Gamma_{2\tau}\chi_2\) is not invariant.

The invariance can be checked without any interfacial model. If the minus-side boundary is displaced within the homogeneous minus bulk by \(\delta h^-\) and the plus-side boundary within the homogeneous plus bulk by \(\delta h^+\), while the selected region continues to contain the full transition zone and both displacements are taken positive when the region is enlarged, then
\[
\tau\mapsto\tau+\delta h^-+\delta h^+,
\qquad
\Gamma_{i\tau}\mapsto
\Gamma_{i\tau}+c_i^-\delta h^-+c_i^+\delta h^+.
\]
The bulk constraints imply
\[
c_1^-\chi_1+c_2^-\chi_2=1,
\qquad
c_1^+\chi_1+c_2^+\chi_2=1,
\]
so every boundary-dependent increment cancels in \cref{eq:tauHcorrected}. Thus \(\tau_{\mathrm H}^{\mathrm{inv}}\) is independent of the arbitrary finite-region boundaries although \(\tau\) and each \(\Gamma_{i\tau}\) are not.

The equivalence with the Gibbs representation can also be shown directly, without invoking any limiting argument. Choose any dividing plane \(b\) inside the finite region and let \(h^-\) and \(h^+\) be its distances to the minus- and plus-side region boundaries, so \(\tau=h^-+h^+\). The Gibbs excess relative to that plane is
\[
\Gamma_i(b)
=\Gamma_{i\tau}-h^-c_i^- -h^+c_i^+.
\]
Using the two preceding bulk identities gives
\[
\tau_{\mathrm H}^{\mathrm{inv}}
=-\Gamma_1(b)\chi_1-\Gamma_2(b)\chi_2,
\]
which is exactly the Gibbs pressure coefficient derived in the next subsection. Thus boundary invariance of the finite-region expression and gauge invariance of the dividing-surface expression are the same cancellation written in two coordinate systems.

The comparison with Hansen and Good concerns the invariant pressure observable \cite{Hansen1962,Good1982II,Good1982III}. In general \(\tau_{\mathrm H}^{\mathrm{inv}}\neq\tau\). The measurable coefficient therefore differs from the physical width of the selected interphase. This distinction is needed for comparison with the geometric-thickness specialization below.

\subsection{Dividing-Surface Representation and Oriented Distance}
\label{app:gibbsinvariant}

The same observable can be written in a Gibbs zero-volume representation. Let a coordinate \(z\) increase from the minus phase to the plus phase, and, assuming that both improper integrals converge, define the surface excess at a dividing plane \(z=b\) by
\[
\Gamma_i(b)
=
\int_{-\infty}^{b}
\left(c_i(z)-c_i^-\right)\dd z
+
\int_b^{\infty}
\left(c_i(z)-c_i^+\right)\dd z.
\]
Then
\[
\frac{\dd\Gamma_i}{\dd b}=c_i^+-c_i^-,
\]
so displacement of the dividing plane is a gauge transformation of the excess coordinates rather than a material evolution.

Using the coexistence coefficients \(\chi_i=\dd\mu_i^{\mathrm m}/\dd p\) defined before \cref{eq:tauHcorrected}, with \(D_c\neq0\), the Gibbs adsorption equation gives
\[
\frac{\dd\widetilde\gamma_{\mathrm{pl}}}{\dd p}
=-\frac{
\Gamma_1(b)(c_2^+-c_2^-)
+\Gamma_2(b)(c_1^- -c_1^+)
}{D_c}.
\]
The combination is independent of \(b\). The gauge terms cancel exactly when the two bulk constraints are used.

To state the distance interpretation with explicit index and orientation conventions, consider components for which \(c_i^+\neq c_i^-\). Because \(\Gamma_i(b)\) is then affine in \(b\) with nonzero slope, there is a unique zero-excess plane \(z_i\) satisfying \(\Gamma_i(z_i)=0\). Define in this paper
\[
\lambda_{ji}^{\mathrm G}:=z_j-z_i.
\]
Since
\[
\Gamma_i(z_j)=(z_j-z_i)(c_i^+-c_i^-),
\]
choosing $b=z_1$ gives the exact zero-excess-plane representation
\begin{equation}
\frac{\dd\widetilde\gamma_{\mathrm{pl}}}{\dd p}
=\mathcal F_c\,\lambda_{21}^{\mathrm G},
\qquad
\mathcal F_c
:=\frac{(c_2^+-c_2^-)(c_1^- -c_1^+)}{D_c}.
\label{eq:gibbs-distance-factor}
\end{equation}
Thus the invariant coefficient and the geometric separation of the two zero-excess planes are not identical at arbitrary finite mutual solubility. In the nondegenerate nearly immiscible limit \(u:=c_1^+/c_1^-\to0\) and \(w:=c_2^-/c_2^+\to0\), one has \(\mathcal F_c=(1-u)(1-w)/(1-uw)\to1\), so
\[
\frac{\dd\widetilde\gamma_{\mathrm{pl}}}{\dd p}
\simeq z_2-z_1
=\lambda_{21}^{\mathrm G}.
\]
Reversing the order of the two zero-excess planes reverses the sign. Use of a historical symbol such as \(\lambda_{12}\) therefore requires the original orientation and index conventions. Good's 1976 paper writes the compact pressure--distance identity in its summary, but the derivation in the body invokes low mutual solubility before obtaining that result \cite{Good1976}. Good's 1982 papers subsequently pursued a broader identification of Gibbsian distance parameters with pressure coefficients and surface-excess-volume quantities \cite{Good1982II,Good1982III}. That geometrical identification later formed part of the Good--Motomura discussion reviewed by Radke \cite{Radke2015}. The present argument therefore does not require an exact identification of the invariant coefficient with the geometric distance between the zero-excess planes. The exact finite-solubility statement is \cref{eq:gibbs-distance-factor}, while the invariant Gibbs excess combination and the finite-region coefficient remain exact.

\subsection{Invariant Comparison and State Selection}

The invariant equivalence of the Gibbs and finite-region adsorption equations \cite{Radke2015} underlies the comparison. The finite-solubility distance factor is given by \cref{eq:gibbs-distance-factor}. Motomura and coworkers used pressure-dependent oil--water measurements to compare Gibbs excess quantities, Hansen-type distance parameters, and interface-formation-volume descriptions \cite{Motomura1983}. A recent molecular simulation provides a complementary example. Sega studies the anisotropic pressure response of a water--carbon-tetrachloride interface and reports interfacial excess-volume and excess-normal-compliance quantities \cite{Sega2026}. These remain excess observables rather than an independent coordinate describing the material content of the sharp surface.

These invariant identities do not identify the classical coordinates with the positive phenomenological state variables of the present theory. The \(c_i^{\Sigma}\) are actual non-negative constituent amount densities by model construction and do not transform under a Gibbs gauge shift. The structural coordinate \(\Vsurf\) is likewise not defined as \(\tau_{\mathrm H}^{\mathrm{inv}}\), as an oriented Gibbsian distance, or as the arbitrarily chosen physical width \(\tau\). The two representations are compared through their predictions for the same observable. On a specified planar equilibrium path with internal descriptors absent, fixed along the path, or independently equilibrated, the enriched material surface predicts
\[
\frac{\dd\widetilde\sigma_{\mathrm{pl}}}{\dd p}
=-\sum_i c_i^{\Sigma}
\frac{\dd\mu_{i}^{\Sigma,\mathrm m}}{\dd p}
+\Vsurf\frac{\dd\PiV}{\dd p},
\]
and consistency with the measured invariant requires comparison of this complete coefficient with \(\tau_{\mathrm H}^{\mathrm{inv}}\), after the thermodynamic path and sign convention have been matched.

This comparison specifies what a measured pressure coefficient can imply for the choice of surface state. A positive coefficient does not indicate a failure of Gibbs thermodynamics, nor does it imply negative physical material inventories. Once equilibrium status has been established and the experimental loading path has been shown to correspond to the theoretical path on which \cref{prop:amount-only-sign} applies, a positive measured coefficient excludes the restricted amount-only model with positive constituent amount densities for that observable. It does not identify a unique additional coordinate. Indeed, \cref{app:additive} shows that any independently retained additive surface content contributes its own term to the area Legendre structure.

The choice of structural volume per area \(\Vsurf\) as an additional state variable is motivated by four properties. A finite interphase has a volume that is omitted from the amount-only sharp state. When retained as an independent additive coordinate, this volume contributes to the Euler relation. Its conjugate has the dimensions and mechanical character of pressure. The classical invariant pressure coefficient also has the form of an excess-volume-like length. The geometric specialization in \cref{sec:volume} adds a mechanical criterion by constructing the conjugate directly from one-sided normal work. These arguments motivate retaining \(\Vsurf\) as one additional additive scalar with a mechanical interpretation. They do not establish uniqueness. State selection concerns which information the chosen phenomenological representation must retain. Classical invariant equivalence concerns how the same observable is encoded in another representation.

\section{Pressure Loading and Structural Compatibility}
\label{app:pressure-loading}

\subsection{Pressure Paths and Curved Structural Loading}

Choose the plus-side thermodynamic pressure trace as a local parameter,
\[
\lambda=\pplus.
\]
For a curved interface subject to a body force, \(p^{\pm}|_{\Sigma}\) and \(\kappas\) are generally spatially varying fields along the surface rather than spatially constant quantities. The parameterization here is therefore understood either locally at a specified reference point of the interface (for example the apex of an axisymmetric drop) or in a setting where those traces are spatially uniform. A complete drop-shape reduction must include the hydrostatic and geometric field relations that connect the chosen reference trace to the rest of the interface. Only after the coupled constraints in \cref{sec:equilibrium} have been solved do we define
\[
\sigmatp(\pplus)
=
\widehat\sigma^{\Sigma}
\left(
T,\widetilde{\bm z}^{\Sigma}(\pplus)
\right).
\]
An equivalent local parameterization by \(\pminus\) gives \(\sigmatm(\pminus)\). Where both parameterizations are regular,
\[
\frac{\dd\sigmatp}{\dd\pplus}
=
\frac{\dd\sigmatm}{\dd\pminus}
\frac{\dd\pminus}{\dd\pplus}.
\]
The intrinsic constitutive law does not take bulk thermodynamic pressure as an independent argument.

Along an equilibrium loading path with internal descriptors absent, fixed along the path, or independently equilibrated, the enriched Gibbs--Duhem relation gives
\[
\frac{\dd\widetilde\sigma}{\dd\lambda}
=
-\eta^{\Sigma}\frac{\dd T}{\dd\lambda}
-\sum_i c_i^{\Sigma}
\frac{\dd\mu_{i}^{\Sigma,\mathrm m}}{\dd\lambda}
+\Vsurf\frac{\dd\PiV}{\dd\lambda}.
\]
For fixed temperature, only the chemical-potential and structural terms remain.

Curvature is part of the same reduced problem. Differentiating
\[
\pminus-\pplus=-\sigma\kappas
\]
along a loading path yields
\[
\frac{\dd\pminus}{\dd\lambda}
-
\frac{\dd\pplus}{\dd\lambda}
=
-\kappas\frac{\dd\widetilde\sigma}{\dd\lambda}
-\sigma\frac{\dd\kappas}{\dd\lambda}.
\]
Thus a pressure-response experiment on a drop or bubble probes a coupled path involving the second pressure trace and, if the shape changes, the curvature response. The planar coefficient at common pressure follows as a special case. Bulk body forces enter through the relation between the imposed and local pressures determined by hydrostatics. If a surface body-force density is retained, it must also enter the normal and tangential momentum balances before pressure-path differentiation.
Curved equilibrium interfaces require a different structural-loading relation. The following reduction assumes that the normal balance contains only isotropic tension and the stated recoil contribution. If a geometric or constraint force \(f_{\mathrm{geom},n}\) is retained, it must be added to every occurrence of \(-\sigma\kappas+\mathcal R_n\) before differentiation. The fixed-curvature structural virtual mode alone does not determine that additional force. If \(\pplus\) is used as a local parameter and isotropic bulk stress gives \(\pi_n^+=\pplus\), then structural equilibrium yields
\[
\frac{\dd\PiV}{\dd\pplus}
=
1
+w^-\frac{\dd}{\dd\pplus}
\left(-\sigma\kappas+\mathcal R_n\right)
+
\left(-\sigma\kappas+\mathcal R_n\right)
\frac{\dd w^-}{\dd\pplus}.
\]
Even with \(\mathcal R_n=0\), this contains both
\[
-\kappas\frac{\dd\sigmatp}{\dd\pplus}
\qquad\text{and}\qquad
-\sigma\frac{\dd\kappas}{\dd\pplus},
\]
as well as the evolution of the geometric loading coefficient. These signs are the same as those obtained by differentiating the normal jump relation \(\pminus-\pplus=-\sigma\kappas\). Curvature therefore enters the structural closure and generally changes the loading derivative from its unit value in the planar case.

\subsection{Conditional Finite-Region Compatibility of the Pressure Coefficient}

The following mapping gives equal pressure coefficients provided that the mapped states form an admissible sharp-surface constitutive family.

\begin{proposition}[Conditional finite-region compatibility of the pressure coefficient]
\label{prop:finite-region-bridge}
Consider a planar binary finite interfacial region of physical width \(\tau\) with actual constituent amounts per area \(\Gamma_{i\tau}\), bounded in homogeneous bulk on both sides so that the complete transition zone is enclosed, on a common-pressure isothermal coexistence path for which the two bulk Gibbs--Duhem constraints are linearly independent. Boundary variations used below are restricted to shifts that keep the respective boundaries in homogeneous bulk while continuing to enclose the transition zone. Map those finite-region contents to the geometric positive-material sharp-surface subclass by
\[
\Vsurf=\tau,
\qquad
c_i^{\Sigma}=\Gamma_{i\tau}.
\]
Assume that these mapped area-specific quantities define a differentiable family of admissible sharp-surface constitutive states along the same coexistence path, and that bulk--surface chemical equilibrium and the planar structural equilibrium of \cref{prop:planar-geometric-loading} hold along that family. Assume also that any additional internal descriptors are absent, fixed along the family, or equilibrated, so that \(\sum_a\mathcal A_a^\Sigma\,\dd\xi_a^\Sigma/\dd p=0\). Then the pressure coefficient of the enriched sharp surface, evaluated on the mapped family, equals the finite-region invariant reconstructed in \cref{eq:tauHcorrected}:
\[
\frac{\dd\widetilde\sigma_{\mathrm{pl}}}{\dd p}
=
\tau-\sum_i\Gamma_{i\tau}\frac{\dd\mu_{i}^{\mathrm m}}{\dd p}
=
\tau_{\mathrm H}^{\mathrm{inv}}.
\]
The finite-region expression on the right is invariant under every admissible homogeneous-bulk boundary shift. If the shifted finite-region contents also define an admissible mapped sharp-surface family satisfying the same equilibrium hypotheses, the sharp-surface response therefore has the same value. By contrast, the amount-only response expression obtained from the same mapped constituent amount densities by omitting the \(\Vsurf\) term changes by the negative of the added finite-region width and is not boundary invariant.
\end{proposition}

\begin{proof}
Because chemical equilibrium holds along the mapped differentiable family, differentiating \(\mu_{i}^{\Sigma,\mathrm m}(p)=\mu_{i}^{\mathrm m}(p)\) equates the surface and bulk coexistence-path chemical-potential derivatives. By \cref{prop:planar-geometric-loading}, planar structural equilibrium along the same family gives \(\dd\PiV/\dd p=1\). The stated internal-descriptor condition makes the corresponding affinity contribution vanish. The enriched Gibbs--Duhem response therefore gives
\[
\frac{\dd\widetilde\sigma_{\mathrm{pl}}}{\dd p}
=-\sum_i c_i^{\Sigma}\frac{\dd\mu_{i}^{\mathrm m}}{\dd p}+\Vsurf
=\tau-\sum_i\Gamma_{i\tau}\chi_i,
\]
where \(\chi_i=\dd\mu_{i}^{\mathrm m}/\dd p\). This is \cref{eq:tauHcorrected}. If the minus and plus finite-region boundaries are enlarged by admissible homogeneous-bulk shifts \(\delta h^-\) and \(\delta h^+\), then
\[
\delta\Vsurf=\delta h^-+\delta h^+,
\qquad
\delta c_i^{\Sigma}=c_i^-\delta h^-+c_i^+\delta h^+.
\]
The bulk Gibbs--Duhem constraints imply \(\sum_i c_i^-\chi_i=\sum_i c_i^+\chi_i=1\), so the total change of the finite-region coefficient expression is zero. Whenever the shifted mapped contents satisfy the same sharp-state admissibility and equilibrium hypotheses, the first part of the proposition identifies that unchanged expression with the corresponding sharp-surface response as well. If the structural term is omitted while the mapped amounts are retained, the resulting amount-only response expression changes instead by
\[
-\sum_i\delta c_i^{\Sigma}\chi_i
=-(\delta h^-+\delta h^+),
\]
which depends on the arbitrary finite-region boundary displacement.
\end{proof}

Changing finite-region boundaries changes the actual contents assigned by \((\tau,\Gamma_{i\tau})\mapsto(\Vsurf,c_i^\Sigma)\). This change is not a Gibbs gauge transformation within one fixed sharp-surface model. Each shifted mapping preserves the equality of pressure coefficients only if it satisfies the same constitutive-admissibility and equilibrium hypotheses. The proposition guarantees neither admissibility for arbitrary boundary choices nor equality of complete free-energy functions or of the tension values themselves. It does not identify \(\Vsurf\) with \(\tau_{\mathrm H}^{\mathrm{inv}}\) or turn material inventories into Gibbs excesses.

The assumed constitutive admissibility imposes a further necessary condition. If the mapped family belongs to a strictly stable independent planar subclass, \cref{prop:pressure-susceptibility} implies, with \(\bm q=(\Gamma_{1\tau},\Gamma_{2\tau},\tau)\) and \(\bm r=(\mu_{1}^{\mathrm m},\mu_{2}^{\mathrm m},-p)\),
\[
(\bm q')^{\mathsf T}\bm r'
=(\bm q')^{\mathsf T}\bm H\bm q'>0.
\]
Equivalently, \(\sum_i\Gamma_{i\tau}'(\mu_{i}^{\mathrm m})'-\tau'>0\) on such a regular branch. This inequality is necessary for the mapped branch to be compatible with strict material-state stability. It does not establish the existence of a free energy whose equilibrium states realize an arbitrary mapped family. It gives a constitutive restriction in addition to the equality of pressure coefficients.

Exact finite-region boundary freedom also imposes a limitation on strict
constitutive convexity. Here let \(a_{\mathrm{eff}}^{\Sigma}(\bm q)\) be the \(C^2\) effective density with additional descriptors held at common fixed values or already smoothly equilibrated, and let \(\bm H=\nabla_{\bm q}^2a_{\mathrm{eff}}^{\Sigma}\). Set
\[
\bm B^{\pm}=(c_1^{\pm},c_2^{\pm},1)^{\mathsf T}.
\]
If all homogeneous-slab extensions at fixed reservoir variables were
distinct equilibria of this effective free energy, then \(\nabla_{\bm q}a_{\mathrm{eff}}^{\Sigma}=\bm r\) would be constant along
both extension directions. Consequently,
\[
\bm H\bm B^-=\bm0,\qquad \bm H\bm B^+=\bm0.
\]
Since \(D_c\ne0\), these directions are linearly independent, and
\(\bm H\) cannot be positive definite on the full three-dimensional
state space. Strict convexity on a convex constitutive domain likewise
makes the gradient injective and excludes distinct equilibria at the same
\(\bm r\). The strict-stability assumption therefore applies to a
specified constitutive realization. Arbitrary finite-region boundary
assignments do not simultaneously define distinct strictly stable states
of that realization.

\subsection{Observable Residuals, Constraints, and Identifiability}

For the planar pure incompressible binary benchmark, \cref{prop:planar-geometric-loading} gives \(\dd\widetilde\sigma_{\mathrm{pl}}/\dd p=\ell_{\mathrm{res}}\), where
\[
\ell_{\mathrm{res}}
=
\Vsurf-c_A^{\Sigma}v_A-c_B^{\Sigma}v_B.
\]
Within the geometric subclass, a planar equilibrium pressure coefficient identifies this combination rather than \(\Vsurf\) separately. It has the same dimensions as the classical invariant \(\tau_{\mathrm H}^{\mathrm{inv}}\). Under the specific finite-region lumping map of \cref{prop:finite-region-bridge},
\[
\ell_{\mathrm{res}}=\tau_{\mathrm H}^{\mathrm{inv}}
\]
holds identically at the pressure-coefficient level. For a general phenomenological surface whose primitive state has not been calibrated by that map, the same equality is instead an observable-level model test. In neither case is it the definition of \(\Vsurf\), nor does it identify the primitive state coordinates across representations.

The inverse problem is correspondingly underdetermined by a single static pressure slope. Even when the bulk molar volumes are independently known, one scalar measurement constrains only the combination \(\Vsurf-\sum_i c_i^{\Sigma}v_i\). It does not separately determine \(\Vsurf\) and the positive constituent amount densities. For real mixtures the more general identified combination is
\[
-\sum_i c_i^{\Sigma}\frac{\dd\mu_{i}^{\mathrm m}}{\dd p}
+\Vsurf\frac{\dd\PiV}{\dd p}.
\]
An independent observable must be mapped to the positive material state before it can resolve this ambiguity. A Gibbs adsorption is an excess and cannot be substituted directly for \(c_i^{\Sigma}\). A surface mole fraction does not determine the total amount per area. A separately calibrated equation of state, measurements related to the constituent amount densities by a specified measurement model, or independent thermodynamic loading directions are needed. The explicit models in \cref{sec:constructive} show that even identical complete pressure--tension curves can coexist with different state responses. Because an enriched free energy may couple composition and structural volume, the equilibrium amounts of an enriched model need not equal those of a restricted amount-only model at the same reservoir state.

\label{sec:holonomic-main}

A static residual does not establish that \(V^{\Sigma}\) is an independent coordinate. For a homogeneous patch, suppose instead that the structural content is constrained by a differentiable degree-one relation
\[
V^{\Sigma}=V^{\Sigma}(N_1^{\Sigma},\ldots,N_N^{\Sigma}),
\qquad
\bar v_i^{\Sigma}:=\frac{\partial V^{\Sigma}}{\partial N_i^{\Sigma}},
\qquad
V^{\Sigma}=\sum_i \bar v_i^{\Sigma}N_i^{\Sigma}.
\]
Then
\[
\dd V^{\Sigma}=\sum_i\bar v_i^{\Sigma}\dd N_i^{\Sigma},
\]
and restriction of the enriched free energy to this manifold changes the intrinsic amount derivative to
\[
\widetilde\mu_{i}^{\Sigma,\mathrm m}=\mu_{i}^{\Sigma,\mathrm m}-\PiV\bar v_i^{\Sigma}.
\]
Elimination of the independent structural variation preserves the induced volume-work contribution. In the literal planar geometric subclass the mechanically appropriate reservoir stationarity condition is
\[
\widetilde\mu_{i}^{\Sigma,\mathrm m}+p\bar v_i^{\Sigma}=\mu_{i}^{\mathrm{bulk},\mathrm m},
\]
so the constrained volume increment still carries hydrostatic work. Consequently a holonomically structured surface may have a positive static pressure coefficient even though it possesses no independent structural relaxation mode. The complete derivation, including the local-integral interpretation on heterogeneous patches and the distinction between intrinsic constrained and mechano-chemical exchange potentials, is given in Appendix~\ref{sec:holonomic}.

The choice of surface state is therefore assessed in successive steps. A pressure-sign experiment can exclude \cref{def:restricted-amount-only}. Static data can then be used to assess models retaining structural information. Establishing whether the structural coordinate is independent requires additional constitutive information, measurements along multiple thermodynamic paths, or dynamical identification.

\section{Pressure Susceptibility and Static Identifiability Examples}
\label{sec:constructive}

\subsection{Stable Pressure Response}
\label{sec:constructive-models}

The following supplementary constitutive models compare the static and dynamic pressure responses of the selected state spaces. The principal mass--thickness constitutive example is given in \cref{sec:mass-thickness-model}. Throughout the next three subsections,
temperature and local area are fixed, the interface is planar, the structural
volume has the specified geometric normalization, and surface superscripts are
suppressed. All statements concern interior states of the material constitutive domain on a regular
single-phase equilibrium branch. They do not constitute a stability theorem
for area, shape, or coupled bulk perturbations. Any non-additive internal
variables are either absent or already minimized locally with a positive
internal-variable Hessian. In the latter case the material Hessian below is
the relaxed Schur complement, not the Hessian with the internal variables
held fixed.

\begin{proposition}[Equilibrium pressure susceptibility]
\label{prop:pressure-susceptibility}
Let \(a_{\mathrm{rel}}(\bm q)\) be a twice continuously differentiable relaxed
free-energy density, with independent material coordinates
\[
\bm q=(c_1,\ldots,c_N,\mathcal V),\qquad
\bm r(p)=(\mu_{1}^{\mathrm m}(p),\ldots,\mu_{N}^{\mathrm m}(p),-p).
\]
Suppose the isothermal reservoir path is twice continuously differentiable and
the planar equilibrium conditions
\[
\nabla a_{\mathrm{rel}}(\bm q(p))=\bm r(p)
\]
hold on an open positive-state branch on which
\(\bm H:=\nabla^2a_{\mathrm{rel}}\succ0\). The implicit function theorem yields a locally unique \(C^1\) equilibrium branch. The tension is \(C^2\) because its first derivative is \(-\bm r'^{\mathsf T}\bm q\). Then
\begin{align}
\bm q'&=\bm H^{-1}\bm r',
\label{eq:pressure-state-response}\\
\sigma'&=\mathcal V-\sum_i c_i(\mu_{i}^{\mathrm m})',
\label{eq:pressure-first-stable}\\
\sigma''&=-\bm r'^{\mathsf T}\bm H^{-1}\bm r'
             -\sum_i c_i(\mu_{i}^{\mathrm m})''.
\label{eq:pressure-second-stable}
\end{align}
In particular, for affine reservoir chemical potentials,
\[
\mu_{i}^{\mathrm m}(p)=\mu_{i0}^{\mathrm m}+v_i(p-p_0),\qquad
\bm u=(v_1,\ldots,v_N,-1),
\]
the equilibrium tension is strictly concave along every such regular branch:
\[
\sigma''=-\bm u^{\mathsf T}\bm H^{-1}\bm u<0.
\]
The pure incompressible reservoir benchmark is one instance of this affine
case. For a general real-fluid path, the corrected quantity constrained by
material-state stability is
\[
\mathcal C(p):=-\sigma''-\sum_i c_i(\mu_{i}^{\mathrm m})''
=\bm r'^{\mathsf T}\bm H^{-1}\bm r'>0;
\]
the sign of \(\sigma''\) alone is then not fixed.
\end{proposition}

\begin{proof}
The fixed-area open-system potential is
\(a_{\mathrm{rel}}-\bm r^{\mathsf T}\bm q\), whose material-state Hessian is
\(\bm H\). Differentiating its stationarity condition gives
\(\bm H\bm q'=\bm r'\). At equilibrium the area-conjugate tension is
\(\sigma=a_{\mathrm{rel}}-\bm r^{\mathsf T}\bm q\). Its first derivative is
\(-\bm r'^{\mathsf T}\bm q\), because the terms proportional to
\(\bm q'\) cancel by stationarity. Differentiating once more proves
\cref{eq:pressure-second-stable}. The quadratic form is strictly positive
because the last component of \(\bm r'\) is \(-1\), so \(\bm r'\ne0\).
\end{proof}

The proposition assumes independent admissible material coordinates. With
additional material constraints, one must differentiate the constrained
stationarity equations and use the appropriate tangent Hessian. A singular
Hessian, a boundary point of the constitutive state domain, a phase transition, or an unstable stationary
branch lies outside this result for a regular branch under strict convexity. Moreover, strict concavity
does not require an experimentally detectable second derivative of the tension. A sufficiently large
but finite material Hessian can make its magnitude arbitrarily small.

For completeness the corresponding calculation can be carried out directly
for a holonomic surface. Let \(\mathcal V=f(\bm c)\) be differentiable,
degree one, and twice continuously differentiable in the interior, and let
\(\bar a(\bm c)\) be its relaxed reduced Helmholtz density. Its open-system
potential is
\[
\omega_{\mathrm{hol}}(\bm c,p)
=\bar a(\bm c)+p f(\bm c)-\bm\mu^{\mathrm m}(p)^{\mathsf T}\bm c.
\]
At a stable stationary point define
\[
\bm K_{\mathrm{hol}}
=\nabla^2\bar a+p\nabla^2f\succ0,
\qquad
\bm g=(\bm\mu^{\mathrm m})'-\nabla f.
\]
Stationarity and degree-one homogeneity give
\(\sigma=\omega_{\mathrm{hol}}\) at equilibrium. Therefore
\begin{align}
\bm c'&=\bm K_{\mathrm{hol}}^{-1}\bm g,\\
\sigma'&=f-\bm c^{\mathsf T}(\bm\mu^{\mathrm m})',\\
\sigma''&=-\bm g^{\mathsf T}\bm K_{\mathrm{hol}}^{-1}\bm g
           -\bm c^{\mathsf T}(\bm\mu^{\mathrm m})''.
\label{eq:holonomic-second-stable}
\end{align}
Indeed, differentiating
\(\nabla\bar a+p\nabla f=\bm\mu^{\mathrm m}\) yields the first equation. The last
two follow by differentiating the stationary potential. Thus strict negative
curvature on an affine reservoir path does not establish the presence of an independent
structural coordinate. In the holonomic case it is strict whenever
\(\bm g\ne0\). A nonzero pressure coefficient implies this condition,
since \(\sigma'=-\bm c^{\mathsf T}\bm g\). For these regular response formulas, it
is \(\bm K_{\mathrm{hol}}\), including \(p\nabla^2f\), that is assumed
positive definite, not merely \(\nabla^2\bar a\).

\subsection{Statically Indistinguishable Constitutive Models}
\label{sec:matched-models}

Consider two hypothetical incompressible reservoirs with constant positive
molar volumes \(v_A,v_B\). Choose positive reference surface amount densities
\(c_{A0},c_{B0}\), reference pressure \(p_0>0\), reference tension
\(\sigma_0>0\), and desired positive pressure coefficient \(\ell_0>0\).
Choose also \(\mathcal C_0>0\), with units \(\mathrm{m\,Pa^{-1}}\), and set
\[
\mathcal V_0=c_{A0}v_A+c_{B0}v_B+\ell_0,
\qquad
\bm q_0=(c_{A0},c_{B0},\mathcal V_0),
\qquad
\bm u=(v_A,v_B,-1).
\]
The following construction matches the complete quadratic function
\begin{equation}
\sigma(p)=\sigma_0+\ell_0\Delta p
                    -\tfrac12\mathcal C_0(\Delta p)^2,
\qquad \Delta p=p-p_0,
\label{eq:matched-quadratic-tension}
\end{equation}
on a nonempty common interval of positive states. It matches reference
surface amount densities and structural volume per area as well as tension, its first pressure derivative, and its second pressure derivative.

For the independent model introduce the dimensional scaling matrix and a
dimensionless positive-definite coupling matrix
\[
\bm D=\operatorname{diag}(c_{A0},c_{B0},\mathcal V_0),
\qquad
\bm G=
\begin{pmatrix}
1&0&\rho\\0&1&0\\\rho&0&1
\end{pmatrix},\qquad 0<\rho<1.
\]
Define
\[
\bm s=\bm D\bm u,
\qquad
a_*=\frac{\bm s^{\mathsf T}\bm G^{-1}\bm s}{\mathcal C_0},
\qquad
\bm H=a_*\bm D^{-1}\bm G\bm D^{-1}.
\]
Each component of \(\bm s\) has units of length, so \(a_*\) has units
\(\mathrm{J\,m^{-2}}\). This scaling is necessary because the surface amount-density and
volume-per-area coordinates have different units. \(\bm G\) is dimensionless, whereas
the entries of \(\bm H\) have the corresponding mixed Hessian units.
With
\(\bm r_0=(\mu_{A0}^{\mathrm m},\mu_{B0}^{\mathrm m},-p_0)\), choose the primitive density
\begin{equation}
a_{\mathrm{ind}}(\bm q)
=\sigma_0+\bm r_0^{\mathsf T}\bm q_0
 +\bm r_0^{\mathsf T}(\bm q-\bm q_0)
 +\tfrac12(\bm q-\bm q_0)^{\mathsf T}\bm H(\bm q-\bm q_0).
\label{eq:explicit-independent-density}
\end{equation}
All coefficients, including \(p_0\), are fixed reference constants. The
varying reservoir pressure is not a primitive constitutive argument.
The equilibrium conditions \(\mu_{i}^{\Sigma,\mathrm m}=\mu_{i}^{\mathrm m}(p)\) and
\(\Pi_{\mathcal V}^{\Sigma}=p\) give
\[
\bm q_{\mathrm{ind}}(p)=\bm q_0+\bm y\Delta p,
\qquad \bm y=\bm H^{-1}\bm u,
\qquad \bm u^{\mathsf T}\bm H^{-1}\bm u=\mathcal C_0.
\]
The Hessian is strictly positive definite, and the nonzero off-diagonal
entries couple composition to structure. Substitution in the area Legendre
relation gives \cref{eq:matched-quadratic-tension} exactly. In this particular
construction \(y_A,y_B>0\) and \(y_{\mathcal V}<0\): the two amount
densities increase while the structural volume decreases with pressure.
These signs follow from \(s_A,s_B>0\), \(s_{\mathcal V}<0\), and the
explicit inverse of the \(A\)--volume block of \(\bm G\).

For the holonomic model put
\[
n_0=c_{A0}+c_{B0},\qquad
\bm d=\frac{\ell_0}{n_0}(1,1),\qquad
\bm w=(v_A,v_B)+\bm d,
\qquad
f(\bm c)=\bm w^{\mathsf T}\bm c.
\]
The components of \(\bm w\) and \(\bm d\) are molar volumes. In particular,
\(\bm w>0\), \(f(\bm c_0)=\mathcal V_0\), and this linear constraint is
both extensive and additive for arbitrary patch combinations. Define
\[
\eta_{\mathrm{hol}}=\frac{\bm d^{\mathsf T}\bm d}{\mathcal C_0},
\qquad
\bm M_{\mathrm{hol}}=\eta_{\mathrm{hol}}\bm I_2,
\qquad
\widetilde{\bm\mu^{\mathrm m}}_0=\bm\mu_{0}^{\mathrm m}-p_0\bm w,
\]
and
\begin{equation}
\bar a_{\mathrm{hol}}(\bm c)
=\sigma_0+\widetilde{\bm\mu^{\mathrm m}}_0^{\mathsf T}\bm c_0
 +\widetilde{\bm\mu^{\mathrm m}}_0^{\mathsf T}(\bm c-\bm c_0)
 +\tfrac12(\bm c-\bm c_0)^{\mathsf T}
       \bm M_{\mathrm{hol}}(\bm c-\bm c_0).
\label{eq:explicit-holonomic-density}
\end{equation}
Here \(\eta_{\mathrm{hol}}\) has units \(\mathrm{J\,m^2\,mol^{-2}}\). The mechanically
loaded stationarity condition is
\(\nabla\bar a_{\mathrm{hol}}+p\bm w=\bm\mu^{\mathrm m}(p)\), so
\[
\bm c_{\mathrm{hol}}(p)
=\bm c_0-\frac{\bm d}{\eta_{\mathrm{hol}}}\Delta p,
\qquad
\mathcal V_{\mathrm{hol}}(p)=\bm w^{\mathsf T}\bm c_{\mathrm{hol}}(p).
\]
Its open-system Hessian is \(\bm M_{\mathrm{hol}}\succ0\), and
\(\bm d^{\mathsf T}\bm M_{\mathrm{hol}}^{-1}\bm d=\mathcal C_0\).
The constrained area conjugate consequently gives
\cref{eq:matched-quadratic-tension} again. Both surface amount densities now decrease with
pressure. An independently calibrated measurement of a surface amount density can therefore
distinguish these two particular models even though their entire local
pressure--tension curves coincide. This does not imply that a composition
measurement distinguishes every possible holonomic and independent model.

The interval of admissibility can be determined explicitly. Positivity of the independent state requires
\[
L_{\mathrm{ind}}<\Delta p<U_{\mathrm{ind}},\qquad
L_{\mathrm{ind}}=\max_{y_j>0}\frac{-q_{j0}}{y_j},\qquad
U_{\mathrm{ind}}=\min_{y_j<0}\frac{-q_{j0}}{y_j}.
\]
Empty maxima and minima mean \(-\infty\) and \(+\infty\), respectively.
The holonomic state is positive whenever
\[
\Delta p<U_{\mathrm{hol}}
:=\min_{i=A,B}\frac{\eta_{\mathrm{hol}}c_{i0}}{d_i};
\]
positive amounts automatically give positive volume because \(\bm w>0\).
Intersect these intervals with \(\Delta p>-p_0\), with
\(\Delta p<\ell_0/\mathcal C_0\) if a positive slope is required, and with
\[
\frac{\ell_0-\sqrt{\ell_0^2+2\mathcal C_0\sigma_0}}{\mathcal C_0}
<\Delta p<
\frac{\ell_0+\sqrt{\ell_0^2+2\mathcal C_0\sigma_0}}{\mathcal C_0}
\]
if positive tension is required. Their intersection contains \(\Delta p=0\)
as an interior point. Thus these are explicit stable positive-state models,
with complete constitutive free energies and admissible state domains. The construction is local and
phenomenological. It does not supply a molecular interpretation, an extension over a global
constitutive state domain, or a calibration to a measured liquid pair.

\begin{figure}[tbp]
\centering
\begin{tikzpicture}
\begin{groupplot}[
 group style={group size=2 by 2,horizontal sep=1.55cm,vertical sep=1.75cm},
 width=0.43\textwidth,height=5.5cm,
 tick label style={font=\footnotesize},label style={font=\small},
 title style={font=\small},legend style={font=\scriptsize,draw=none,fill=white},
 axis lines=left,grid=major,grid style={gray!20},
 every axis plot/.append style={thick,no marks},
 xmin=0,xmax=50,xtick={0,10,20,30,40,50},
 xlabel={$\Delta p\;(\mathrm{MPa})$},
 scaled ticks=false,
]
\nextgroupplot[
 title={(a) Equilibrium tension},
 ylabel={$\sigma\;(\mathrm{mN\,m^{-1}})$},
 ymin=49.98,ymax=50.95,ytick={50,50.2,50.4,50.6,50.8},
 yticklabel style={/pgf/number format/fixed,/pgf/number format/precision=1},
 legend pos=north west,
]
\addplot[blue!70!black,domain=0:50,samples=65] {50+0.02*x-0.00005*x^2};
\addlegendentry{Independent}
\addplot[orange!80!black,dashed,domain=0:50,samples=65] {50+0.02*x-0.00005*x^2};
\addlegendentry{Holonomic}

\nextgroupplot[
 title={(b) Relative constituent amounts},
 ylabel={$c_i/c_{i0}$},ymin=0.34,ymax=1.04,
 ytick={0.4,0.6,0.8,1},legend pos=south west,
]
\addplot[blue!70!black,domain=0:50] {1+0.0001062371489*x};
\addlegendentry{A independent}
\addplot[blue!70!black,dashed,domain=0:50] {1-0.003125*x};
\addlegendentry{A holonomic}
\addplot[orange!80!black,domain=0:50] {1+0.0001285126800*x};
\addlegendentry{B independent}
\addplot[orange!80!black,dashed,domain=0:50] {1-0.0125*x};
\addlegendentry{B holonomic}

\nextgroupplot[
 title={(c) Structural volume per area},
 ylabel={$\mathcal V\;(\mathrm{nm})$},ymin=0.15,ymax=0.31,
 ytick={0.16,0.2,0.24,0.28,0.3},legend pos=south west,
]
\addplot[blue!70!black,domain=0:50] {0.3-0.00006579849212*x};
\addlegendentry{Independent}
\addplot[orange!80!black,dashed,domain=0:50] {0.3-0.00285*x};
\addlegendentry{Holonomic}

\nextgroupplot[
 title={(d) Independent structural dynamics},
 xmode=log,xmin=0.001,xmax=1000,
 xtick={0.001,0.1,10,1000},
 xlabel={$\omega\tau_{\mathrm{op}}$},
 ylabel={$G_{\sigma p}\;(\mathrm{pm})$},
 ymin=-320,ymax=40,ytick={-300,-200,-100,0},
 legend pos=south west,
]
\addplot[blue!70!black,domain=-3:3,samples=181]
 ({10^x},{-300+320/(1+10^(2*x))});
\addlegendentry{$\operatorname{Re}G_{\sigma p}$}
\addplot[orange!80!black,dashed,domain=-3:3,samples=181]
 ({10^x},{-320*10^x/(1+10^(2*x))});
\addlegendentry{$\operatorname{Im}G_{\sigma p}$}
\end{groupplot}
\end{tikzpicture}
\caption{Constructive comparison using the synthetic parameters in \cref{sec:matched-models}. No experimental data are fitted. The independent and holonomic models have exactly coincident pressure--tension curves (a), but distinct relative constituent amount densities (b) and structural volumes per area (c). Panel (d) shows the independent model's linear pressure-to-tension response from \cref{eq:dynamic-tension-response}, with frequency normalized by its unspecified structural relaxation time. Its limiting coefficients are \(G_{\sigma p}(0)=20\,\mathrm{pm}\) and \(G_{\sigma p}(\infty)=-300\,\mathrm{pm}\) within the assumed fast-exchange frequency window. Static coincidence therefore does not imply equality of state response or of a specified kinetic realization.}
\label{fig:constructive-comparison}
\end{figure}
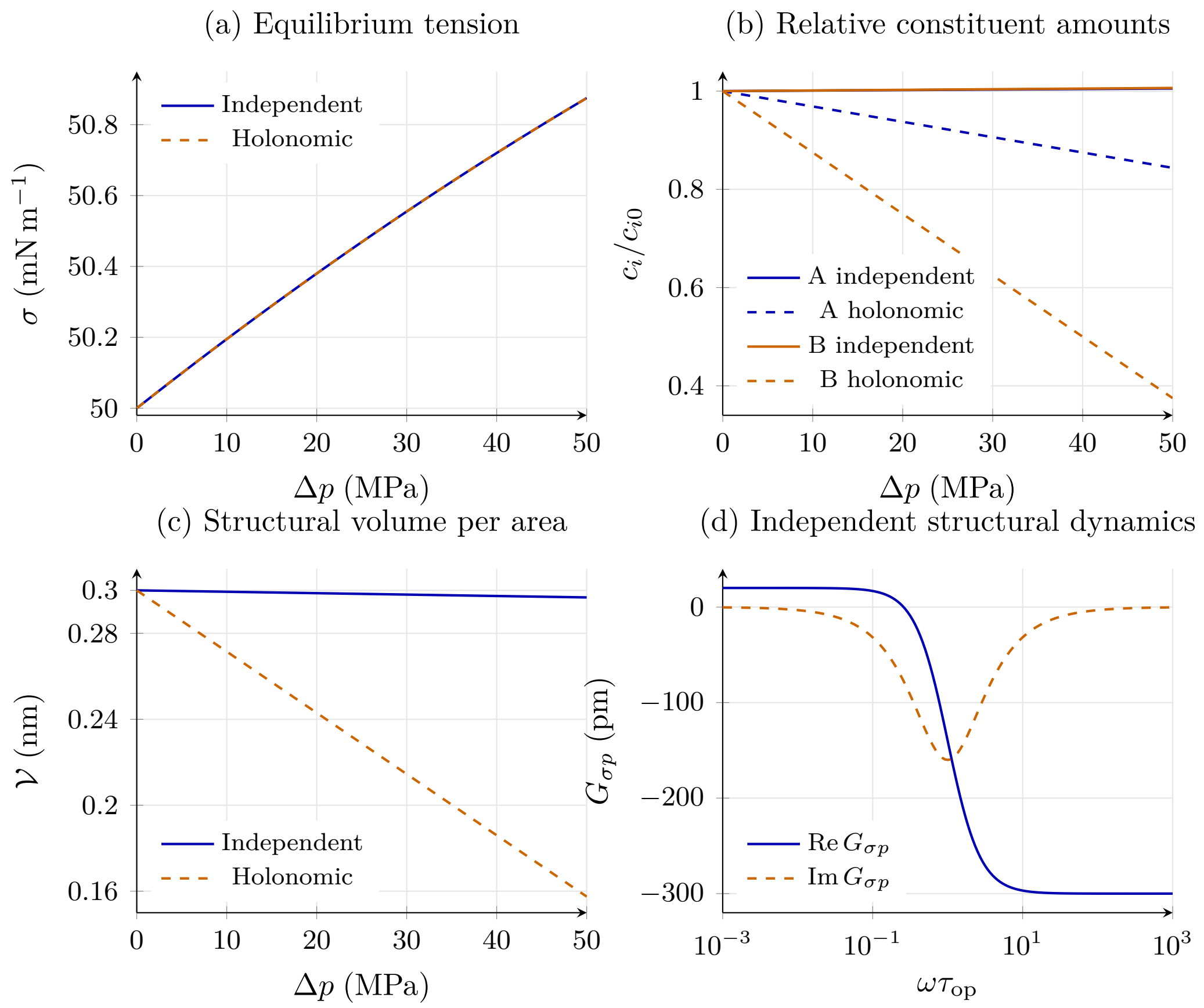

\paragraph{Dimensioned illustration.}
For a synthetic example, take
\[
\begin{gathered}
p_0=0.1\,\mathrm{MPa},\qquad
\sigma_0=50\,\mathrm{mN\,m^{-1}},\qquad
(v_A,v_B)=(20,200)\,\mathrm{cm^3\,mol^{-1}},\\
(c_{A0},c_{B0})=(4,1)\,\mathrm{\mu mol\,m^{-2}},\qquad
\ell_0=20\,\mathrm{pm},\qquad
\mathcal C_0=10^{-19}\,\mathrm{m\,Pa^{-1}},\qquad \rho=\tfrac14.
\end{gathered}
\]
These values give \(\mathcal V_0=0.300\,\mathrm{nm}\),
\(a_*=1.5562667\,\mathrm{J\,m^{-2}}\), and
\(\eta_{\mathrm{hol}}=3.20\times10^8\,\mathrm{J\,m^2\,mol^{-2}}\).
The independent-state positivity interval is approximately
\((-7781.33,4559.38)\,\mathrm{MPa}\) in \(\Delta p\). The first holonomic surface
amount density reaches zero at \(\Delta p=80\,\mathrm{MPa}\). Including positive
absolute pressure, tension, and slope gives the common open interval
\(-0.1<\Delta p<80\,\mathrm{MPa}\). These algebraic bounds are not a claim
of realistic incompressibility over the full interval. A conservative
illustration over \(0\le\Delta p\le50\,\mathrm{MPa}\) remains strictly
inside both material domains. At its upper endpoint,
\[
\begin{array}{c|ccc}
&c_A\;(\mathrm{\mu mol\,m^{-2}})&c_B\;(\mathrm{\mu mol\,m^{-2}})
 &\mathcal V\;(\mathrm{nm})\\\hline
\text{independent}&4.021247&1.006426&0.296710\\
\text{holonomic}&3.375000&0.375000&0.157500
\end{array}
\]
while both predict \(\sigma=50.875\,\mathrm{mN\,m^{-1}}\). No experimental
surface-tension or adsorption data enter this illustration.
Figure~\ref{fig:constructive-comparison}(a)--(c) compares the coincident tension curves with the distinct surface amount-density and volume-per-area responses. Panel (d) uses the dynamic reduction derived below.

\subsection{Projected Susceptibility and Identifiability}

The second-derivative restriction identifies one projected material
susceptibility, not the individual entries of the constitutive Hessian.
Writing the relaxed Hessian at a reference state as
\[
\bm H=
\begin{pmatrix}\bm M&\bm b\\\bm b^{\mathsf T}&k\end{pmatrix},
\qquad
K_{\mathrm{op}}=k-\bm b^{\mathsf T}\bm M^{-1}\bm b>0,
\qquad \bm\chi=(\bm\mu^{\mathrm m})'(p),
\]
block elimination gives
\begin{align}
\mathcal V'
&=-\frac{1+\bm b^{\mathsf T}\bm M^{-1}\bm\chi}{K_{\mathrm{op}}},
\label{eq:open-structural-path-response}\\
\mathcal C(p)
&=\bm\chi^{\mathsf T}\bm M^{-1}\bm\chi
 +\frac{(1+\bm b^{\mathsf T}\bm M^{-1}\bm\chi)^2}{K_{\mathrm{op}}}.
\label{eq:pressure-susceptibility-split}
\end{align}
Here \(k\) and \(K_{\mathrm{op}}\) have units \(\mathrm{Pa\,m^{-1}}\),
whereas \(\bm b^{\mathsf T}\bm M^{-1}\bm\chi\) is dimensionless.
The negative structural derivative at fixed constituent amount densities therefore does not,
alone, imply \(\mathcal V'<0\) on a coexistence path. Composition--volume
coupling enters through the numerator in
\cref{eq:open-structural-path-response}. Nor is the second term in
\cref{eq:pressure-susceptibility-split} independently identified by the
observed second pressure derivative of tension unless the amount-density Hessian block and the coupling coefficients are already
constrained. For the synthetic independent model above the two terms are,
respectively, \(2.98149\times10^{-20}\) and
\(7.01851\times10^{-20}\,\mathrm{m\,Pa^{-1}}\).

At the same reference state, the susceptibilities satisfy an ordering relation. If freezing the
non-additive internal variables gives \(\bm H_{\mathrm{fr}}\succ0\) and
allowing their stable relaxation gives
\(0\prec\bm H_{\mathrm{rel}}\preceq\bm H_{\mathrm{fr}}\), then
\[
\bm r'^{\mathsf T}\bm H_{\mathrm{rel}}^{-1}\bm r'
\ge
\bm r'^{\mathsf T}\bm H_{\mathrm{fr}}^{-1}\bm r'.
\]
Relaxation can only increase this quasistatic projected susceptibility when
the reference state, direction of the imposed loading, and variable material coordinates
are held fixed. This comparison is not an identification of a finite-frequency
measurement with either limiting Hessian unless the relevant separation of relaxation time scales
has been established.

In the affine benchmark the first two tension derivatives constrain only
\(-\bm q^{\mathsf T}\bm u\) and \(\bm u^{\mathsf T}\bm H^{-1}\bm u\).
The explicit pair above shows that even an exactly known complete quadratic
curve need not determine the number of independent material coordinates.
Repeated measurements along the same coexistence path do not provide additional
independent loading directions at a fixed state. Additional observables require the measurement-to-state relations discussed in \cref{app:pressure-loading}. Additional thermodynamic paths help only to the extent that
they provide independently calibrated directions or constrain a common
constitutive model.

\subsection{Observable Structural Relaxation under Fast Exchange}
\label{sec:dynamic-response}

An independently retained coordinate can be distinguished dynamically only if the imposed load excites it and the measured observable detects it. Consider a locally planar patch at fixed area and temperature, neglect tangential transport, and impose a time-scale ordering in which chemical exchange is fast compared with the structural relaxation under study. Thus the constituent surface amount densities remain on the chemical-equilibrium manifold while \(\Vsurf\) may depart from structural equilibrium. This reduction differs from imposing fixed surface composition. Any additional internal variables are either frozen or already eliminated consistently in the free energy used below.

Linearize about an interior equilibrium at \(p_0\). Let \(\bm v=(\dd\mu_{i}^{\mathrm{bulk},\mathrm m}/\dd p)_0\), and write the relevant material-state Hessian as
\[
\bm H_0=
\begin{pmatrix}
\bm M&\bm b\\
\bm b^{\mathsf T}&d
\end{pmatrix}\succ0,
\qquad
\bm M=\left(\frac{\partial\mu_{i}^{\Sigma,\mathrm m}}{\partial c_j^{\Sigma}}\right)_0,
\quad
\bm b=\left(\frac{\partial\mu_{i}^{\Sigma,\mathrm m}}{\partial\Vsurf}\right)_0,
\quad
d=-\left(\frac{\partial\PiV}{\partial\Vsurf}\right)_0.
\]
Define the open-exchange structural stiffness and dimensionless coupling
\[
K_{\mathrm{op}}=d-\bm b^{\mathsf T}\bm M^{-1}\bm b>0,
\qquad
B=\bm b^{\mathsf T}\bm M^{-1}\bm v.
\]
Here \(K_{\mathrm{op}}\) has units \(\mathrm{Pa\,m^{-1}}\). The chemical-equilibrium linearization gives
\[
\delta\bm c=\bm M^{-1}(\bm v\,\delta p-\bm b\,\delta\Vsurf),
\qquad
\delta\PiV=-B\,\delta p-K_{\mathrm{op}}\,\delta\Vsurf.
\]
For the structural constitutive relation in the interior of the state domain
\(\dot{\Vsurf}=L_{\mathcal V}(\PiV-p)\), with \(L_{\mathcal V}>0\),
\[
\delta\dot{\Vsurf}
=-L_{\mathcal V}K_{\mathrm{op}}\,\delta\Vsurf
-L_{\mathcal V}(1+B)\,\delta p,
\qquad
\tau_{\mathrm{op}}=\frac{1}{L_{\mathcal V}K_{\mathrm{op}}}.
\]
The relevant stiffness is the Schur complement after chemical equilibration, not the stiffness at fixed surface amount densities \(d\).

The intrinsic Gibbs--Duhem differential holds also during this departure from structural equilibrium. Because the surface chemical potentials remain equal to those of the reservoirs, it gives
\[
\delta\sigma
=-\bigl(\bm c_0^{\mathsf T}\bm v+\Vsurf_0 B\bigr)\delta p
-\Vsurf_0K_{\mathrm{op}}\,\delta\Vsurf.
\]
For a small harmonic perturbation with convention \(\exp(i\omega t)\), the pressure-to-tension response is consequently
\begin{equation}
\label{eq:dynamic-tension-response}
G_{\sigma p}(\omega)
=-\bigl(\bm c_0^{\mathsf T}\bm v+\Vsurf_0 B\bigr)
+\frac{\Vsurf_0(1+B)}{1+i\omega\tau_{\mathrm{op}}}.
\end{equation}
Its low-frequency limit is
\[
G_{\sigma p}(0)=\Vsurf_0-\bm c_0^{\mathsf T}\bm v,
\]
as required by the equilibrium residual. Its high-frequency limit relative to structural relaxation is the first term in \cref{eq:dynamic-tension-response}, understood only within a frequency window still slow enough for chemical equilibration and the planar quasistatic mechanical loading to remain valid. If \(\tau_{\mathrm{ex}}\) and \(\tau_{\mathrm{mech}}\) denote the unresolved chemical and mechanical adjustment times, the structural high-frequency plateau requires \(\omega\tau_{\mathrm{op}}\gg1\), \(\omega\tau_{\mathrm{ex}}\ll1\), and \(\omega\tau_{\mathrm{mech}}\ll1\).

For a pressure step \(\delta p=u_0\) from an equilibrated state, the same reduction gives
\[
\delta\sigma(t)
=u_0\left[
\Vsurf_0-\bm c_0^{\mathsf T}\bm v
-\Vsurf_0(1+B)e^{-t/\tau_{\mathrm{op}}}
\right],\qquad t\ge0,
\]
after the unresolved faster exchange transient.

The pole associated with structural relaxation is not excited by pressure when \(1+B=0\), even though the independent coordinate and its restoring stiffness remain present.

In this scalar model the structural output coefficient is \(-\Vsurf_0K_{\mathrm{op}}\ne0\), so this cancellation is an excitation degeneracy rather than loss of structural observability under every possible forcing. Otherwise the amplitude of its relaxing tension response is \(\Vsurf_0(1+B)\), and the measured relaxation time identifies the product \(L_{\mathcal V}K_{\mathrm{op}}\). A tension response alone does not separate mobility from stiffness or volume from coupling. An independent structural-response measurement, a measurement-to-state relation for the surface amount densities, or additional kinetic constraints are needed to make those quantities identifiable. The susceptibility decomposition in \cref{eq:pressure-susceptibility-split} shows that this same excitation factor controls the structural contribution to the static second derivative.

For a holonomic model under the same assumption of fast exchange, no additional structural pole remains. Its amount response is already equilibrated on the retained time scale. This statement does not preclude relaxation in a holonomic model with finite exchange, thermal dynamics, or other internal variables. A measured pole distinguishes these two kinetic reductions only when the competing processes are independently constrained. It does not assign a unique microscopic interpretation to the relaxation time.

\section{Archival Experiments as Prospective State-Selection Tests}
\label{sec:selection}

This prospective analysis distinguishes three questions concerning the planar amount-only sign at a specified coexistence state, the persistence of its finite-solubility inequalities along a high-pressure path, and the derivative measured in a curved-drop experiment. The reported slopes may be used to test the model once equilibrium status and the relation between the imposed pressure and the local interfacial loading have been established. The slopes alone do not demonstrate an independent structural coordinate.

\subsection{Archival Measurements and the Water--n-Decane Benchmark}

Matubayasi and coworkers measured water against several organic liquids at \(303.15\,\mathrm K\) to \(150\,\mathrm{MPa}\) and reported increasing interfacial tension with pressure together with a positive volume change on interface formation \cite{Matubayasi1977}. Susnar, Hamza, and Neumann studied the nominally pure \(n\)-decane--water system over \(0\)--\(35\,\mathrm{MPa}\) by axisymmetric drop-shape analysis and reported highly linear positive slopes, including \((22.37\pm0.35)\times10^{-12}\,\mathrm m\) at \(21.2^{\circ}\mathrm C\) and \((23.29\pm0.48)\times10^{-12}\,\mathrm m\) at \(21.5^{\circ}\mathrm C\) \cite{Susnar1994}. For the liquid-hydrocarbon comparison series, Wiegand and Franck report measurements to elevated temperatures and pressures up to \(3000\,\mathrm{bar}\) for water with \(n\)-hexane, \(n\)-decane, and toluene, with increasing tension with pressure. The article title's \(2800\,\mathrm{bar}\) limit refers to the broader study, while the abstract states \(3000\,\mathrm{bar}\) for these liquid comparisons \cite{WiegandFranck1994}. Earlier measurements by Michaels and Hauser and by Jennings, and the later thermodynamic analysis of Motomura and coworkers, provide additional independent series and volumetric context \cite{MichaelsHauser1951,Jennings1967,Motomura1983}.

The pressure variable reported in an experiment need not equal \(\pplus\), \(\pminus\), or a planar common thermodynamic pressure. Denote the actual experimental control by \(q\). The measured response should first be written
\[
\frac{\dd\widetilde\sigma}{\dd q}.
\]
A rigorous reanalysis then determines whether \(q\) equals a local one-sided trace, a remote chamber pressure, or another hydraulic variable related to the interface by hydrostatics and apparatus geometry. In an axisymmetric-drop experiment the local pressure difference is constrained simultaneously by the normal jump balance and by the fitted shape, so curvature may evolve with \(q\).

Preparation and measurement time must be included in the interpretation of archival pressure coefficients. Georgiadis et al. report substantial scatter among published near-ambient water--\(n\)-decane tensions and identify purification, cleaning, and measurement protocol as relevant factors \cite{Georgiadis2011}. Their comparison records decreasing tension with time and an immediate-time value for Susnar et al. Their accompanying discussion also reports previous mutual saturation of the liquids. Thus immediate-time sampling does not, in isolation, demonstrate incomplete bulk equilibration. Its interpretation requires the original pressure-step and time-extrapolation protocol, including whether the preparation and delay after each pressure change support the equilibrium branch used in the present analysis.

For the different \(\mathrm{CO_2}\)--water system, Hinton and Alvarez demonstrate that bulk saturation can materially affect high-pressure tension measurements \cite{HintonAlvarez2021}. This provides methodological context rather than evidence of inadequate saturation in the water--\(n\)-decane experiments. An archival comparison must distinguish contamination drift, saturation transients, density reduction, and uncertainty in the hydraulic loading path.

\begin{table}[t]
\centering
\small
\caption{Selected archival liquid--liquid pressure measurements relevant to state selection. A quantitative comparison requires reconstruction of the experimental control pressure, the two one-sided thermodynamic traces, and the curvature path.}
\label{tab:archival-pressure}
\begin{tabularx}{\textwidth}{L{0.18\textwidth}L{0.17\textwidth}L{0.19\textwidth}Y}
\toprule
System & Source & Range & Role in the present analysis \\
\midrule
Water with several hydrocarbons and carbon tetrachloride & Matubayasi et al. \cite{Matubayasi1977} & \(303.15\,\mathrm K\), to \(150\,\mathrm{MPa}\) & early high-pressure slopes and explicit interpretation in terms of the volume associated with interface formation \\
Water--\(n\)-decane & Susnar et al. \cite{Susnar1994} & about room temperature, \(0\)--\(35\,\mathrm{MPa}\) & reported positive pressure coefficient. Later literature describes presaturation and an immediate-time protocol, requiring reconstruction of pressure-step timing, local pressure loading, and curvature before quantitative equilibrium comparison \\
Water with \(n\)-hexane, \(n\)-decane, toluene & Wiegand and Franck \cite{WiegandFranck1994} & to \(473\,\mathrm K\) and \(3000\,\mathrm{bar}\) & wide thermodynamic range and independent test of temperature dependence \\
Water--benzene and water--\(n\)-decane & Jennings \cite{Jennings1967} & \(25\)--\(176^{\circ}\mathrm C\), \(1\)--\(817\,\mathrm{atm}\) & pendant-drop data accompanied by density information for mutually saturated phases \\
Several oil--water systems & Motomura et al. \cite{Motomura1983} & pressure-dependent series & direct comparison of pressure coefficients with classical invariant thermodynamic descriptions \\
\bottomrule
\end{tabularx}
\end{table}

At the representative room-temperature water--\(n\)-decane states, \cref{sec:mutual,res:finite-solubility-sign} establish a negative planar amount-only response for positive material surface amount densities. Extension over a pressure interval requires verification of the same inequalities along its coexistence path.

The two bulk Gibbs--Duhem equations determine the chemical-potential first derivatives from coexistence compositions and phase molar volumes. Individual partial molar volumes can support a chosen constitutive model or extrapolation, but are not an additional input required for that inversion. A reference-volume decomposition is optional and must use a reference fixed independently of the two phase equations. 

\subsection{Experimental Loading-Path Reconstruction}

A curved equilibrium experiment should be represented by a coupled system
\[
\bm G
\left(
\pminus,\pplus,\kappas,
\bm x^-,\bm x^+,\bm z^{\Sigma},q
\right)=\bm0,
\]
containing the hydraulic relation between \(q\) and the measured bulk pressure, hydrostatic corrections where needed, bulk equations of state and phase equilibrium, one-sided bulk--surface chemical equilibrium, the normal momentum jump balance, the enriched surface constitutive equations, structural equilibrium if \(\Vsurf\) is retained, and the geometric relations used by the drop-shape analysis. Only after this system has been solved may the measured function be interpreted as a reduced thermodynamic response.

Differentiating the simple isotropic equilibrium jump balance along \(q\) gives
\[
\frac{\dd\pminus}{\dd q}
-
\frac{\dd\pplus}{\dd q}
=
-\kappas\frac{\dd\widetilde\sigma}{\dd q}
-\sigma\frac{\dd\kappas}{\dd q}
\]
for the sign convention used in \cref{sec:equilibrium}. The curvature derivative therefore belongs to the measured response problem whenever the drop shape evolves with imposed pressure.

For a gravity-affected drop, choose a fixed reference location such as the apex. If \(\dd\pplus/\dd q>0\) along the path, then
\[
\frac{\dd\widetilde\sigma}{\dd q}
=\frac{\dd\sigmatp}{\dd\pplus}
\frac{\dd\pplus}{\dd q}
\]
preserves the response sign. Two sufficient curved sign tests are then available. A pathwise bound \(R_{\min}\le r_p\le R_{\max}\) invokes \cref{res:curved-pressure-ratio} without a surface-composition hypothesis. Alternatively, the tie-line-interior hypothesis invokes \cref{res:finite-solubility-curvature} and the integrated necessary condition in \cref{cor:curvature-reversal}. Its premise is pointwise nonnegative response throughout the interval, not merely a positive fitted regression slope. A precise comparison of pressure-coefficient magnitudes still requires the complete reconstruction of the pressure, density, and geometry. A sign test may require only conservative bounds.

The high-precision Susnar series is a candidate for a pressure-sign test, subject to verification of its equilibrium interpretation. In addition to the exact pressure-control location, phase assignment in the drop geometry, and sufficient raw or tabulated shape information to recover the curvature path, the reconstruction must recover the preparation and timing conditions identified by the later literature review, namely the original presaturation procedure reported by Georgiadis et al., whether that procedure applies at each pressure state, the cleaning or purification treatment, the waiting time after pressure changes, the measured \(\gamma(t)\) behavior at each state, and whether a common \(t=0\) extrapolation procedure generated the reported pressure series. Georgiadis et al. also report that Susnar et al. regarded density changes caused by mutual diffusion as negligible and therefore used pure-compound phase densities in the ADSA reduction \cite{Georgiadis2011}. This choice is physically plausible at low solubility but should be reproduced or updated explicitly in a precise reanalysis. These questions are especially important because a pressure coefficient extracted consistently from immediate-time extrapolates can be highly reproducible without necessarily being the equilibrium derivative assumed by the present sign theorem. The same loading-reconstruction requirements apply to the Wiegand--Franck data.

A quantitative scale for the uncertainty assessment follows directly from the reported \(21.2^{\circ}\mathrm C\) coefficient. Over \(35\,\mathrm{MPa}\), the fitted tension increment is \(0.783\,\mathrm{mN\,m^{-1}}\), and the quoted slope uncertainty corresponds to \(0.0123\,\mathrm{mN\,m^{-1}}\). These are propagated values from the reported fit, not an independent reanalysis. Absolute offsets between laboratories need not bias a slope determined within a single experimental series. Pressure-correlated timing, density, or preparation effects can. The reconstruction should distinguish these error sources.

\begin{table}[tbp]
\centering\small
\caption{Sequence of steps in the archival-data analysis. Each stage answers a different question. A precise constitutive fit is not needed to perform a bounded sign test.}
\label{tab:archival-gates}
\begin{tabularx}{\textwidth}{L{0.20\textwidth}YY}
\toprule
Stage & Required information & Permitted inference \\
\midrule
Equilibrium classification & Saturation, purity, waiting times, and meaning of immediate-time extrapolation & Whether equilibrium chemical-potential matching is applicable \\
Bounded sign test & Local pressure loading, coexistence compositions and phase volumes, and a verified sign criterion & Whether the restricted amount-only class is excluded on that path \\
Constitutive comparison & Mapped surface observables, full pressure--tension data, uncertainty, and an independently constrained state model & Calibration and prediction of a complete structural model \\
State-dimension test & Independent state response or kinetically resolved observables & Discrimination of specified holonomic and independent realizations \\
\bottomrule
\end{tabularx}
\end{table}

An immediate-time extrapolation can estimate a clean equilibrium tension when the bulk phases are mutually saturated and slower contamination is the principal source of subsequent drift. The account of Susnar et al. given by Georgiadis et al. is consistent with that interpretation, although it does not independently determine equilibration after every pressure step. In an initially unsaturated system, the early-time pressure response need not satisfy equilibrium chemical-potential matching and requires a finite-rate model. These are distinct thermodynamic classifications. Time dependence alone does not determine which applies.

\subsection{Model Discrimination}

Once an isothermal equilibrium experimental loading path has been reconstructed, a \emph{quantitative} amount-only prediction requires more than the bulk chemical-potential derivatives. A specified amount-only surface equation of state, or independent adsorption information, must also determine the surface amount densities on that same path. Only then is
\[
K_{\mathrm{amt}}(q)
=-\sum_i c_{i,\mathrm{amt}}^{\Sigma}(q)
\frac{\dd\mu_{i}^{\mathrm m}}{\dd q}
\]
a model prediction rather than an adjustable decomposition. The sign theorem requires only positivity of the material surface amount densities and therefore permits model discrimination without determining their magnitudes.

After the real-fluid and geometric corrections have been evaluated and an amount-only constitutive model has been fixed independently, one may define the diagnostic residual
\[
R_{\sigma}(q)
=
\frac{\dd\sigma_{\mathrm{exp}}}{\dd q}
-K_{\mathrm{amt}}(q).
\]
Assessment of the \emph{full enriched model}, including its coupled surface amount densities, requires evaluating the structural contribution
\[
\Vsurf\frac{\dd\PiV}{\dd q}
\]
together with the second-pressure-trace and curvature couplings implied by the mechanical closure. In general, the enriched equilibrium values \(c_i^{\Sigma}(q)\) need not equal the amount-only values because the enriched free energy may couple composition and structural volume. Therefore \(R_\sigma\) can be attributed to the structural term only when the constituent-state closure is independently shared or otherwise constrained across the two model classes. Otherwise, comparison must use complete model predictions, since a fitted decomposition into amount and structural contributions is not uniquely determined. In the planar geometric-thickness specialization the structural loading factor is fixed exactly to unity, so one fewer adjustable factor remains. In curved configurations the full structural loading formula must be used.

Static archival data constrain equilibrium combinations such as \(\ell_{\mathrm{res}}\), but they cannot by themselves determine all primitive surface coordinates or prove that \(\Vsurf\) is an independently relaxing dynamic state variable. A successful equilibrium fit may equally represent a rapidly equilibrating coordinate or, for static data, a mechanically loaded holonomic amount--volume relation. Predictive consistency across independent data sets additionally requires calibration of the minimum number of structural parameters on one series and testing at independent temperatures, pressure ranges, or fluid pairs without changing the meaning of \(\Vsurf\). This procedure avoids circular parameter identification. Evidence for a particular enrichment requires exclusion of the restricted model after the available real-fluid and geometric effects have been assessed, together with successful predictions for independent observations using the same enriched constitutive structure.

The dynamic distinction is developed in \cref{sec:dynamic-response} for a planar patch on a time scale with fast chemical exchange and finite structural relaxation. Its measured amplitude, relaxation time, and excitation conditions must be compared with independent kinetic information. Once thermal, transfer, or other internal-variable dynamics overlap, measured poles belong to the coupled system and need not correspond one-to-one to named state coordinates.

\section{Geometric Thickness and Complete Structural Work}
\label{app:geometric}

\subsection{Geometric-Thickness Specialization}

The planar result below determines the mechanical load conjugate to a scalar thickness variation.
For a curved interface, the one-parameter formulas determine the load
conjugate to structural variations at fixed instantaneous curvature.
Their extension to unrestricted shape variations requires the complete
external work and the chosen thickness kinematics. Curvature derivatives
generated by rewriting a scalar volume in geometric coordinates are not,
by themselves, independent bending constitutive terms.

Although \(\Vsurf\) is not generally defined as a thickness, a geometric realization defines a constitutive subclass. Let \(\bm n_\Sigma\) point from minus to plus and let the shape operator be
\[
\bm L=-\grads\bm n_\Sigma.
\]
Denote its principal curvatures by \(k_1^{\Sigma}\) and \(k_2^{\Sigma}\), so that
\[
\kappas=k_1^{\Sigma}+k_2^{\Sigma}=-\divs\bm n_\Sigma,
\qquad
\Ks=k_1^{\Sigma}k_2^{\Sigma}.
\]
Introduce notional normal extents \(h^-\ge0\) and \(h^+\ge0\). The area ratio of a parallel surface at signed distance \(\xi\) is
\[
J(\xi)
=(1-k_1^{\Sigma}\xi)(1-k_2^{\Sigma}\xi)
=1-\kappas\xi+\Ks\xi^2.
\]
The geometric construction is restricted to an admissible tubular neighborhood in which the normal-offset map remains one-to-one and
\[
J(\xi)>0
\qquad
\text{for }-h^-\le\xi\le h^+,
\]
so that no focal singularity or local orientation reversal is crossed. The tubular volume per unit area of the reference sharp surface is therefore
\[
\Vsurf_{\mathrm{geom}}
=
\int_{-h^-}^{h^+}J(\xi)\,\dd\xi,
\]
that is,
\[
\Vsurf_{\mathrm{geom}}
=
h^-+h^+
-\frac{\kappas}{2}
\left[(h^+)^2-(h^-)^2\right]
+\frac{\Ks}{3}
\left[(h^+)^3+(h^-)^3\right].
\]
For a symmetric layer \(h^-=h^+=h/2\),
\[
\Vsurf_{\mathrm{geom}}
=h+\frac{\Ks h^3}{12},
\]
and for a planar interface it reduces to \(h^-+h^+\).

At fixed curvature,
\[
\delta\Vsurf_{\mathrm{geom}}
=J(h^+)\delta h^+
+J(-h^-)\delta h^-.
\]
The virtual work per unit reference area done \emph{by the external compressive tractions} on an expanding notional layer is
\[
\delta W_{\mathrm{ext}}
=-\pi_n^+J(h^+)\delta h^+
-\pi_n^-J(-h^-)\delta h^-.
\]
Here \(\pi_n^\pm\) are the one-sided normal-traction traces at the reference sharp surface, used as localized loads for the notional side displacements. Thus this formula is a constitutive sharp-surface realization, not an exact integral of a resolved traction field across a finite layer. The specified local structural mode uses reference-normal displacements and localized normal loads. Actual offset-boundary work for spatially varying widths additionally requires the offset normals, tangential traction contributions, and their area factors. If the normal traction varies appreciably over the retained offset distance, the work must instead be constructed from the corresponding offset tractions or a normal integral. Throughout the calculation above and below, the structural virtual mode is taken at fixed instantaneous curvature. Allowing curvature to vary in the same mode requires the complete external-work calculation and coordinate-transformation analysis below.

This sign is opposite to the positive resisting-work convention used in \(\mathcal W_{\mathcal V,\mathrm{mech}}^{\Sigma}\) above. Accordingly define
\[
\delta W_{\mathrm{res}}:=-\delta W_{\mathrm{ext}}.
\]
For a one-parameter structural mode
\[
\delta h^+=g^+\delta q,
\qquad
\delta h^-=g^-\delta q,
\]
assume the structural volume increment is nondegenerate,
\[
J(h^+)g^+ + J(-h^-)g^-\neq0.
\]
Then \(\delta W_{\mathrm{res}}=\Pi_{\mathrm{mech}}^{\Sigma}\delta\Vsurf\) gives
\[
\Pi_{\mathrm{mech}}^{\Sigma}
=
\frac{
\pi_n^+J(h^+)g^+
+\pi_n^-J(-h^-)g^-
}{
J(h^+)g^+
+J(-h^-)g^-
}.
\]
Thus the mechanical weighting coefficients are determined by the chosen geometry and structural mode rather than postulated independently. The same formula would acquire the opposite sign if \(\Pi_{\mathrm{mech}}^{\Sigma}\) were defined as external work per positive expansion rather than as positive resisting work.

Define the normalized coefficients
\[
w^+
=
\frac{J(h^+)g^+}
{J(h^+)g^+ + J(-h^-)g^-},
\qquad
w^-=1-w^+.
\]
Then
\[
\Pi_{\mathrm{mech}}^{\Sigma}
=w^+\pi_n^+ + w^-\pi_n^-.
\]
The coefficients satisfy \(w^++w^-=1\). With the nondegeneracy condition above, they are convex weights precisely when the two signed mode contributions \(J(h^+)g^+\) and \(J(-h^-)g^-\) have the same sign or one vanishes. If both are negative, reversing the orientation of the scalar mode parameter makes both positive without changing the physical mode. For a mode involving contraction on one side and expansion on the other, the two contributions have opposite signs and the normalized coefficients need not lie in \([0,1]\).

If the retained surface mechanics consists only of the complete isotropic
tension and the stated momentum-flux terms, define the recoil contribution
\(\mathcal R_n\) by
\[
\pi_n^--\pi_n^+=-\sigma\kappas+\mathcal R_n.
\]
The fixed-curvature structural load then becomes
\[
\Pi_{\mathrm{mech}}^{\Sigma}
=\pi_n^++w^-(-\sigma\kappas+\mathcal R_n).
\]
This substitution is conditional on that normal momentum balance.
If additional geometric actions or forces enforcing a shape constraint
are present, write their net contribution with the convention
\[
\pi_n^--\pi_n^+
=-\sigma\kappas+\mathcal R_n+f_{\mathrm{geom},n}.
\]
The same contribution must then be retained in
\(\Pi_{\mathrm{mech}}^{\Sigma}
=\pi_n^++w^-(-\sigma\kappas+\mathcal R_n+f_{\mathrm{geom},n})\).
The fixed-curvature virtual mode determines a structural conjugate.
It does not independently establish the normal balance for unrestricted
shape changes.

For geometric coordinates \((h^-,h^+,\kappas,\Ks)\), the exact
differential at fixed reference area is
\[
\delta\Vsurf_{\mathrm{geom}}
=J(h^+)\delta h^++J(-h^-)\delta h^-
-\frac{(h^+)^2-(h^-)^2}{2}\,\delta\kappas
+\frac{(h^+)^3+(h^-)^3}{3}\,\delta\Ks.
\]
The last two terms arise from the coordinate transformation. If a
nondegenerate scalar thickness parameter \(q\) is used, write
\(\Vsurf_{\mathrm{geom}}=\Phi(q,\kappas,\Ks)\), with
\(\Phi_q\ne0\). Along the selected one-parameter family of side widths, replacing \(\Vsurf\)
by \(q\) is locally invertible at each fixed geometry. This replacement
does not eliminate a second independently variable asymmetry coordinate. The occurrence
of \(-\PiV\Phi_{\kappa}\,\delta\kappas\) in that representation
therefore does not alone demonstrate an independent bending response.

To identify the remaining generalized mechanical forces, let the complete resisting-work
one-form per reference area, restricted for this calculation to fixed
temperature, area, amounts, and internal descriptors, be
\[
\delta w_{\mathrm{res}}
=W_q\,\delta q+W_{\kappa}\,\delta\kappas
+W_K\,\delta\Ks.
\]
The sum of intrinsic free-energy variation and resisting work contains
\[
(W_q-\PiV\Phi_q)\,\delta q
+(W_{\kappa}-\PiV\Phi_{\kappa})\,\delta\kappas
+(W_K-\PiV\Phi_K)\,\delta\Ks.
\]
Structural equilibrium gives \(W_q=\PiV\Phi_q\). The residual
curvature coefficients are consequently
\[
M_{\kappa}^{\mathrm{res}}=W_{\kappa}-\PiV\Phi_{\kappa},
\qquad
M_K^{\mathrm{res}}=W_K-\PiV\Phi_K.
\]
If this mechanical work is conservative, the same calculation follows
by adding its potential to the intrinsic free energy before varying.
For example, the local work specialization
\(\delta w_{\mathrm{res}}=p\,\delta\Phi\) gives
\(\PiV=p\) from the nondegenerate structural variation and cancels
both displayed residual curvature coefficients. This cancellation verifies the
mechanical-power representation. It does not establish a common-pressure Laplace equilibrium for an unsupported curved interface with
nonzero tension.

Conversely, nonzero residual coefficients can remain when the one-sided work,
the admissible thickness kinematics, or additional constitutive energies
prevent that cancellation. Their shape variation must then be combined
with area variation, reference-surface displacement, bulk work, and
boundary terms to obtain the actual surface forces and any moments.
Only that complete calculation decides whether higher-order surface
mechanics is required and whether the ordinary isotropic normal jump
remains valid. The present paper determines the planar geometric loading and
the structural load for a curved interface at fixed curvature. It does not claim a closed
theory for arbitrary free variations of a finite tubular layer.

\subsection{Planar Equilibrium Specialization}

The geometric subclass determines the unit structural loading factor under explicit assumptions.

\begin{proposition}[Planar geometric structural loading]
\label{prop:planar-geometric-loading}
Consider a differentiable common-pressure family of admissible geometric-thickness states in planar hydrostatic equilibrium, with isotropic bulk stresses and a nondegenerate structural mode \(g^-+g^+\neq0\). Then structural equilibrium along that family implies
\[
\PiV=p,
\qquad
\frac{\dd\PiV}{\dd p}=1.
\]
The result is independent of how the structural increment is distributed between the two sides.
\end{proposition}

\begin{proof}
Planarity gives \(k_1^{\Sigma}=k_2^{\Sigma}=0\) and \(J=1\). Isotropic equilibrium gives \(\pi_n^-=\pminus\) and \(\pi_n^+=\pplus\), while the planar normal momentum jump balance gives \(\pminus=\pplus=p\). Hence
\[
\Pi_{\mathrm{mech}}^{\Sigma}
=
\frac{p g^+ + p g^-}{g^++g^-}
=p.
\]
Structural equilibrium \(\PiV=\Pi_{\mathrm{mech}}^{\Sigma}\) and differentiation along the common-pressure family complete the proof.
\end{proof}

The result is exact within the stated specialization but is not a universal constitutive identity: curvature, normal recoil, anisotropic stresses, or another structural virtual mode require re-evaluation of the load and do not generally preserve this reduction.

\begin{remark}[Open-reservoir ensemble consistency in the planar geometric subclass]
The definition of \(\sigma\) as the area derivative at fixed total additive contents is also consistent with allowing both surface composition and structural volume to relax against bulk reservoirs. Restrict attention to the \emph{literal planar geometric-volume specialization}, where \(V^{\Sigma}\) is the volume swept by the structural thickness mode and \(pV^{\Sigma}\) is therefore the external hydrostatic-work term. At a bulk equilibrium state on the admissible coexistence manifold, let the reservoir intensives \(T\), \(p\), and \(\mu_{i}^{\mathrm m}\) have their constrained equilibrium values. They are not asserted to be independently tunable controls. Define
\[
\mathcal G_{p,\bm\mu^{\mathrm m}}^{\Sigma}
=F^{\Sigma}+pV^{\Sigma}-\sum_i\mu_{i}^{\mathrm m}N_i^{\Sigma}.
\]
At fixed area, stationarity with respect to the open surface amounts and the structural content gives
\[
\frac{\partial\mathcal G_{p,\bm\mu^{\mathrm m}}^{\Sigma}}{\partial N_i^{\Sigma}}
=\mu_{i}^{\Sigma,\mathrm m}-\mu_{i}^{\mathrm m}=0,
\qquad
\frac{\partial\mathcal G_{p,\bm\mu^{\mathrm m}}^{\Sigma}}{\partial V^{\Sigma}}
=-\PiV+p=0.
\]
If internal variables are also equilibrated, stationarity with respect to them adds \(\mathcal A_a^{\Sigma}=0\). Thus the reservoir potential reproduces both bulk--surface chemical equilibrium and the planar geometric structural condition \(\PiV=p\). At such a stationary state the Euler relation gives the identity
\[
\frac{\mathcal G_{p,\bm\mu^{\mathrm m}}^{\Sigma}}{A}
=a^{\Sigma}-\sum_i c_i^{\Sigma}\mu_{i}^{\mathrm m}+p\Vsurf
=\sigma.
\]
If the stationary surface state \((\bm N_*^{\Sigma},V_*^{\Sigma},\bm\xi_*^{\Sigma})\) varies differentiably with area at fixed reservoir state, the envelope theorem gives
\[
\left.\frac{\dd}{\dd A}
\mathcal G_{p,\bm\mu^{\mathrm m}}^{\Sigma}
\bigl(A,\bm N_*^{\Sigma}(A),V_*^{\Sigma}(A),\bm\xi_*^{\Sigma}(A)\bigr)
\right|_{T,p,\bm\mu^{\mathrm m}}
=
\left.\frac{\partial F^{\Sigma}}{\partial A}\right|_{T,\bm N^{\Sigma},V^{\Sigma},\bm\xi^{\Sigma}}
=\sigma.
\]
Local thermodynamic stability requires the corresponding second variation of this reservoir potential to be non-negative on the admissible open-system variations. In particular, with composition and internal variables frozen, its structural curvature is
\[
\left.\frac{\partial^2\mathcal G_{p,\bm\mu^{\mathrm m}}^{\Sigma}}{\partial (V^{\Sigma})^2}\right|_{A,T,p,\bm\mu^{\mathrm m},\bm N^{\Sigma},\bm\xi^{\Sigma}}
=-\frac{1}{A}
\left.\frac{\partial\PiV}{\partial\Vsurf}\right|_{T^{\Sigma},\bm c^{\Sigma},\bm\xi^{\Sigma}},
\]
consistent with the constrained convexity condition above. This is a consistency result for the planar geometric subclass, not a generic Legendre transform for every abstract patch-additive structural coordinate: the term \(pV^{\Sigma}\) uses the additional geometric identification of structural content with pressure work. Curved loading likewise requires the appropriate geometry-dependent external mechanical potential rather than the simple expression above.
\end{remark}

\begin{remark}[Closed-control-volume balance for literal planar volume]
The same condition \(\PiV=p\) can be obtained without introducing an open surface reservoir and also clarifies why the structural pressure work is not double counted. Consider a closed planar control volume in the literal geometric specialization with
\[
V_{\mathrm{tot}}=V^-+V^++V^{\Sigma}
\]
fixed. At fixed bulk constituent amounts and temperature, a structural variation satisfies \(\delta V^-+\delta V^++\delta V^{\Sigma}=0\). At common hydrostatic pressure \(p\), the two bulk Helmholtz contributions are
\[
\delta F^-+\delta F^+
=-p(\delta V^-+\delta V^+)
=p\,\delta V^{\Sigma},
\]
whereas the intrinsic surface structural contribution at fixed area and amounts is \(-\PiV\,\delta V^{\Sigma}\). Hence
\[
\delta F_{\mathrm{tot}}
=(p-\PiV)\,\delta V^{\Sigma},
\]
and stationarity gives \(\PiV=p\). The bulk and surface terms refer to complementary volume changes and are therefore not duplicate work contributions. Away from the planar common-pressure specialization, the same balance must retain the side on which volume is displaced and the corresponding one-sided mechanical loads.
\end{remark}

\section{Enriched Transfer and Coupled Jump Conditions}
\label{sec:transfer}

\subsection{One-Sided Transfer and Structural Dissipation}

This appendix supplies the detailed one-sided transfer extension of the enriched framework. It separates independent structural relaxation, transfer-induced structural production, and holonomic state reduction. The simplified force--rate structure in \cref{sec:coupled-evolution} is recovered only under its declared thermal and mechanical assumptions.

The equilibrium manifold used above is the zero-resistance limit of a finite-rate transfer theory. It is therefore important to retain the complete one-sided cofactor before reducing it to a chemical-potential equality. This ordering follows the general Class-II strategy developed by Bothe and Dreyer, in which the constituent balance structure is retained until the entropy-production terms have been identified. Reduction is performed afterwards because premature summation of constituent momenta can eliminate the information needed to identify mechanical cofactors and cross-couplings \cite{BotheDreyer2015,Bothe2025ClassII}. Here $\mu_i^\alpha$ is mass-specific and $\dot m_i^{\alpha,\Sigma}$ is positive from bulk phase $\alpha$ into the surface, consistently with \cref{sec:notation}. The established Class-II one-sided cofactor has the structure \cite{Bothe2025ClassII}
\[
\mathcal F_{i,0}^{\alpha,\Sigma}
=
\frac{\mu_i^{\alpha}}{T^{\alpha}}
-
\frac{\mu_i^{\Sigma}}{T^{\Sigma}}
+
\frac1{T^{\Sigma}}
\left[
\frac12\left|\bm v_i^{\alpha}-\bm v_i^{\Sigma}\right|^2
-
\frac{
\bm n^{\alpha}\cdot
\bm S_i^{\alpha,\mathrm{visc}}
\bm n^{\alpha}
}{\rho_i^{\alpha}}
\right].
\]
Here \(\bm S_i^{\alpha,\mathrm{visc}}\) is the irreversible partial-stress contribution in the stated Class-II stress decomposition. The outward-normal convention is that of \cref{sec:equilibrium}. This quotient representation applies to transfer processes with \(\rho_i^{\alpha}>0\). At a vanishing partial density, a limiting constitutive prescription is required. The displayed formula alone does not define the boundary value. The corresponding entropy-production contribution is
\[
\zeta_{\mathrm{tr}}^{\Sigma}
=
\sum_{\alpha=\pm}\sum_i
\dot m_i^{\alpha,\Sigma}
\mathcal F_{i,0}^{\alpha,\Sigma}.
\]
The normal mechanical term is not an optional correction. It results from the same mass, momentum, energy, and entropy transmission structure that produces the chemical-potential and kinetic terms. In a Class-I reduction the organization of stresses and energy fluxes changes, but a normal irreversible-traction contribution remains in the complete cofactor \cite{Bothe2022}.

A finite-rate closure may be linear near equilibrium or nonlinear farther away.

Structural-variable elimination already provides one route to nonlinear effective resistivities \cite{StruchtrupOttinger2023}. For liquid--vapor transfer, Struchtrup \cite{Struchtrup2026} extends Hertz--Knudsen--Schrage relations to nonideal vapor properties and nonlinear mass/heat transfer. These kinetic closures do not follow from a surface free energy alone, and their thermal and vapor-regime assumptions must be retained when they are used to close the present balances. For kinetically accessible transfer processes with coercive exchange kinetics, bounded rates in the zero-resistance limit require the complete cofactor to vanish. Chemical-potential equality is recovered under common temperature and vanishing relative kinetic-energy and irreversible normal-stress corrections.

At the primitive enriched level, one-sided mass transfer and structural-volume relaxation are independent irreversible processes. Before factorizing the structural mechanical power, their contribution is
\[
T^{\Sigma}\zeta_{\mathrm{tr+V}}^{\Sigma}
=
T^{\Sigma}\sum_{\alpha,i}
\dot m_i^{\alpha,\Sigma}
\mathcal F_{i,0}^{\alpha,\Sigma}
+
\PiV r_{\mathcal V}^{\Sigma}
-
\mathcal W_{\mathcal V,\mathrm{mech}}^{\Sigma}.
\]
For a single factorized structural mode, \(\mathcal W_{\mathcal V,\mathrm{mech}}^{\Sigma}=\Pi_{\mathrm{mech}}^{\Sigma}r_{\mathcal V}^{\Sigma}\), this reduces to separate binary products for transfer and structural relaxation. The flux-force pairs may then be kinetically cross-coupled without being kinematically identified. Linearization about an \emph{interior} equilibrium with positive surface amount densities and \(\Vsurf_*>0\) may be written schematically as
\[
\begin{pmatrix}
\dot{\bm m}\\[0.3em]
r_{\mathcal V}^{\Sigma}
\end{pmatrix}
=
\bm L
\begin{pmatrix}
\bm{\mathcal F}_0\\[0.3em]
(\PiV-\Pi_{\mathrm{mech}}^{\Sigma})/T^{\Sigma}
\end{pmatrix}.
\]
This is a local tangent-space closure, not a global kinetic law on the boundary of the constitutive state domain defined by the positivity constraints. A coupled constitutive relation extended to the full admissible state domain must preserve all admitted positivity constraints. In particular, when tangential structural transport is absent its total structural component must satisfy \(r_{\mathcal V}^{\Sigma}\ge0\) at \(\Vsurf=0\). Cross-coupling coefficients admissible in an interior Onsager linearization may therefore be inadmissible at the boundary. For this linearized schematic closure the entropy principle requires only the symmetric part
\[
\bm L_{\mathrm s}
=\frac12(\bm L+\bm L^{\mathsf T})
\succeq0
\]
on the admissible force space. An antisymmetric part is nondissipative. Onsager or Onsager--Casimir symmetry supplies additional restrictions only when the corresponding microscopic reversibility and time-reversal-parity hypotheses apply. Thermodynamic equilibrium is characterized by vanishing independent transfer and structural forces. If \(\bm L_{\mathrm s}\) is positive definite on the relevant force space, zero entropy production implies this locally. With only semidefiniteness, zero entropy production may also occur along dissipation-null force directions, which can carry nondissipative rates and must be distinguished from equilibrium.

The primitive representation makes every mechanical-power contribution explicit before reduction. In particular, normal traction work already present in \(\mathcal F_{i,0}^{\alpha,\Sigma}\) must not be counted a second time in the structural power.

\subsection{Rate-Level Transfer-Induced Volume}

A scalar specialization allows part of the structural production to be induced kinematically by transfer,
\[
r_{\mathcal V}^{\Sigma}
=
r_{\mathcal V,\mathrm{rel}}^{\Sigma}
+
\sum_{\alpha,i}
\upsilon_{i\mathcal V}^{\alpha,\Sigma}
\dot m_i^{\alpha,\Sigma}.
\]
The coefficient \(\upsilon_{i\mathcal V}^{\alpha,\Sigma}\) has dimensions of specific volume, m$^3$\,kg$^{-1}$. It specifies the increment of phenomenological surface volume per unit mass transferred through the corresponding side. It is not, at this level, identified with a bulk specific volume or a surface partial specific volume. This rate-level decomposition is likewise an interior constitutive specialization unless supplemented by a domain-preserving boundary rule. If \(\Vsurf=0\) is admitted, the \emph{total} right-hand side---including both the relaxing and transfer-induced parts---must obey the inward-pointing condition required by the structural balance. Arbitrary choices of the \(\upsilon_{i\mathcal V}^{\alpha,\Sigma}\) do not guarantee that property.

The mechanical power must be decomposed consistently with the same kinematics. Let the independently relaxing part work against \(\Pi_{\mathrm{mech,rel}}^{\Sigma}\), while the transfer-induced volume increment associated with \((\alpha,i)\) works against a side- and species-specific load \(\pi_{i\mathcal V}^{\alpha,\Sigma}\):
\[
\mathcal W_{\mathcal V,\mathrm{mech}}^{\Sigma}
=
\Pi_{\mathrm{mech,rel}}^{\Sigma}
 r_{\mathcal V,\mathrm{rel}}^{\Sigma}
+
\sum_{\alpha,i}
\pi_{i\mathcal V}^{\alpha,\Sigma}
\upsilon_{i\mathcal V}^{\alpha,\Sigma}
\dot m_i^{\alpha,\Sigma}.
\]
Regrouping the entropy production then gives
\[
\zeta_{\mathrm{tr+V}}^{\Sigma}
=
\sum_{\alpha,i}
\dot m_i^{\alpha,\Sigma}
\mathcal F_{i,\mathrm{enr}}^{\alpha,\Sigma}
+
\frac{\PiV-\Pi_{\mathrm{mech,rel}}^{\Sigma}}{T^{\Sigma}}
 r_{\mathcal V,\mathrm{rel}}^{\Sigma},
\]
where
\[
\mathcal F_{i,\mathrm{enr}}^{\alpha,\Sigma}
=
\mathcal F_{i,0}^{\alpha,\Sigma}
+
\frac{\upsilon_{i\mathcal V}^{\alpha,\Sigma}}{T^{\Sigma}}
\left(
\PiV-\pi_{i\mathcal V}^{\alpha,\Sigma}
\right).
\]
The additional term gives the structural contribution to the one-sided transfer cofactor. The enlargement follows from an additional rate-level kinematic specialization. It is not forced merely by including \(\Vsurf\) in the surface state.

For a literal one-sided thickness increment, a geometrically consistent choice is
\[
\pi_{i\mathcal V}^{\alpha,\Sigma}=\pi_n^{\alpha}.
\]
In a one-component simple-fluid specialization, take additionally \(\upsilon_{\mathcal V}^{\alpha,\Sigma}=1/\rho^{\alpha}\) and split the total bulk stress as \(\bm S^{\alpha}=-p^{\alpha}\bm I+\bm S^{\alpha,\mathrm{visc}}\). Since \(\pi_n^{\alpha}=p^{\alpha}-\bm n^{\alpha}\!\cdot\bm S^{\alpha,\mathrm{visc}}\bm n^{\alpha}\), the primitive viscous normal-traction term and the transfer-induced structural term combine exactly as
\[
-\frac{\bm n^{\alpha}\!\cdot\bm S^{\alpha,\mathrm{visc}}\bm n^{\alpha}}{\rho^{\alpha}T^{\Sigma}}
+\frac{\PiV-\pi_n^{\alpha}}{\rho^{\alpha}T^{\Sigma}}
=\frac{\PiV-p^{\alpha}|_{\Sigma}}{\rho^{\alpha}T^{\Sigma}}.
\]
This cancellation is therefore exact under the stated one-component stress split, but it is not a multicomponent identity. In a Class-II mixture the primitive cofactor contains partial stresses divided by partial densities, whereas a structural volume may couple to partial molar or partial specific volumes. The relation must be specified by the mechanical constitutive model for the mixture.

\subsection{Holonomic Amount--Volume Constraint}
\label{sec:holonomic}

A holonomic state constraint differs from kinetic
cross-coupling or a rate-level decomposition. For a homogeneous surface
patch, consider the particular constraint class
\[
V^{\Sigma}=\mathscr V(N_1^{\Sigma},\ldots,N_N^{\Sigma}),
\qquad
\mathscr V(\lambda\bm N^{\Sigma})
=\lambda\mathscr V(\bm N^{\Sigma})
\quad(\lambda>0),
\]
with \(\mathscr V\ge0\) on its admissible amount domain. Assume that the constraint map and the primitive free energy are \(C^2\) on their interior state domains. All constrained states below are restricted to portions of the constraint graph lying in the primitive free energy's \(C^2\) state domain. Equivalently,
the local density relation is
\(\Vsurf=\phi(\bm c^{\Sigma})\), where \(\phi\) is degree-one
homogeneous. This choice is extensive and preserves the primitive area
conjugate under reduction. For a heterogeneous surface patch \(\omega^{\Sigma}\),
patch additivity is enforced locally by
\[
V^{\Sigma}(\omega^{\Sigma})=\int_{\omega^{\Sigma}}\phi(\bm c^{\Sigma}(\bm x))\,\dd A.
\]
Degree-one homogeneity alone does not imply that this integral equals
\(\mathscr V(\int_{\omega^{\Sigma}}\bm c^{\Sigma}\,\dd A)\) when compositions vary.
Requiring the same aggregate formula for arbitrary unions of
differently composed patches would additionally impose
\(\mathscr V(\bm N+\bm M)=\mathscr V(\bm N)+\mathscr V(\bm M)\).
Under the stated differentiability this restricts the aggregate map to a
linear form with constant partial structural volumes. Nonlinear
degree-one relations remain admissible local constitutive constraints, with
the preceding integral providing their additive extension.

With
\(
\bar v_i^{\Sigma}:=\partial V^{\Sigma}/\partial N_i^{\Sigma}
\)
for an extensive differentiable constraint on the interior amount domain, Euler's theorem and differentiation give
\[
V^{\Sigma}=\sum_i\bar v_i^{\Sigma}N_i^{\Sigma},
\qquad
\Vsurf=\sum_i c_i^{\Sigma}\bar v_i^{\Sigma},
\]
and
\[
\dd V^{\Sigma}
=
\sum_i\bar v_i^{\Sigma}\dd N_i^{\Sigma}.
\]
The partial structural volumes may depend on composition. The homogeneity condition, not constancy of the \(\bar v_i^{\Sigma}\), is what preserves extensivity. Differentiating the Euler identity and comparing it with the exact differential above also gives the associated integrability condition
\[
\sum_i N_i^{\Sigma}\,\dd\bar v_i^{\Sigma}=0
\]
for amount variations within this stated constraint. Thus the \(\bar v_i^{\Sigma}\) cannot be prescribed as mutually independent composition functions. At fixed internal descriptors \(\bm\xi^{\Sigma}\), reduction onto this constraint gives the total free-energy differential
\[
\dd F^{\Sigma}
=
-S^{\Sigma}\dd T
+\sigma\dd A
+
\sum_i
\left(
\mu_{i}^{\Sigma,\mathrm m}-\PiV\bar v_i^{\Sigma}
\right)
\dd N_i^{\Sigma}.
\]
The intrinsic constrained amount conjugate is therefore
\[
\widetilde\mu_{i}^{\Sigma,\mathrm m}
=
\mu_{i}^{\Sigma,\mathrm m}-\PiV\bar v_i^{\Sigma}.
\]
This intrinsic constrained chemical potential is the amount derivative of the reduced Helmholtz free energy. The complete exchange potential also includes the induced mechanical work. If a transfer event is kinematically tied to the structural increment, the accompanying mechanical work must be reduced on the same constraint. For a molar transfer increment with structural increment \(\bar v_i^{\Sigma}\), the corresponding mass-specific coefficient is \(\upsilon_{i\mathcal V}^{\alpha,\Sigma}=\bar v_i^{\Sigma}/M_i\). Under common temperature and vanishing relative kinetic-energy and irreversible normal-stress corrections, the reduced one-sided equilibrium combination is
\[
\mu_{i,\mathrm{ex}}^{\Sigma,\alpha,\mathrm m}
:=\widetilde\mu_{i}^{\Sigma,\mathrm m}
+\pi_{i\mathcal V}^{\alpha,\Sigma}\bar v_i^{\Sigma}
=\mu_{i}^{\Sigma,\mathrm m}
-\left(\PiV-\pi_{i\mathcal V}^{\alpha,\Sigma}\right)\bar v_i^{\Sigma}.
\]
In this specialization, zero resistance equates the bulk chemical potential with the \emph{mechano-chemical} constrained potential. Equality with \(\widetilde\mu_{i}^{\Sigma,\mathrm m}\) alone additionally requires zero mechanical work conjugate to the induced structural increment. Outside the stated thermal and mechanical specialization, the complete one-sided cofactor remains the exchange force.

\begin{proposition}[Holonomic reduction and mechanically loaded exchange]
\label{prop:holonomic-reduction}
Under the \(C^2\) degree-one amount--volume constraint and free-energy assumptions above, define the reduced free-energy density by restricting \(a^{\Sigma}\) to the constrained structural state. Then, on the interior amount domain,
\[
\dd \bar a^{\Sigma}
=-\eta^{\Sigma}\dd T^{\Sigma}
+\sum_i\widetilde\mu_{i}^{\Sigma,\mathrm m}\dd c_i^{\Sigma}
-\sum_a\mathcal A_a^{\Sigma}\dd\xi_a^{\Sigma},
\]
while the area conjugate satisfies
\[
\sigma
=\bar a^{\Sigma}-\sum_i c_i^{\Sigma}\widetilde\mu_{i}^{\Sigma,\mathrm m}.
\]
Consequently, the \emph{intrinsic reduced} Gibbs--Duhem relation has the amount-only algebraic form
\[
\dd\sigma
=-\eta^{\Sigma}\dd T^{\Sigma}
-\sum_i c_i^{\Sigma}\dd\widetilde\mu_{i}^{\Sigma,\mathrm m}
-\sum_a\mathcal A_a^{\Sigma}\dd\xi_a^{\Sigma}.
\]

For the literal planar geometric-volume subclass in a common imposed hydrostatic pressure \(p\), the mechanically appropriate open-system potential for a constrained surface patch is
\[
\mathcal G_{\mathrm{hol}}^{\Sigma}
=\bar F^{\Sigma}+pV^{\Sigma}(\bm N^{\Sigma})
-\sum_i\mu_{i}^{\mathrm{bulk},\mathrm m}N_i^{\Sigma}.
\]
At a regular equilibrium state, stationarity with respect to the surface amounts gives
\[
\widetilde\mu_{i}^{\Sigma,\mathrm m}+p\bar v_i^{\Sigma}
=\mu_{i}^{\mathrm{bulk},\mathrm m},
\]
or, in the primitive variables inherited from the enriched embedding,
\[
\mu_{i}^{\Sigma,\mathrm m}
+\left(p-\PiV\right)\bar v_i^{\Sigma}
=\mu_{i}^{\mathrm{bulk},\mathrm m}.
\]
Because the holonomic constraint removes independent variations of \(V^{\Sigma}\) at fixed amounts, it does not alone imply \(\PiV=p\). The mismatch \(p-\PiV\) may act as the generalized reaction enforcing the constraint. Imposing \(\PiV=p\) as an additional condition recovers primitive chemical-potential equality. Along a regular planar isothermal equilibrium branch with equilibrated internal variables,
\[
\frac{\dd\widetilde\sigma_{\mathrm{pl}}}{\dd p}
=\Vsurf
-\sum_i c_i^{\Sigma}
\frac{\dd\mu_{i}^{\mathrm{bulk},\mathrm m}}{\dd p}.
\]
Hence a holonomic constraint removes independent structural relaxation, but it does not in general remove the structural contribution to a mechanically loaded equilibrium pressure response.
\end{proposition}

\begin{proof}
At fixed area the constraint gives
\[
\dd\Vsurf
=\sum_i\bar v_i^{\Sigma}\dd c_i^{\Sigma}.
\]
Restricting the primitive differential of \(a^{\Sigma}\) to that tangent space therefore replaces \(\mu_{i}^{\Sigma,\mathrm m}\) by \(\mu_{i}^{\Sigma,\mathrm m}-\PiV\bar v_i^{\Sigma}=\widetilde\mu_{i}^{\Sigma,\mathrm m}\), proving the first relation. The homogeneous constraint also gives \(\Vsurf=\sum_i c_i^{\Sigma}\bar v_i^{\Sigma}\). Substitution into the primitive area Legendre relation yields
\[
\sigma
=\bar a^{\Sigma}
-\sum_i c_i^{\Sigma}
\left(\mu_{i}^{\Sigma,\mathrm m}-\PiV\bar v_i^{\Sigma}\right),
\]
which is the reduced Euler form. Differentiating it and using the reduced free-energy differential gives the displayed intrinsic reduced Gibbs--Duhem relation.

For the planar geometric subclass, differentiation of \(\mathcal G_{\mathrm{hol}}^{\Sigma}\) with respect to \(N_i^{\Sigma}\) at fixed \(T,A,p\) and at a fixed reservoir state gives \(\widetilde\mu_{i}^{\Sigma,\mathrm m}+p\bar v_i^{\Sigma}-\mu_{i}^{\mathrm{bulk},\mathrm m}\). This proves the mechano-chemical stationarity condition. Equivalently, if the same constrained problem is written in the primitive variables with a multiplier \(\lambda\) for \(V^{\Sigma}-V^{\Sigma}(\bm N^{\Sigma})=0\), variation in \(V^{\Sigma}\) gives \(\lambda=\PiV-p\), while variation in \(N_i^{\Sigma}\) gives \(\mu_{i}^{\Sigma,\mathrm m}-\mu_{i}^{\mathrm{bulk},\mathrm m}-\lambda\bar v_i^{\Sigma}=0\). Thus \(p-\PiV\) is precisely the generalized reaction carried by the holonomic constraint. Since an independently variable \(V^{\Sigma}\) is absent from the constrained state manifold, no separate free variation is available from which to infer \(\PiV=p\). That equality is an additional specialization, not a consequence of holonomic reduction. Finally, differentiating the reduced Gibbs--Duhem relation along the isothermal equilibrium branch gives
\[
\frac{\dd\widetilde\sigma_{\mathrm{pl}}}{\dd p}
=-\sum_i c_i^{\Sigma}
\frac{\dd}{\dd p}
\left(\mu_{i}^{\mathrm{bulk},\mathrm m}-p\bar v_i^{\Sigma}\right).
\]
The degree-one constraint implies \(\sum_i c_i^{\Sigma}\dd\bar v_i^{\Sigma}=0\) for amount variations at fixed temperature, so the terms proportional to \(p\dd\bar v_i^{\Sigma}\) cancel and \(\sum_i c_i^{\Sigma}\bar v_i^{\Sigma}=\Vsurf\) gives the stated result.
\end{proof}

\paragraph{Dependence on the enriched embedding.}
The primitive reaction multiplier is defined relative to a specified
enriched extension of the constrained energy. Replacing the
primitive density by
\[
a_{\mathrm{new}}(\bm c,q)
=a(\bm c,q)+b(\bm c,q)[q-\phi(\bm c)]
\]
with smooth $b$ leaves the restricted energy unchanged. On the constraint
it gives
\[
\mu_{i,\mathrm{new}}^{\mathrm m}=\mu_{i}^{\mathrm m}-b\phi_{c_i},
\qquad \Pi_{\mathrm{new}}=\Pi-b.
\]
Consequently, both
$\widetilde\mu_{i}^{\mathrm m}=\mu_{i}^{\mathrm m}-\Pi\phi_{c_i}$ and
$\mu_{i,\mathrm{ex}}^{\mathrm m}=\mu_{i}^{\mathrm m}-(\Pi-p)\phi_{c_i}$ are invariant.
The quantity $p-\Pi$ therefore represents the constraint reaction in the
selected enriched embedding. It is not an additional observable of the
holonomic theory. The equilibrium condition is the invariant loaded
combination. For a degree-one constraint, the primitive area conjugate is invariant as well, since its change is \(b(\sum_i c_i\phi_{c_i}-q)=0\) on the constraint.

\paragraph{Local compatibility of the constrained balances.}
The state restriction also constrains structural production and transport.
Let $\phi$ be $C^2$ on an open cone of positive surface amount densities and let all fields and
fluxes be sufficiently smooth for the following local derivatives. For
this calculation suppress surface superscripts, write $D_t=D_t^{\bm u}$, and set $q=\phi(\bm c)$,
$v_i=\phi_{c_i}$, and write the molar balances as
\[
D_t c_i+c_i\theta+\divs\bm j_i=S_i,
\qquad \theta=\divs\bm u.
\]
Here $S_i$ includes the retained molar exchange and reaction supplies, and
$\bm j_i$ is a tangential molar flux relative to the selected transport
field $\bm u$. The structural balance implies the exact compatibility
identity
\[
r_q=\sum_i v_i S_i-\sum_i v_i\divs\bm j_i
 +\left(\phi-\sum_i c_i v_i\right)\theta
 +\divs\bm J_q.
\]
For the degree-one subclass the dilation coefficient vanishes. Choosing
the induced structural flux
\[
\bm J_q=\sum_i v_i\bm j_i
\]
then gives
\[
r_q=\sum_i v_iS_i+\sum_i\bm j_i\cdot\grads v_i.
\]
This follows directly from the chain rule and the product rule. The
second term generally survives for composition-dependent partial
structural volumes. A source consisting only of the amount supplies is
recovered for constant partial structural volumes or under an additional
homogeneous zero-diffusion restriction. After imposing the holonomic
constraint, $r_q$ is determined by this compatibility condition and cannot
also be assigned an independent relaxation relation.

The same substitution is consistent with the isothermal available-free-energy
flux when non-additive descriptors are absent or materially fixed. For constant planar load $p$, set $A_q=\Pi-p$ and
$\mu_{i,\mathrm{load}}^{\mathrm m}=\mu_{i}^{\mathrm m}-A_qv_i$. The primitive flux
$\sum_i\mu_{i}^{\mathrm m}\bm j_i-A_q\bm J_q$ becomes
$\sum_i \mu_{i,\mathrm{load}}^{\mathrm m}\bm j_i$. Moreover, with
$\Delta_\phi=\phi-\sum_i c_i v_i$, the local identity reduces to
\begin{align*}
D_t\bar f+\bar f\theta
 +\divs\left(\sum_i\mu_{i,\mathrm{load}}^{\mathrm m}\bm j_i\right)
={}&(\sigma-A_q\Delta_\phi)\theta
 +\sum_i\mu_{i,\mathrm{load}}^{\mathrm m}S_i\\
&+\sum_i\bm j_i\cdot\grads\mu_{i,\mathrm{load}}^{\mathrm m},
\end{align*}
where
\[
\bar f=a(\bm c,\phi(\bm c))+p\phi(\bm c).
\]
Thus no additional independent available-free-energy flux is introduced
by the restriction. The structural flux and its power are transformed
with the same constraint. This is an isothermal free-energy identity.
A nonisothermal energy and entropy reduction must additionally transform
the heat and entropy fluxes and the complete mechanical power.

\begin{remark}[Scope of the degree-one holonomic subclass]
A general smooth local constraint
\(\Vsurf=\phi(\bm c^{\Sigma})\) can also define an additive content
by local integration without degree-one homogeneity. For a homogeneous
patch it gives \(V^{\Sigma}=A\phi(\bm N^{\Sigma}/A)\), which is
extensive jointly in \((A,\bm N^{\Sigma})\). In that more general class,
however,
\[
\left.\frac{\partial V^{\Sigma}}{\partial A}\right|_{\bm N^{\Sigma}}
=\phi-\sum_i c_i^{\Sigma}\phi_{c_i},
\]
so the intrinsic constrained area conjugate is
\[
\bar\sigma
=\sigma-\PiV\left(\phi-\sum_i c_i^{\Sigma}\phi_{c_i}\right).
\]
For the literal planar common-pressure realization, differentiation of
the mechanically loaded potential on the same constraint gives
\[
\begin{aligned}
\sigma_{\mathrm{load}}
&=\left.\frac{\partial}{\partial A}
\left[\bar F+pA\phi(\bm N/A)\right]\right|_{T,p,\bm N}
\\
&=\bar\sigma+p\left(\phi-\sum_i c_i\phi_{c_i}\right)
=\sigma+(p-\Pi)
 \left(\phi-\sum_i c_i\phi_{c_i}\right).
\end{aligned}
\]
The degree-one hypothesis removes the additional area term exactly.
For a general constraint, the intrinsic constrained area derivative and
the mechanically loaded area derivative must therefore be distinguished.

Degree-one homogeneity is a substantive specialization in \cref{prop:holonomic-reduction}, not a consequence of patch additivity. The loading potential must be reduced on the same constraint before interpreting a measured tension.
\end{remark}

In the pure/incompressible planar benchmark, where \(\dd\mu_{i}^{\mathrm{bulk},\mathrm m}/\dd p=v_i^{\mathrm{bulk}}\), the constrained response becomes
\[
\frac{\dd\widetilde\sigma_{\mathrm{pl}}}{\dd p}
=\sum_i c_i^{\Sigma}
\left(\bar v_i^{\Sigma}-v_i^{\mathrm{bulk}}\right).
\]
Its sign is therefore not fixed by positivity of the surface amount densities alone. This does not contradict Proposition~\ref{prop:amount-only-sign}. The holonomic model retains additional structural-mechanical information through the amount--volume relation and its pressure work, even though that information no longer appears as an independent state coordinate.

This constitutive elimination must not be confused with solving the equilibrium equation for an originally independent structural coordinate and only then substituting the equilibrium solution into a reduced observable. In the latter case the solved state may take the form
\[
\Vsurf_* = \Vsurf_*(T^{\Sigma},\bm c^{\Sigma},\Pi_{\mathrm{mech}}^{\Sigma},\ldots),
\]
so pressure, curvature, recoil, or other mechanical loading information remains in the reduced equilibrium map. Such equilibrium elimination does not convert the primitive theory into the holonomic amount-only state of Proposition~\ref{prop:holonomic-reduction}. The distinction is between removing a coordinate from the constitutive state space and eliminating an independent coordinate only after its own equilibrium condition has been imposed.

In contrast, the rate-level constitutive relation of the preceding subsection leaves \(\Vsurf\) independent. An algebraic work product shared by the two reductions does not identify their state spaces.

\subsection{Joint Use of the Jump Balances}

The transfer cofactor is only one part of the jump-condition system. The normal mechanical quantities appearing in it are constrained by the momentum jump balance. The energy jump balance determines how mechanical power and transmitted enthalpy are partitioned. Tangential slip constitutive jump relations must be combined with the tangential momentum jump balance. This coupled use of jump conditions is essential for avoiding apparent constitutive freedom that disappears once the balances are enforced.

The normal part may be represented abstractly as
\[
\mathcal J_n
\left(
\pi_n^-,\pi_n^+,\sigma,\kappas,
\dot m_i^{\pm,\Sigma},\bm v_i^{\pm},\bm v_i^{\Sigma},\ldots
\right)=0.
\]
For static equilibrium in the first-order isotropic surface subclass with no residual geometric or shape-constraint force, it reduces to the Laplace-type relation of \cref{sec:equilibrium}. Additional geometric or constraint forces must be retained in the normal balance whenever the chosen mechanics supplies them. With finite transfer, recoil and momentum-flux terms enter. The existing Class-I derivation shows explicitly that the traction expression for the transfer cofactor can be rewritten as an equivalent pressure--curvature--recoil expression by using the normal momentum jump balance. Setting the irreversible normal traction to zero is a different constitutive approximation and must not be confused with that algebraic elimination.

The structural Euler contribution \(\PiV\Vsurf\) in \eqref{eq:general-enriched-euler} modifies normal mechanics even when the primitive transfer cofactor is not kinematically enlarged. Conversely, a constitutive relation for transfer-induced structural-volume production places the structural force directly in an enlarged transfer cofactor. A holonomic amount--volume constraint changes the thermodynamic exchange potential itself. These are three different routes and must remain distinguishable.

To avoid counting a mechanical-power term twice, the enriched mass, momentum, and energy jump balances are first retained together with the structural balance and its additional mechanical power. The entropy inequality is then evaluated to identify the primitive binary products. Only then are the jump balances used to derive algebraically equivalent reduced cofactors. A geometric, rate-level, or constrained-state specialization is imposed only after these steps. This is the Class-II procedure of retaining the constituent balance structure until the thermodynamic role of each mechanical contribution has been determined.

\section*{Statements and Declarations}





\paragraph*{Use of generative AI}
ChatGPT (OpenAI) was used to assist with literature searches, checks of mathematical and thermodynamic arguments, numerical verification, and manuscript and LaTeX editing. Responsibility for the scientific content and the final manuscript rests with the author.

\end{document}